\documentclass[traditabstract,a4paper]{aa}

\usepackage{graphicx}
\usepackage{epsfig}
\usepackage{boxedminipage}
\usepackage{rotating}
\usepackage[varg]{txfonts}
\usepackage{natbib}
\usepackage{amssymb}
\usepackage{flafter}
\usepackage{fixltx2e}

\def\Msun{{\rm\thinspace M_{\odot}}}

\bibpunct{(}{)}{;}{a}{}{,}

\begin{document}

\title{The Astrophysical Multipurpose Software Environment}

\author{F.I. Pelupessy\inst{1}\thanks{pelupes@strw.leidenuniv.nl} \and
        A. van Elteren\inst{1} \and
        N. de Vries\inst{1} \and
        S.L.W. McMillan\inst{2} \and \\
        N. Drost\inst{3} \and 
        S.F. Portegies Zwart\inst{1}
        }

\institute{Leiden Observatory, Leiden University, PO Box 9513, 2300 RA, Leiden, The Netherlands \and 
           Department of Physics, Drexel University, Philadelphia, PA 19104, USA \and
           Netherlands eScience Center, Science Park 140, Amsterdam, The Netherlands 
           }
           
\date{\today}

\abstract{
We present the open source Astrophysical Multi-purpose Software 
Environment (AMUSE, www.amusecode.org), a component library for 
performing astrophysical simulations involving different physical 
domains and scales. It couples existing codes within a Python framework 
based on a communication layer using MPI. The interfaces are 
standardized for each domain and their implementation based on MPI guarantees 
that the whole framework is well-suited for distributed computation. It 
includes facilities for unit handling and data storage. Currently it 
includes codes for gravitational dynamics, stellar evolution, 
hydrodynamics and radiative transfer. Within each domain the interfaces to 
the codes are as similar as possible. We describe the design and 
implementation of AMUSE, as well as the main components and community codes 
currently supported and we discuss the code interactions facilitated 
by the framework. Additionally, we demonstrate how AMUSE can be used 
to resolve complex astrophysical problems by presenting example 
applications.
}


\titlerunning{AMUSE}
\authorrunning{F.I. Pelupessy \textit{et al.}} 

\maketitle
 
\section{Introduction\label{sec:intro}}

Astrophysical simulation has become an indispensable tool to understand 
the formation and evolution of astrophysical systems. Problems ranging 
from planet formation to stellar evolution and from the formation of 
galaxies to structure formation in the universe have benefited from the 
development of sophisticated codes that can simulate the physics 
involved. The development of ever faster computer hardware, as `mandated' 
by Moore's law, has increased the computational power available 
enormously, and this trend has been reinforced by the emergence of 
Graphic  Processing Units as general purpose computer engines precise 
enough for scientific applications.  All this has meant that codes are 
able to handle large realistic simulations and generate enormous amounts 
of data.

However, most of the computer codes that are used have been developed 
with a  particular problem in mind, and may not be immediately usable 
outside the domain of that application. For example, an N-body code may 
solve the gravitational dynamics of hundreds of thousand of stars, but 
may not include algorithms for stellar evolution or the dynamics of the 
gas between the stars. The development of the latter needs a whole 
different set of expertise that developers of the former may 
not have. This presents a barrier as the trend is to move beyond 
solutions to idealized problems to more realistic scenarios that take 
into account more complex physical interactions. This is not 
restricted to astrophysics: fields as diverse as molecular dynamics, 
aerospace engineering, climate and earth system modelling are 
experiencing a similar trend towards multiphysics simulation \citep
{Groen2012}.

Within astrophysics, different ways of combining physical solvers are 
being pursued: examples are collaborations on large monolithic codes 
such as Flash~ \citep{Fryxell2000} or GADGET~\citep{Springel2001, 
Springel2005}, or the aggregation of codes in heterogeneous  packages, 
e.g. NEMO~\citep{Teuben1995}. These approaches are limited by the effort 
needed to assemble and maintain evolving software packages (for 
monolithic codes) or by the lack of integration of their component  
software base (in the case of code packages).  More recently the MUSE 
\citep {PortegiesZwart2009,PortegiesZwart2013b} framework was developed to 
overcome these problems by binding existing codes into a modern and 
flexible scripting language. 

\begin{figure*}
 \centering
 \epsfig{file=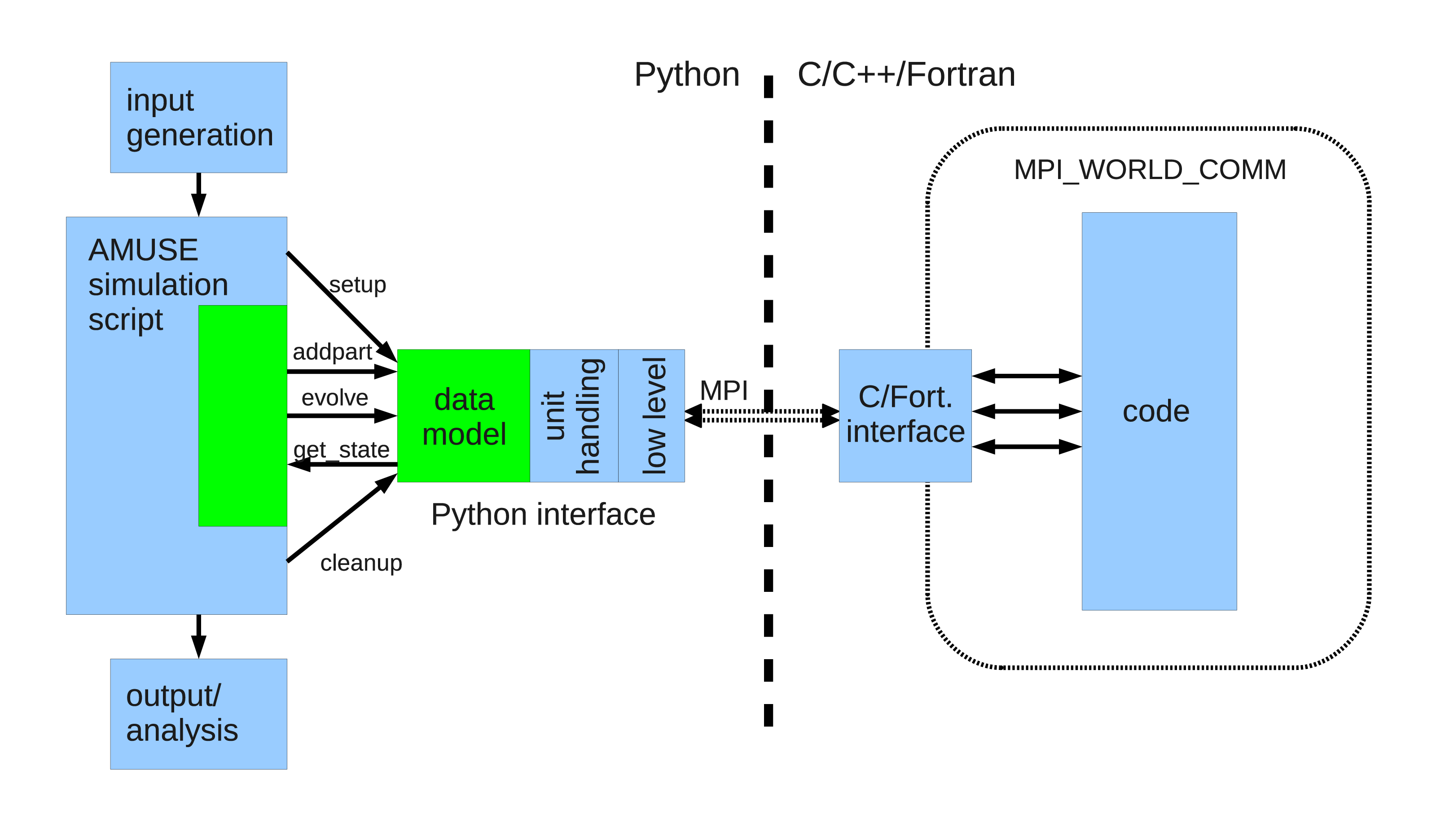,width=.9\textwidth}
 \caption{Design of the AMUSE interface. This diagram represents the way in
which a community code (``code'') is accessed from the AMUSE framework. The
code has a thin layer of interface functions in its native language which
communicate through an MPI message channel with the Python host process. On
the Python side the user script (``AMUSE simulation script'')
only accesses generic calls (``setup,'' ``evolve'' etc.) to a high level 
interface. This high level interface calls the low level interface functions,
hiding details about units and the code implementation. The communication
through the MPI channel does not interfere with the code's own
parallelization because the latter has its own MPI\_WORLD\_COMM context.}
 \label{fig:interface}
\end{figure*}

AMUSE is an astrophysical implementation of the general principles of 
MUSE~\citep{PortegiesZwart2013b}. It presents a coherent interface and 
simplified access to astrophysical codes allowing for multi-physics 
simulations. The main characteristics of MUSE that make this possible are:
\begin{itemize}
\item{physical coupling:} by defining physically based interfaces
  to the modules, AMUSE allows tight coupling between them, 
\item{consistency over domains:} AMUSE has consistent
  data handling and consistent interfaces over different domains,
\item{unit module:} AMUSE aims to be transparent with regard to
  the physical units employed by the different simulation codes,
\item{IO facilities:} AMUSE includes file IO and converters for 
  legacy formats. Code specific input and output routines are
  bypassed,
\item{error handling:} codes can detect that a system evolves outside their 
  domain of applicability, and can alert the framework to this fact (The 
  capabilities of AMUSE handling code crashes - another MUSE requirement - 
  is limited at the moment),
\item{multiple domains:} MUSE does not define the actual application domain. 
  The current capabilities of AMUSE are focused on four domains of astrophysics:  
  gravitational dynamics, stellar evolution, hydrodynamics and radiative 
  transfer.
\end{itemize}
In section~\ref{sec:arch} we present a technical overview of the design 
and architecture of AMUSE. The astrophysical domains and included codes 
are presented in section~\ref{sec:mod}. The use of AMUSE for 
multi-physics astrophysical simulations is described in section~\ref
{sec:couple}. Applications and ongoing research are presented in section~
\ref{sec:res}. We discuss the performance and testing aspects and some 
of the present limitations of AMUSE in section \ref{sec:disc}.

\section{Design and Architecture\label{sec:arch}}

Building multiphysics simulation codes becomes increasingly complex with 
each new physical ingredient that is added. Many monolithic codes grow  
extending the base solver with additional physics. Even when this 
is successful, one is presented with the prospect of duplicating much of 
this work when a different solver or method is added, a situation that 
often arises when a slightly different regime is accessed than originally 
envisaged or when results need to be verified with a different method.

In order to limit the complexity we need to compartmentalize the codes. 
The fundamental idea of AMUSE is the abstraction of the functionality of 
simulation codes into physically motivated interfaces that hide their 
complexity and numerical implementation. AMUSE presents the user with 
optimized building blocks that can be combined into applications for 
numerical experiments. This means that the requirement of the high level 
interactions is not so much performance but one of algorithmic 
flexibility and ease of programming, and thus the use of a modern 
interpreted scripting language with object oriented features suggests 
itself. For this reason we have chosen to implement AMUSE in Python. In 
addition, Python has a large user and developer base, and many libraries 
are available. Amongst these are libraries for scientific computation, 
data analysis and visualization. 

An AMUSE application consists roughly speaking of a \emph{user script}, 
an \emph{interface layer} and the \emph{community code base} \citep [][
fig \ref {fig:interface}]{PortegiesZwart2013b} . The user script is 
constructed by the user and specifies the initial data and the simulation 
codes to use. It may finish with analysis or plotting functions, in 
addition to writing simulation data to file. The setup and communication 
with the community code is handled by the interface layer, which 
consists of a communication interface with the community code as well as 
unit handling facilities and an abstract data model.

\subsection{Remote function interface\label{sec:mpi}}

\begin{figure*}
\centering
\begin{boxedminipage}{14cm}
{
\small
\begin{verbatim}
(1) gravity=PhiGRAPE()
(2) gravity=PhiGRAPE(number_of_workers=32)
(3) gravity=PhiGRAPE(mode="GPU")
(4) gravity=PhiGRAPE(channel_type="estep",hostname="remote.host.name")
\end{verbatim}
}
\end{boxedminipage}
\caption{
Starting a community code interface. (1) simple local startup, (2) 
startup of an MPI parallel worker with 32 worker processes, (3) startup 
of the community code with GPU support, (4) remote startup of worker.
}
\label{fig:workerstart}
\end{figure*}

The interface to a community code provides the equivalent functionality 
of importing the code as a shared object library. The default 
implementation in AMUSE is a remote function call protocol based on MPI. 
A community code is started by the straightforward instantiation of an 
interface object (fig.~\ref{fig:workerstart}) transparent to this. 
Python provides the possibility of linking directly Fortran or 
C/C++ codes, however we found that a remote protocol provides two 
important benefits: 1) built-in parallelism 2) separation of memory 
space and thread safety. The choice for an intrinsically parallel 
interface is much preferable over an approach where parallelism is added 
a posteriori, because unless great care is taken in the design, features 
can creep in that preclude easy parallelization later on. The second 
benefit comes as a bonus: during early development we found that many 
legacy simulation codes make use of global storage that makes it either 
impossible or unwieldy to instantiate multiple copies of the same code - 
using MPI interfaces means that the codes run as separate executables, 
and thus this problem cannot occur. An additional advantage of 
the interface design is that the MPI protocol can be easily 
replaced by a different method, two of which are available: a channel 
based on sockets and one based on the eSTeP library for distributed 
computing\footnote {this is the production environment version of a research
software project named Ibis.}. The sockets channel is currently mainly 
useful for cases were all the component processes are to be run on one 
machine. As its name implies it is based on standard TCP/IP sockets. The 
eSTeP channel is described below.

In practice the interface works as follows: when an instance of an 
imported simulation code is made, an MPI process is spawned as a 
separate process somewhere in the MPI cluster environment. This process 
consists of a simple event loop that waits for a message from the python 
side, executes the subroutines on the basis of the message ID and any 
additional data that may follow the initial MPI message, and sends the 
results back~\citep[see also][]{PortegiesZwart2013b}. The messages 
from the python script may instruct the code to perform a calculation 
(like evolving for a specified amount of time), or may be a request to 
send or receive data. Calls can be executed synchronously (with the 
python script waiting for the code to finish) or asynchronously. The 
simulation code itself will continue executing the event loop, 
maintaining its memory state, until instructed to stop.

The benefit of using MPI must be weighed against the disadvantages, of 
which the two most important are: 1) it is more involved to 
construct an MPI interface and 2) there is no shared memory space, and 
thus also no direct access to simulation data. However, linking  
legacy C or Fortran codes (using SWIG or f2py) is actually not 
straightforward either and this concern is further alleviated by 
automating much of the interface construction (in AMUSE the source code 
for the main program on the client side is generated by a Python script 
and all the actual communication code is hidden). This means that the 
codes of the interfaces actually become more similar for codes of 
different target languages. The second issue is potentially more 
serious: communication using MPI is less direct than a memory reference. 
As a consequence the interfaces must be carefully designed to ensure all 
necessary information for a given physical domain can be retrieved. 
Additionally, the communication requirements between processes must not 
be too demanding. Where this is not the case (e.g. when a strong 
algorithmic coupling is necessary) a different approach may be more 
appropriate.

An obvious additional concern is the correct execution of MPI parallel 
simulation codes: the communication with the python host must not 
interfere with the  communication specified in the code. This is 
guaranteed with the recursive parallelism mechanism in MPI-2. The 
spawned processes share a standard MPI\_WORLD\_COMM context, which 
ensures that an interface can be build around an existing MPI code with 
minimal adaptation (fig. \ref{fig:interface}). In practice, for the 
implementation of the interface for an MPI code one has to reckon with 
similar issues as for the stand-alone MPI application. The socket and 
eSTeP channels also accomodate MPI parallel processes.

\subsubsection{Distributed Computing}

Current computing resources available to researchers are more varied 
than simple workstations: clusters, clouds, grids, desktop grids, 
supercomputers and mobile devices complement stand-alone workstations, 
and in practice one may want to take advantage of this ecosystem, which 
has been termed Jungle computing~\citep{Seinstra2011}, to quickly scale 
up calculations beyond local computing resources. 

To run in a Jungle computing environment the eSTeP channel is available 
in AMUSE~\citep{Drost2012}. Instead of using MPI this channel connects 
with the eSTeP daemon to start and communicate with remote workers. This 
daemon is aware of local and remote resources and the middleware over 
which they communicate (e.g. SSH).  The daemon is started locally before 
running any remote code, but it can be re-used for different simulation 
runs. The eSTeP software is written in Java. Once the daemon is running, 
and a simulation requests a worker to be started, the daemon uses a 
deployment library to start the worker on a remote machine, executing 
the necessary authorization, queueing or scheduling. Because AMUSE 
contains large portions of C, C++, and Fortran, and requires a large 
number of libraries, it is not copied automatically, but it is assumed 
to be installed on the remote machine. A binary release is available for 
a variety of resources, such as clouds, that employ virtualization. With 
these modifications, AMUSE is capable of starting remote workers on any 
resource the user has access to, without much effort required from the 
user. In short, to use the distributed version of AMUSE one must (1) 
ensure that AMUSE is installed on all machines and (2) specify for each 
resource used some basic information, such as hostname and type of 
middleware,  in a configuration file. Once (1) and (2) are done, any 
AMUSE script can be distributed by simply adding properties to each 
worker instantiation in the script, specifying the channel used, as well 
as the name of the resource, and the number of nodes required for this 
worker (see figure \ref {fig:workerstart}). 

\subsubsection{AMUSE Job server\label{sec:jobserver}}

Python scripts can also be interfaced with AMUSE using either the MPI or 
sockets/eSTeP channel. One unintended but pleasant consequence of this 
is that it is very easy to run an AMUSE script remotely, and to build a 
framework to farm out AMUSE scripts over the local network machines - or 
using eSTeP over any machines connected to the internet - in order to 
marshall computing resources to quickly do e.g. a survey of runs. AMUSE 
includes a module, the JobServer, to expedite this.

\subsection{Unit conversion\label{sec:unit}}

\begin{figure*}
\centering
\begin{boxedminipage}{14cm}
{
\small
\begin{verbatim}
(1) m= 1 | units.MSun
(2) m= [ 1., 2.] | units.MSun
(3) def escape_velocity(mass,distance,G=constants.G):
      return sqrt(G*mass/distance)
    v=escape_velocity(m, 1.| units.AU)
(4) dt=1.| units.hour
    (v*dt).in_(units.km)
\end{verbatim}
}
\end{boxedminipage}

 \centering
\begin{boxedminipage}{14cm}
{
\small
\begin{verbatim}
(1) converter=nbody_system.nbody_to_si( 1. | units.MSun, 1.| units.AU )
(2) converter.to_si(1. | nbody_system.time)
(3) converter.to_nbody(1. | units.kms)
\end{verbatim}
}
\end{boxedminipage}
\caption{
The AMUSE units module. The upper panel illustrates the use of the unit 
algebra module. (1) definition of a scalar quantity using the $|$ 
operator, (2) definition of a vector quantity (3) use of units in 
functions and (4) conversion of units. Lower panel illustrates the use 
of a converter between N-body units and SI. (1) defines a converter from 
units with G=1 (implicit), $\Msun=1$ and AU=1 to SI units, (2) conversion to SI units, 
(3) conversion from SI units.
}
\label{fig:units}
\end{figure*}

Keeping track of different systems of units and the various conversion 
factors when using different codes quickly becomes tedious and prone to 
errors. In order to simplify the handling of units, a unit algebra module is 
included in AMUSE (figure~\ref{fig:units}). This module wraps standard 
Python numeric types or Numpy arrays, such that the resulting quantities 
(i.e. a numeric value together with a unit) can transparently be used as 
numeric types (see the function definition example in figure~\ref
{fig:units}).  Even high level algorithms, like e.g. ODE solvers, 
typically do not need extensive modification to work with AMUSE 
quantities. 

AMUSE \emph{enforces} the use of units in the high level interfaces. The 
specification of the unit dimensions of the interface functions needs to 
be done only once, at the time the interface to the code is programmed. 
Using the unit-aware interfaces, any data that is exchanged within 
modules will be automatically converted without additional user input, 
or - if the units are not commensurate - an error is generated. 

Often codes will use a system of units that is underspecified, e.g. most 
N-body codes use a system of units where $G=1$. Within AMUSE these codes 
can be provided with data in this system of units (N-body units) or, 
using a helper class specifying the scaling, in physical units. Figure~
\ref {fig:units} shows an example of such a \emph{converter}. Converters 
are often necessary when a code that uses scale-free units is coupled 
to another code which uses a definite unit scale (e.g. a gravitational 
dynamics code coupled to a stellar evolution code, where the latter 
has definite input mass scales). 

\subsection{Data model\label{sec:datam}}

\begin{figure*}
\centering
\begin{boxedminipage}{10cm}
{
\small
\begin{verbatim}
(1) cluster=new_plummer_model(100)
(2) cluster.x+=10| units.kpc
(3) cluster[0:5].velocity=0. | units.kms
(4) galaxy.add_particles( cluster )
(5) gravity.particles.add_particles( cluster )
(6) channel=gravity.particles.new_channel_to( cluster )
(7) channel.copy_attributes( ["position","velocity"] )
\end{verbatim}
}
\end{boxedminipage}
\caption{
Example usage of high level particles sets. (1) initialize a set, (2) 
shift particle positions, (3) indexing of particle sets, (4) joining two 
particle sets, (5) sending particle data to a code, (6) definition of 
an explicit channel from in-code storage to a particle set in memory (7)
update of particle attributes over the channel. Similar data 
structures are available for grids as needed for e.g. grid hydrodynamic 
solvers.
}
\label{fig:datamodel}
\end{figure*}

The low level interface works with ordinary arrays for input and output. 
While this is simple and closely matches the underlying C or Fortran 
interface, its direct use entails a lot of duplicated bookkeeping 
code in the user script. Therefore in order to simplify work with the 
code, a data model is added to AMUSE based on the construction of \emph{ 
particle sets}, see figure~\ref{fig:datamodel}. Here a  particle is a an 
abstract object with different attributes (eg. mass, position  etc) and 
a unique ID. The data model and the unit interface are combined in a 
wrapper to the plain (low level) interface. The particle sets also 
include support code for a set of utility functions commonly needed 
(e.g. calculation of center of mass, kinetic energy, etc). 

Particle sets store their data in Python memory space or reference 
the particle data in the memory space of the community code. A particle 
set with storage in the community code copies the data from the code 
transparently upon access. The data of a memory set can be synchronized 
explicitly by defining a \emph{channel} between the memory particle set 
and the code particle set ((6) and (7) in figure~\ref{fig:datamodel}). 
Hence, data is only transferred from a code by accessing the attributes 
of its in-code particle set, or by a copy operation of a channel between 
such a set and another set.

Subsets can be defined on particle sets without additional storage and 
new sets can be constructed using simple operations (addition, 
subtraction). The same methods of the parent set are available for 
the subsets (but only affecting the subset).

A similar datastructure is defined for grids. A Grid set consists of a 
description of the grid properties and a set of grid points with 
attributes.

\subsection{State Model\label{sec:state}}

\begin{figure}
 \centering
 \epsfig{file=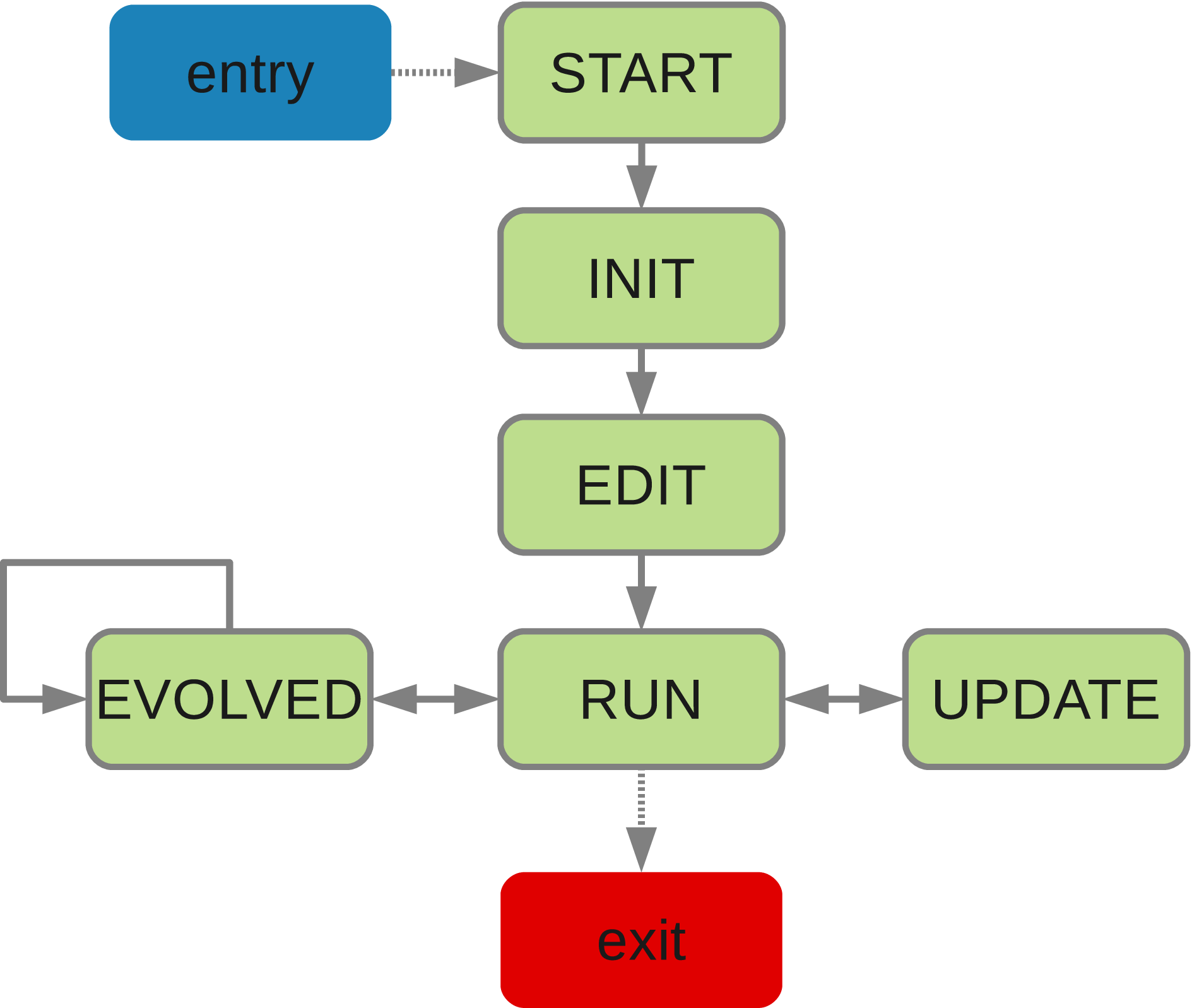, width=.49\textwidth}
 \caption{
 State model of gravitational dynamics codes. The diagram gives the state 
 that a gravitational dynamics code can be in. Transitions can be 
 effected explicitly (by calling the corresponding function, e.g. 
 \texttt{initialize\_code} from `start' to `init,' or implicitly (e.g. 
 calling \texttt{get\_position} to get the position of a particle is only 
 allowed in the `run' state - the framework will call the necessary 
 functions to get in that state, which will be a difference sequence of 
 functions depending on the starting state). 
 }
 \label{fig:state}
\end{figure}

When combining codes that use very different algorithms to represent the 
same physics one quickly runs into the problem that different codes have 
different work-flows. For example, the gravitational dynamics interface 
defines methods to add and remove particles and the calculation of 
gravitational forces obviously needs to track these changes. However, 
for, for example, a gravitational Barnes-Hut type tree code the data 
structure of the tree has to be updated before the new gravitational 
forces can be calculated. Such a tree update is an expensive operation. 
It is undesirable to add explicit tree update functions to the 
gravitational interface, as this functionality only makes sense for tree 
codes. One way to solve this would be program the interface code of 
the add-particle method of a tree code to update the tree each time, but 
this would be highly inefficient if a large group of particles is added to 
the code. The more efficient approach in this case is to flag the state 
of the code and do a tree update once all particle transactions have 
been completed. The only problem with this is that it is error prone if 
under control of the user. We have added facilities to AMUSE to keep 
track of the state a code is in, and to change state automatically if an 
operation is requested that is not allowed in the current state and to 
return an error if the state transition is not allowed. For the example 
above, this means that in the user script the addition and removal of 
particles is the same for all gravitational codes, but that in the case 
of a tree code an automatic call is made to the tree update routine once 
the user script proceeds to request, for example, a gravitational force.

The state model is flexible: states can be added and removed as required,
but in practice similar codes share similar state models. e.g. all the 
gravitational dynamics codes included in AMUSE conform to the state model 
with six states shown in fig.~\ref {fig:state}.

\subsection{Object oriented interfaces\label{sec:hlint}}

\begin{figure*}
\centering
\begin{boxedminipage}{10cm}
{
\small
\begin{verbatim}
 gravity=BHTree()
 gravity.initialize_code()
 gravity.set_eps2( 0.00155 )
 gravity.commit_parameters()
 for i in range(n):
   gravity.add_particle(mass[i],radius[i], 
           x[i],y[i],z[i],vx[i],vy[i],vz[i])
 gravity.commit_particles()
 gravity.evolve_model( 6.707 )
\end{verbatim}
}
\end{boxedminipage}

\centering
\begin{boxedminipage}{10cm}
{
\small
\begin{verbatim}
 gravity=BHTree()
 gravity.parameters.epsilon_squared=(1 | units.pc)**2 
 gravity.particles.add_particles(stars)
 gravity.evolve_model( 100. | units.Myr)
\end{verbatim}
}
\end{boxedminipage}
\caption{
Low level vs high level object oriented interface. The upper panel shows the 
schematic calling sequence for the low level interface of a gravity 
module, The lower panel shows the equivalent using the high level 
interface.
}
\label{fig:oointerface}
\end{figure*}

The object oriented, or high level, interfaces are the recommended 
way of interacting with the community codes. They consist of the low-level 
MPI interface to a code, with the unit handling, data model and state 
model on top of it. At this level the interactions with the code are 
uniform across different codes and the details of the code are hidden as 
much as possible. The difference in calling sequence is illustrated in 
figure~ \ref {fig:oointerface}. A lot of the bookkeeping 
(arrays/ unit conversion) is absent in the high level interface 
formulation. This makes the high level interface much easier to work 
with and less prone to errors: the user does not need to know what 
internal units the code is using, and does not need to remember the 
calling sequence nor the specific argument order of calls.

\subsection{IO conversion\label{sec:io}}

The community codes that are included into AMUSE normally contain 
subroutines to read in and write simulation data. This functionality is 
not used within AMUSE. Instead, all simulation data is written and read 
from within the AMUSE script (if tabulated data sets are needed by a 
community code, e.g. opacity tables, these are read by the code in the 
course of the initialization process). AMUSE includes a default output 
format based on hdf5\footnote{\texttt http://www.hdfgroup.org} that 
writes out all data pertaining to a data set. In order to simplify 
import and export of data, AMUSE contains a framework for generic I/O to 
and from different file formats. A number of common file formats  are 
implemented (Starlab, Gadget, Nemo), as well as generic table format 
file readers.

\subsection{Initial conditions and data analysis\label{sec:anic}}

Although not a main objective for AMUSE, some provisions are made to 
facilitate the generation or import of initial conditions and the 
analysis of data generated by a simulation. A number of N-body 
specific analysis tools are available by default (such as energy 
diagnostics, lagrangian radii, binary detection).

The generation of initial conditions for astrophysical simulations is 
often non-trivial, with specialized initial condition codes being used 
to construct initial density distributions or stellar structure models. 
As such, a limited selection of common initial condition generators is 
included in AMUSE. A number of the more common N-body particle 
generators (Plummer and King models), a selection of basic grid 
generators (random, regular cubic  grid and regular body centered cubic 
grids), useful for the generation of SPH initial conditions and zero age 
main sequence (ZAMS) stellar evolution model generators are available. 
This allows simple tests and experiments to be quickly developed. A number 
of more advanced initial condition generators, such as the GalactICS 
code for galaxy models~\citep{Widrow2008} or the FractalCluster code~
\citep {Goodwin2004}, are also interfaced.

After a simulation, the generated data needs to be analyzed. A limited 
collection of analysis tools is distributed with AMUSE. Python has good 
numerical and plotting libraries, such as Numpy and Matplotlib, 
available and data analysis can be easily incorporated into the AMUSE 
workflow. The nature of the data sets often means that visualization can 
be very helpful for the interpretation. Blender\footnote{ a free and 
open-source 3D computer graphics software package, \texttt
{www.blender.org}} is python scriptable and can be used to visualize 3D 
data sets. In addition, simple  OpenGL utilities can be coupled to the 
community codes to inspect the simulation data at runtime, especially 
useful in the development and debugging stage.

\subsection{Importing codes\label{sec:import}}

Bringing a new code into the framework involves a number of steps. The 
complete procedure (along with examples) is described in detail in the 
documentation section of the source distribution and the project website.
Here we outline the procedure.

To import a community code the code one first creates a directory in the 
AMUSE community code base directory with the name of the module. The 
original source tree is imported in a subdirectory (by convention named `
\texttt{src}'). The top-level directory also contains the Python side 
of the interface (`\texttt {interface.py}'), the interface in the 
native language of the code (e.g. `\texttt{interface.c}') and a file 
for the build system (`\texttt{Makefile}'). 

The Python interface (the file \texttt {interface.py}) typically defines 
two classes which define the low-level interface functions (as much as 
possible by inheritance from one of the base interface classes) and the 
high level interface. The low level interface then contains the function 
definitions of the calls which are redirected through the MPI 
communications channel (see section \ref{sec:mpi}) to the corresponding 
call defined in the native interface file (\texttt{interface.c}). The 
high level interface defines the units of the arguments of the function 
calls where appropriate (see section~\ref {sec:unit}). In addition it 
specifies the parameters of the code, the state model (section~\ref
{sec:state}) and the mapping of the object oriented data types to the 
corresponding low-level calls. By default the data of the simulation is 
maintained in the community code's memory (and accessed transparently as 
described in section~\ref{sec:datam}).

The modifications to the code itself (in `\texttt{src}') that are 
necessary fall in the following loosely defined categories: 1) In the most 
favourable case no modifications to the community code's source base are 
necessary and only the Python and native interfaces need to be added. 
For modern and modular codes this is often the case. 2) In addition to the 
interface code, it may be necessary to add a small library of helper functions, 
either to provide some secondary functionality that was not part of the community 
code (e.g. a gravity code may not have had functionality to change the number of 
particles in memory) or to reorganize previously existing code (e.g. the 
initialization, read-in and simulation loop might have been written in a 
single main program, where in AMUSE they need to be separated). 
3) or a code may need significant source code changes to implement functionality 
that was not present but required for the AMUSE interface (e.g. externally 
imposed boundary conditions for grid hydrodynamics).

It is clear that the actual effort needed to interface a community code 
increases as succesively more changes are necessary. Apart from this, it 
also becomes more difficult from the viewpoint of code maintenance: when 
a new version of a community code is brought into AMUSE, case 1) above 
may necessitate no additional steps beyond copying the updated version, 
while in case 3), the adaptations to the community code need to be 
carefully ported over.

\section{Component modules\label{sec:mod}}

The target domains for the initial release of AMUSE were gravitational 
dynamics, stellar evolution, hydrodynamics (grid and particle based) and 
radiative transfer. For a given physical domain, the community codes 
that are included in AMUSE share a similar interface that is tailored to 
its domain. An overview of all the codes that are included in the 
current release (version 7.1) is given in Table~\ref{tab:codes}. This 
selection of community codes is the result of a somewhat organic growth 
process where some codes were selected early on in the development of 
AMUSE to demonstrate proof of principle, others were contributed by 
their respective authors as a result of collaborations. We have taken 
care though that each domain is represented by at least 2 different 
codes, as this is the minimum to be able to conduct cross verification 
of simulations \citep[cf. ``Noah's Ark'' objective,][]{PortegiesZwart2009}. 
AMUSE can be extended by including other codes and domains, which can 
then be contributed to the AMUSE repository. 

\begin{table*}
\centering
\begin{tabular}{ l l l l}  
\hline
code     & language                    & short description           & main reference\\
\hline
Hermite0 & C++                         & Hermite N-body              & \cite{Hut1995} \\
PhiGRAPE & Fortran, MPI/GPU            & Hermite N-body              & \cite{Harfst2007}\\
ph4      & C++, MPI/GPU                & Hermite N-body              & McMillan, in prep.\\
BHTree   & C++                         & Barnes-Hut treecode         & \cite{Barnes1986}\\
Octgrav  & C++, CUDA                   & Barnes-Hut treecode         & \cite{Gaburov2010}\\
Bonsai   & C++, CUDA                   & Barnes-Hut treecode         & \cite{Bedorf2012} \\  
Twobody  & Python                      & Kepler solver               & \cite{Bate1971} \\
Huayno   & C, OpenMP/OpenCL            & Hamiltonian splitting       & \cite{Pelupessy2012c} \\
SmallN   & C++                         & regularized solver          & \cite{PortegiesZwart1999}\\
Mercury  & Fortran                     & symplectic planetary integrator    & \cite{Chambers1999} \\
Mikkola  & Fortran                     & relativistic regularization        & \cite{Mikkola2008} \\
MI6      & C++, MPI/GPU                & Hermite with Post-Newtonian terms  & \cite{Iwasawa2011}\\
Pikachu  & C++, CUDA                   & Hybrid Barnes-Hut/Hermite          & Iwasawa et al., in prep.\\
Brutus   & C++, MPI                    & arbitrary precision Bulirsch-Stoer & Boekholt \& Portegies Zwart, in prep. \\
HiGPUs   & C++, CUDA                   & Hermite N-body                     & \cite{Capuzzo2013} \\
Tupan    & Python, OpenCL              & Symplectic N-body, Post-Newtonian  & Ferrari, in prep. \\
MMC      & Fortran                     & Monte-Carlo gravitational dynamics & \cite{Giersz2006}\\
\hline
SSE      & Fortran                     & stellar evolution fits             & \cite{Hurley2000}\\
Evtwin   & Fortran                     & Henyey stellar evolution           & \cite{Glebbeek2008}\\
MESA     & Fortran, OpenMP             & Henyey stellar evolution           & \cite{Paxton2011}\\
BSE      & Fortran                     & binary evolution                   & \cite{Hurley2000}\\
SeBa     & C++                         & stellar and binary evolution       & \cite{PortegiesZwart2001}\\
\hline
Fi       & Fortran, OpenMP             & TreeSPH                            & \cite{Pelupessy2005}\\
Gadget-2 & C, MPI                      & TreeSPH                            & \cite{Springel2005} \\
Capreole & Fortran, MPI                & Finite volume grid hydrodynamics   & \cite{Mellema1991} \\
Athena3D & C, MPI                      & Finite volume grid hydrodynamics   & \cite{Stone2008} \\
MPIAMRVAC& Fortran, MPI                & AMR code for conservation laws     & \cite{Keppens2011} \\
\hline
Simplex  & C++, MPI                    & Rad. transport on Delaunay grid    & \cite{Paardekooper2010} \\ 
SPHRAY   & Fortran                     & Monte Carlo on SPH particles       & \cite{Altay2008} \\ 
Mocassin & Fortran                     & Monte Carlo, steady state          & \cite{Ercolano2003} \\
\hline
MMAMS    & C++                         & stellar mergers by entropy sorting & \cite{Gaburov2008} \\
Hop      & C++                         & particle group finder              & \cite{Eisenstein1998} \\
FractalCluster & Fortran               & Fractal cluster generator          & \cite{Goodwin2004}\\
Halogen        & C                     & Halo distribution functions        & \cite{Zemp2008} \\
GalactICS      & C                     & Galaxy model generator             & \cite{Widrow2008} \\
\end{tabular}
\caption{
Overview of community codes included in the public release of AMUSE. 
Table lists the name of the code, the language it is written in and the 
parallelization model, a short description and a reference to either the 
code paper or a description of the method.
}
\label{tab:codes}
\end{table*}

\subsection{Gravitational dynamics}

The gravity codes included in AMUSE span most dynamic regimes except 
that support for large scale cosmological simulations is 
limited. Fully relativistic metric solvers are also not included at present. 
Hermite0, PhiGRAPE, ph4 and HiGPUs are direct N-body integrators 
with Hermite timestepping schemes, where the latter three are tooled to 
use hardware (GRAPE or GPU) accelerated force calculations. All are MPI 
parallelized. Huayno is a (semi-)symplectic integrator based on 
Hamiltonian splitting. BHTree, Octgrav and Bonsai are Barnes-Hut 
treecodes (Octree is partly GPU accelerated, Bonsai runs completely on 
the GPU). Twobody, SmallN and Mikkola are integrators that calculate the 
dynamics for small N systems using analytic solutions and regularization 
respectively (with relativistic corrections in the case of Mikkola). 
Such solvers are also potentially useful as components in compound 
gravity solvers where one of the other solvers is used for the general 
dynamics and close passages and binaries are handled by a small N solver 
(see section~\ref {sec:couple}). Tupan is a symplectic integrator with 
post-newtonian correction terms. Mercury and MI6 are examples of a 
specialized class of gravitational integrators, geared towards long time 
evolution of systems dominated by a central object: planetary systems 
for Mercury, black hole dominated systems for MI6. Brutus and MMC 
represent two completely opposite approaches in a sense to solving 
gravitational dynamics: Brutus uses arbitrary precision arithmetic 
libraries to obtain converged solutions, while MMC uses Monte Carlo 
gravitational dynamics to solve globular cluster dynamics. Finally, the 
N-body/SPH codes included in AMUSE (Fi and Gadget) can also be used in 
mixed gravity/hydrodynamics or purely gravitational mode (see section~ 
\ref {sec:hydro}). They also conform to the gravitational dynamics 
interface.

\begin{table*}
\centering
\begin{tabular}{ l | c c c c c c}  
\hline
code      & small N & intermediate N & large N & large N         & centrally dominated & post-Newtonian \\
          & (collisional) & (collisional) & (collisional) & (collisionless) &                     &  \\
\hline
Hermite0  & $+$     & $+$            & $\circ$ & $\circ$       & $\circ$             & $-$  \\
PhiGRAPE  & $+$     & $+$            & $+$     & $\circ$       & $\circ$             & $-$  \\
ph4       & $+$     & $+$            & $+$     & $\circ$       & $\circ$             & $-$  \\
BHTree    & $-$     & $-$            & $-$     & $+$           & $-$                 & $-$  \\
Octgrav   & $-$     & $-$            & $-$     & $+$           & $-$                 & $-$  \\
Bonsai    & $-$     & $-$            & $-$     & $+$           & $-$                 & $-$  \\
Twobody   & N=2     & $-$            & $-$     & $-$           & $+$                 & $-$  \\
Huayno    & $+$     & $+$            & $+$     & $\circ$       & $\circ$             & $-$  \\
SmallN    & $+$     & $-$            & $-$     & $-$           & $-$                 & $-$  \\
Mercury   & $-$     & $-$            & $-$     & $-$           & $+$                 & $-$  \\
Mikkola   & $+$     & $-$            & $-$     & $-$           & $-$                 & $+$  \\
MI6       & $+$     & $+$            & $+$     & $-$           & $+$                 & $+$  \\
Pikachu   & $+$     & $+$            & $+$     & $+$           & $-$                 & $-$  \\
Brutus    & $+$     & $-$            & $-$     & $-$           & $+$                 & $-$  \\
HiGPUs    & $-$     & $+$            & $+$     & $\circ$       & $-$                 & $-$  \\
Tupan     & $+$     & $+$            & $\circ$ & $\circ$       & $\circ$             & $+$  \\
MMC       & $-$     & $-$            & $+$     & $-$           & $-$                 & $-$  \\
Fi        & $-$     & $-$            & $-$     & $+$           & $-$                 & $-$  \\
Gadget2   & $-$     & $-$            & $-$     & $+$           & $-$                 & $-$  \\
\hline
\end{tabular}
\caption{
Overview of the suitability of gravitational dynamics solvers for 
different gravitational regimes. A $+$ indicates that a codes is well 
suited to a particular regime, a $-$ indicates that the code will fail or run very 
inefficiently, a $\circ$ indicates a limited capability; 
``small N'' means a problem with a small number of bodies, N$\approx 
2-100$, ``intermediate N'' means N$\approx 100-10^4$ , ``large N'' means N
$\gtrsim 10^4$ , ``collisionless'' indicates gravitational dynamics with 
some form of softening, ``collisional'' means simulations without softening. 
``Centrally dominated'' means problems were the dynamics is dominated by a 
massive central object, such as a solar system or cluster with a 
supermassive black hole. ``Post-Newtonian'' indicates whether the code 
includes post-Newtonian correction terms. Note this table does 
not address special requirements set by the timescales or the dynamic 
range of a problem, nor the choice of parameters that may affect the 
solutions.
}
\label{tab:grav}
\end{table*}

The interface to the different gravitational dynamics codes is the same 
and changing the core integrator in a script is trivial. However, the 
gravitational dynamics integrators that are included in AMUSE are geared 
towards different types of problem: choosing a correct and efficient 
integrator for a particular problem is the responsibility of the AMUSE 
user, however table~\ref{tab:grav} gives a rough overview of the 
suitability of the various integrators in different regimes.

\subsection{Stellar evolution}

\begin{figure}
\begin{boxedminipage}{9cm}
{
\small
\begin{verbatim}
0: deeply or fully convective low mass main seq. star
1: main sequence star
2: hertzsprung gap
3: first giant branch
4: core helium burning
5: first asymptotic giant branch
6: second asymptotic giant branch
7: main sequence naked helium star
8: hertzsprung gap naked helium star
9: giant branch naked helium star
10: helium white dwarf
11: carbon/oxygen white dwarf
12: oxygen/neon white dwarf
13: neutron star
14: black hole
15: massless supernova
16: unknown stellar type
17: pre-main-sequence star
\end{verbatim}
}
\end{boxedminipage}
\caption{
Stellar types used in AMUSE. 
}
\label{fig:stellartypes}
\end{figure}

The stellar evolution codes currently included fall into two 
categories: table or parametrized fits based and Henyey-type solvers (1D 
evolution code). SSE implements simple heuristics and table look-up to 
determine approximate stellar evolution tracks. BSE and SeBa are the 
equivalent of SSE for binary system evolution. Evtwin and MESA are 
(Henyey-type) solvers for stellar evolution. They calculate the full 
internal evolution of a 1D model for a star~\cite[][]{Henyey1959, 
Henyey1964, Eggleton1971,Paxton2011}. They share however the same basic 
interface as table based stellar evolution codes (with the details of 
the internal evolution hidden) and can be used interchangeably. In 
addition, for the full stellar evolution codes the internal (1D) 
structure of the stars can be accessed. The internal structure 
(basically radial profiles) can be used to construct 3D hydrodynamic 
models of stars (see section~\ref{sec:couple}). During the evolution of 
a stellar model a stellar evolution code may encounter a phase where 
numerical instabilities are encounterd (e.g. during the Helium flash). 
Below we describe our strategy to handle this error condition.

\subsubsection{Robust stellar evolution with fallback\label{sec:fallback}}

\begin{figure}
 \centering
 \epsfig{file=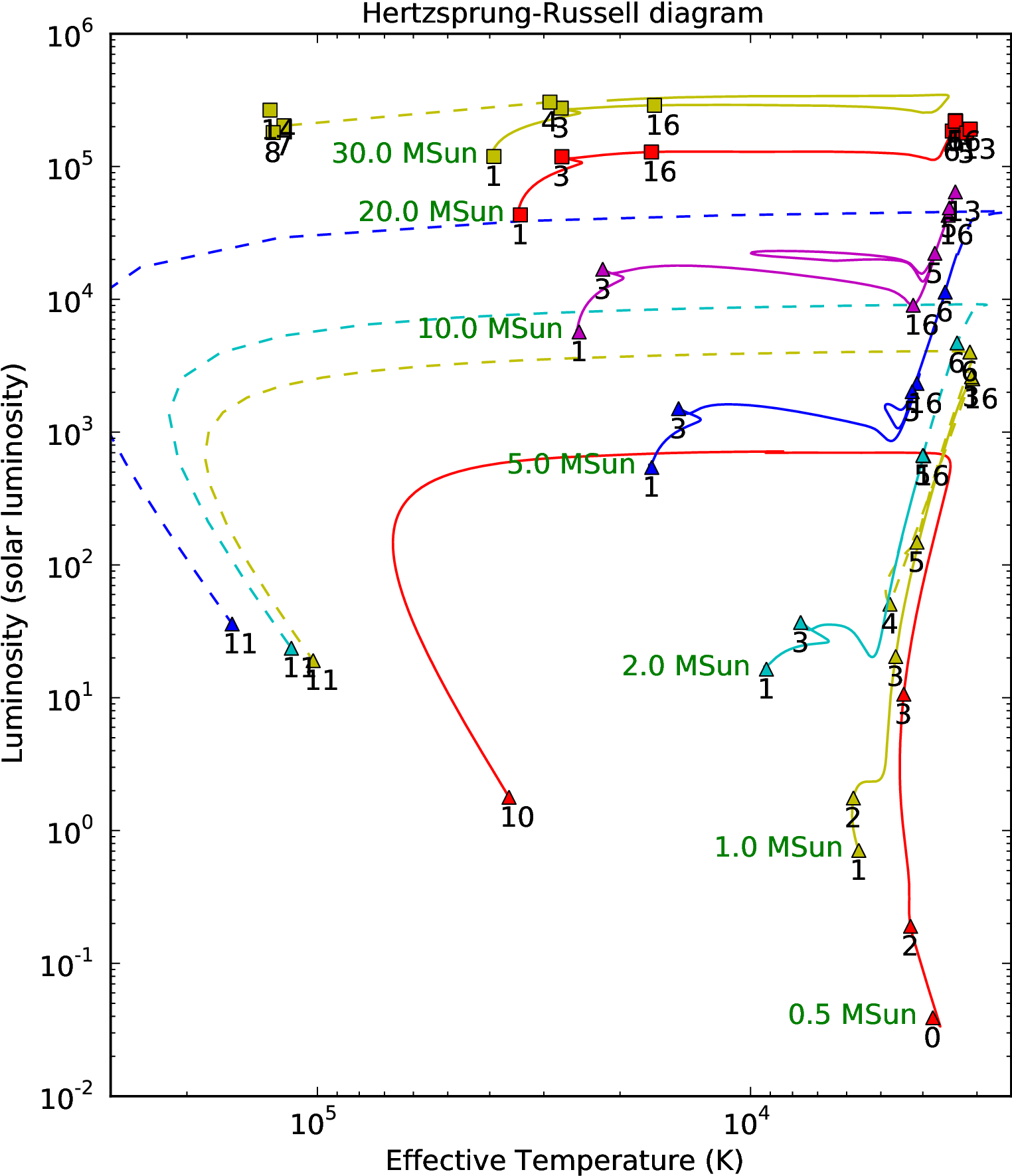, width=.43\textwidth}
 \caption{
 Stellar evolution with SSE Fallback. Shown are evolutionary tracks 
 calculated with EVTWIN (drawn lines), with SSE fallback (dashed lines) 
 in case the full stellar evolution code could not progress. For this 
 particular version of EVTWIN this happened at the Helium or Carbon flash. 
 Stellar type labels are given in figure~\ref{fig:stellartypes}.
 }
 \label{fig:se_fallback}
\end{figure}

The typical use case of a stellar evolution code in AMUSE may be  
as a subcode in a larger script, for example when 
calculating the evolution of a star cluster with stellar collisions, 
where the collision cross section depends on the radii of the stars (and 
hence on its evolution). However, stellar evolution physics is considerably 
more complicated than gravitational dynamics in the sense that 
numerical instabilities may be encountered, which may slow down these 
codes tremendously or even require manual intervention. Within AMUSE we 
have the option to use a code based on lookup tables like SSE as a 
fallback code, since such codes will always find a solution. This 
pragmatic solution may be justified if the fallback option is only 
rarely needed. As an example we show the Herzpsrung-Russel diagram of a 
set of stars evolved using EVtwin, with SSE as a fallback option. The 
actual implementation of the fallback minimizes the $\chi^2$ error on the 
luminosity, radius and mass to find the closest matching SSE model at a 
given switchpoint.

\subsection{Hydrodynamics\label{sec:hydro}}

\begin{figure*}
 \centering
 \epsfig{file=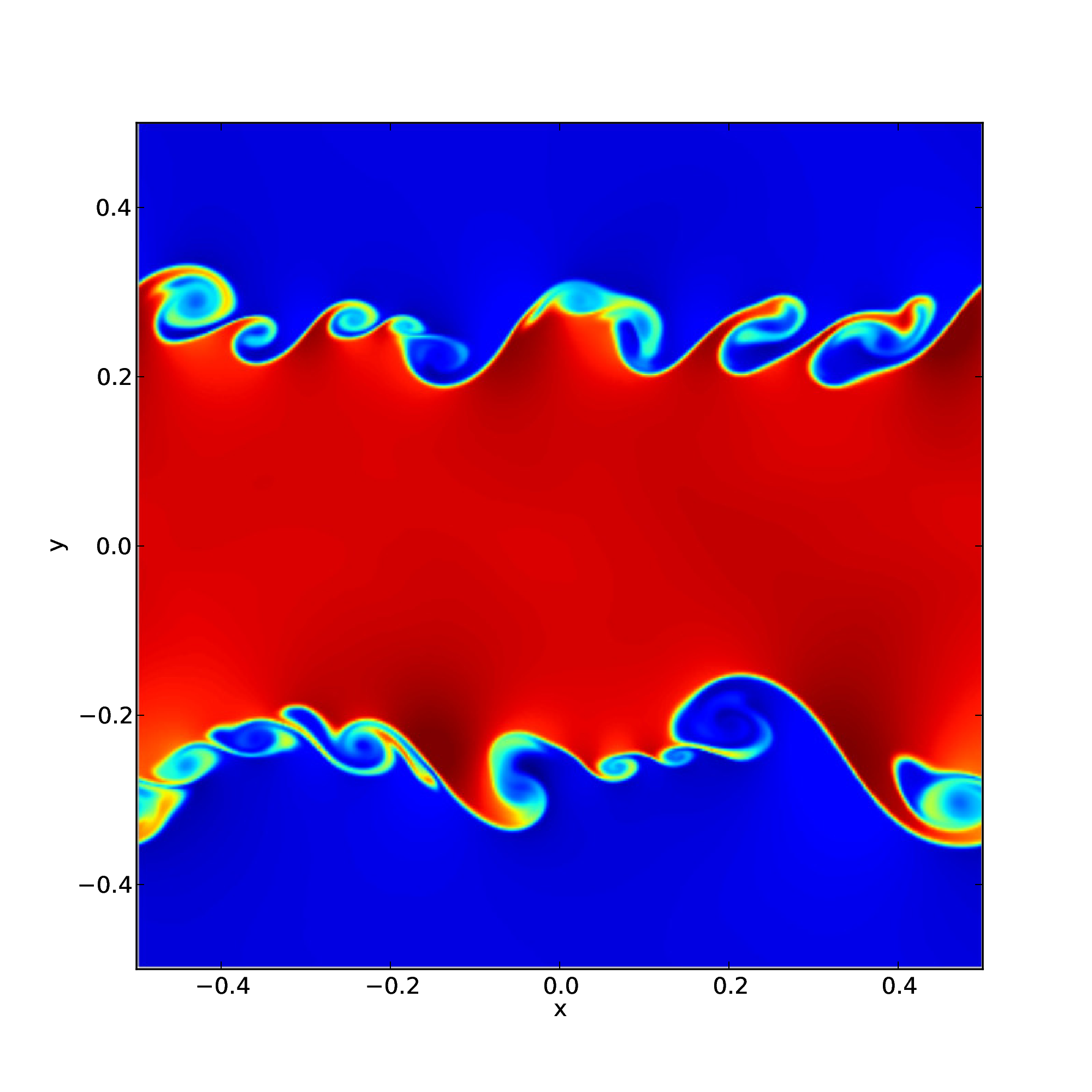, width=.32\textwidth}
 \epsfig{file=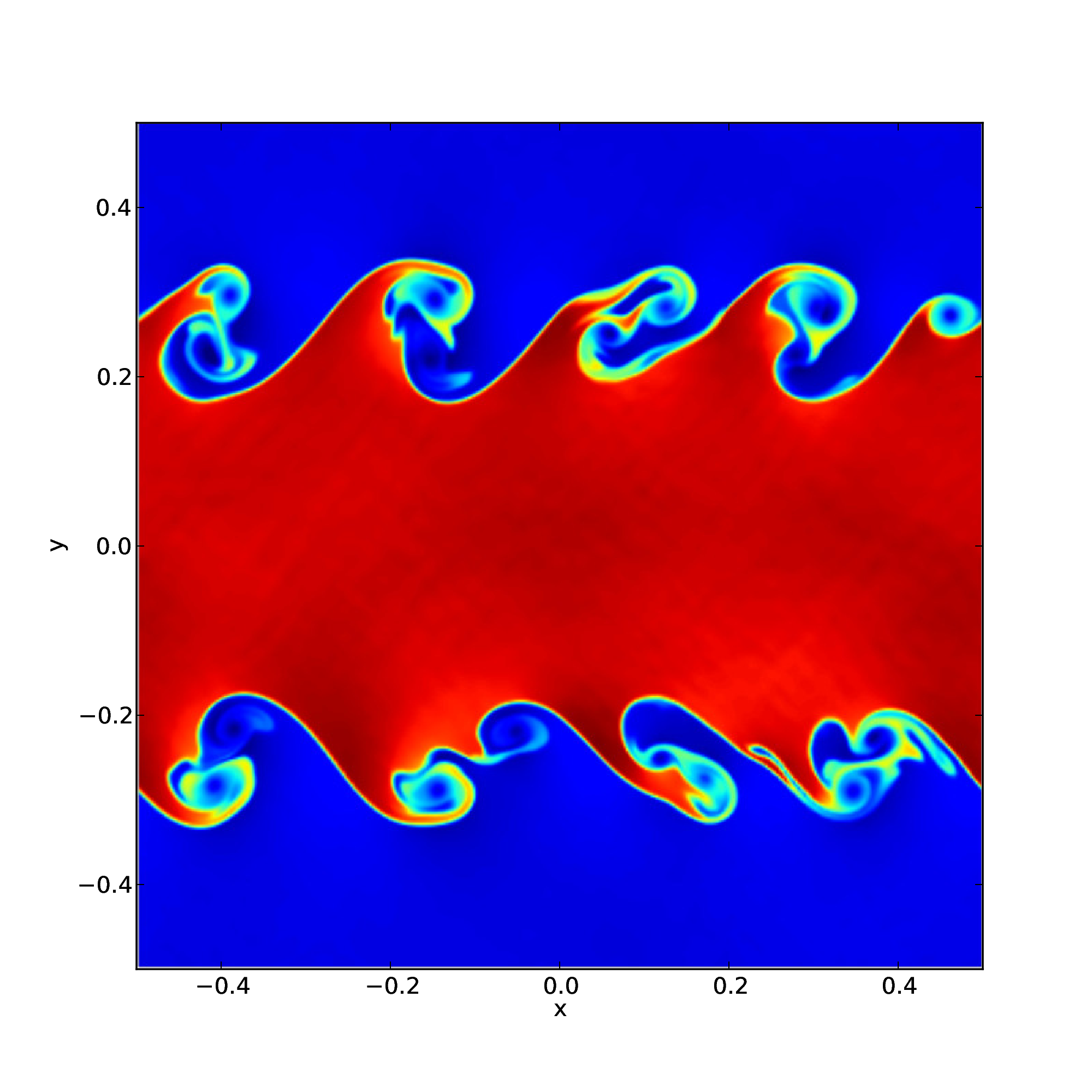, width=.32\textwidth}
 \epsfig{file=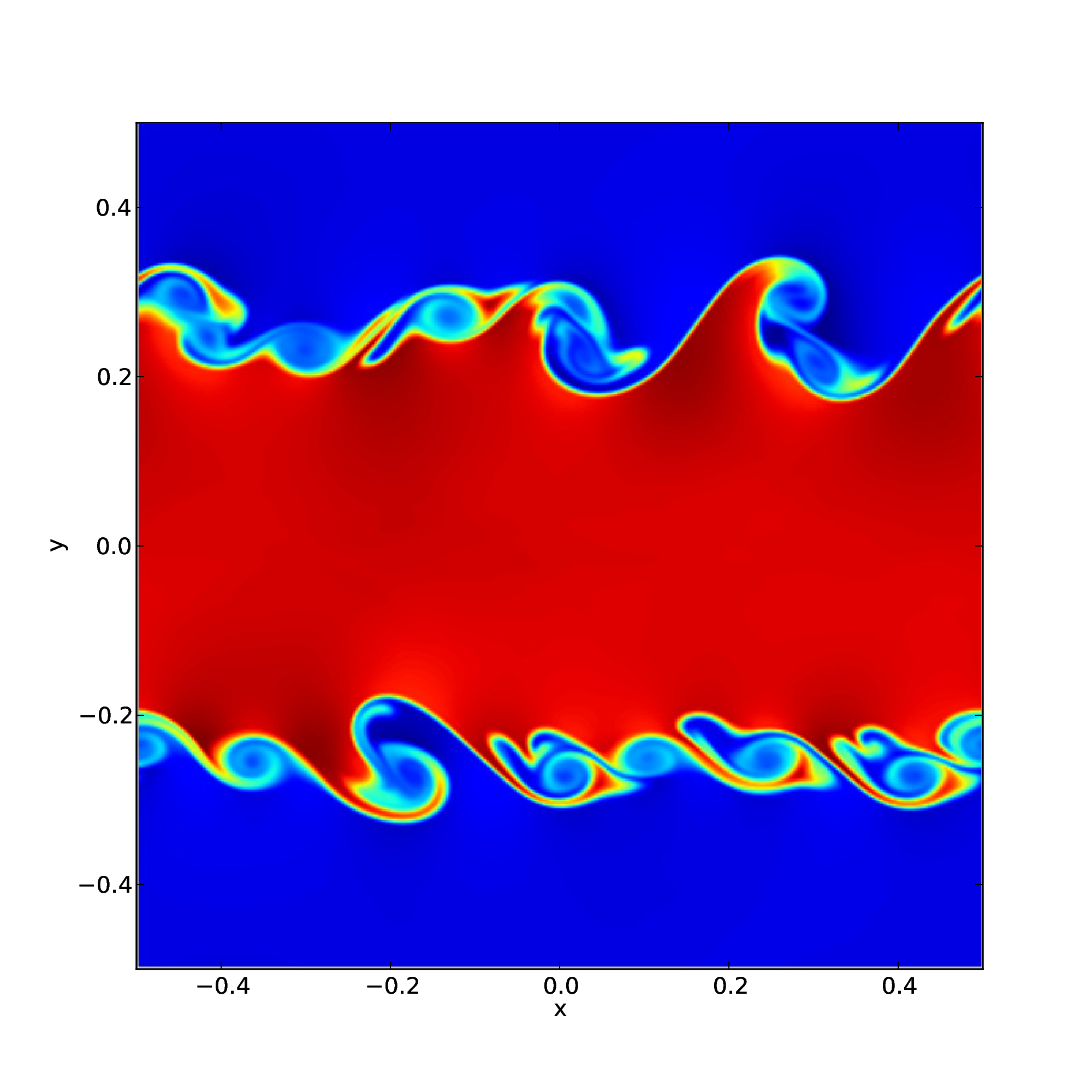, width=.32\textwidth}
 \caption{
 Kevin-Helmholtz test. Shown are the density distribution at $t=1$ for 
 Athena (left panel), Capreole (middle panel) and MPIAMRVAC (right 
 panel). Random velocities were seeded with an amplitude of $0.01 c_s$
 }
 \label{fig:kh}
\end{figure*}

\begin{figure*}
 \centering
 \epsfig{file=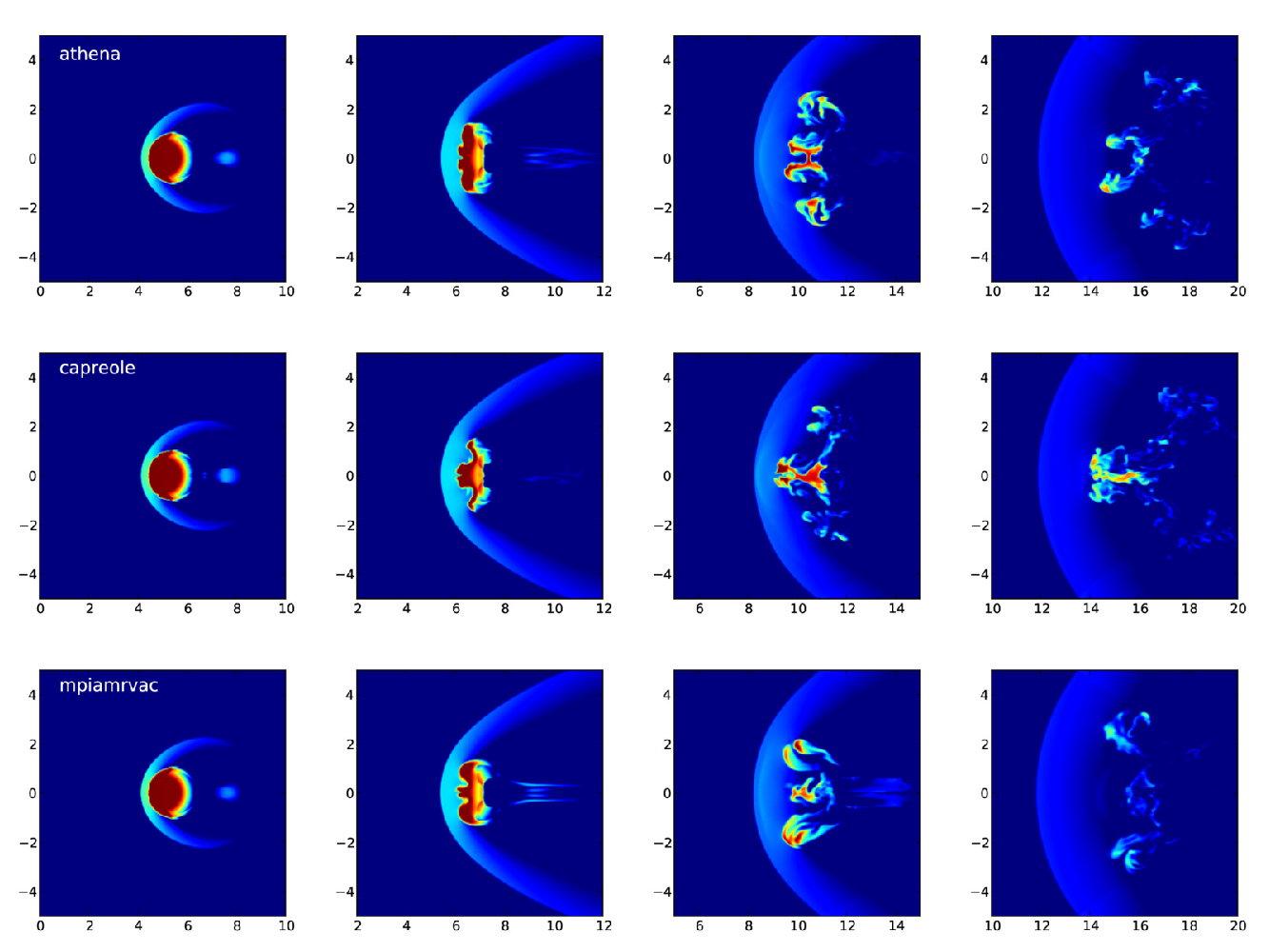, width=.99\textwidth}
 \caption{
 Cloud-shock test. Comparison of the results of the cloud-shock test 
 \citep{Agertz2007} for 3 grid hydrodynamic codes (each row showing 
 results for respectively Athena, Capreole and MPIAMRVAC). Panels show 
 slices through the 3D density distribution for times $t={0.25, 1., 
 1.75, 2.5} \times \tau_{\rm KH}$. The resolution for Athena and 
 Capreole is 320x320x1280, for MPIAMRVAC it is 80x80x320 with 2 levels 
 of refinement for the same effective resolution.
 }
 \label{fig:shocked}
\end{figure*}

In hydrodynamics a distinction is made between two types of 
code: particle based codes and grid based codes. Particle 
based methods include for example Smooth Particle Hydrodynamics (SPH) 
codes. Grid based codes may be further subdivided into Eulerian and 
Langrangian (comoving) grid codes. AMUSE is limited at the moment to 
Eulerian grid solvers on regular Cartesian grids, which may be 
statically or dynamically refined.

Most current astrophysical SPH codes contain solvers for gravitational 
dynamics (using the ~\cite{Hernquist1989} TreeSPH scheme) and can evolve 
collisionless particles in addition to gas particles. Hence the 
interface to these codes define at least two different sets of 
particles, (\texttt{particles} and \texttt{gas\_particles}). The 
interface of these codes contains a subset restricted to the \texttt 
{particles} that conforms to the specification of the gravitational 
dynamics interface. The hydrodynamic part of the interface is very 
similar to the gravitational part, adding hydrodynamic 
state variables (density, internal energy) to the gas particles.

The fundamental difference of the grid hydrodynamics interface with 
respect to the particle hydrodynamics and gravitational interfaces is 
that it contains functions to specify the grid properties and boundary 
conditions and that the complete grid must be initialized before use. A 
complication for the inclusion of grid hydrodynamics is that these codes 
typically are compiled for a specific problem setup, which entails 
picking different source files for e.g. different grids or boundary 
conditions. Within AMUSE, this is hidden from the user by choosing 
cartesian grids and limiting the options with respect to boundary 
conditions (simple inflow/outflow, reflecting and periodic). This 
can be extended as needed. 

Presently two parallel TreeSPH codes, Fi and Gadget, are included. 
In addition the grid hydrocodes Capreole, Athena3D and MPIAMRVAC are 
included. Capreole is a finite volume eulerian grid code based on Roe's 
Riemann solver~\citep{Mellema1991}. Athena3D is a finite volume Eulerian 
grid code for hydrodynamics and magnetohydrodynamics using different 
Riemann solvers (limited support for magnetohydrodynamics is implemented 
in AMUSE) and allows for static mesh refinement. MPIAMRVAC is an 
MPI-parallelized Adaptive Mesh Refinement code for hyperbolic partial 
differential equations geared towards conservation laws. Within AMUSE 
the hydrodynamic solver of MPIAMRVAC is implemented, including full AMR 
support.

\subsubsection{Hydrodynamic test cases}

As an example we show the results of two common hydrodynamic tests using
AMUSE, where the exact same script for different codes is used. 
The first test is the well known Kelvin-Helmholtz (KH) instability test 
(figure~\ref{fig:kh}) which demonstrates the ability of the grid codes 
to model the KH instabilities. The second test, the cloud-shock test 
\citep{Agertz2007}, consists of a cloud compressed by a high mach number 
shock, representative of the shock-induced star formation process (figure~
\ref{fig:shocked}).

\subsection{Radiative transfer}

Due to the high computational cost of radiative transfer caluclations, 
most radiative transfer codes are limited and optimized for a particular 
type of transport, e.g.  photo-ionisation, dust or molecular line 
transfer. Within AMUSE the main focus is on photo-ionisation codes. 
They calculate the time dependent propagation of UV photons as well as 
the ionization and thermal balance of gas irradiated by hot stars or an AGN. 
Dust radiation transfer is in development. The latter is more 
important as a post-processing tool to generate realistic virtual 
observations or diagnostics.

AMUSE currently includes the photo-ionisation codes Simplex and SPHRay. 
SimpleX is based on the transport of radiation along the vertices of a 
Delaunay grid. It includes heating and cooling as well as H and He 
chemistry and contains optional treatment of diffuse recombination 
radiation. SPHRay is a Monte-Carlo photo-ionisation code which works on 
(SPH) particle distributions directly - it solves similar physics as 
SimpleX. 

In addition, a proto-type interface for the Monte Carlo code Mocassin is 
available. Mocassin calculates a more extensive chemical network than SimpleX
or SPHRay, albeit in the steady state approximation.

\begin{figure*}
 \centering
 \epsfig{file=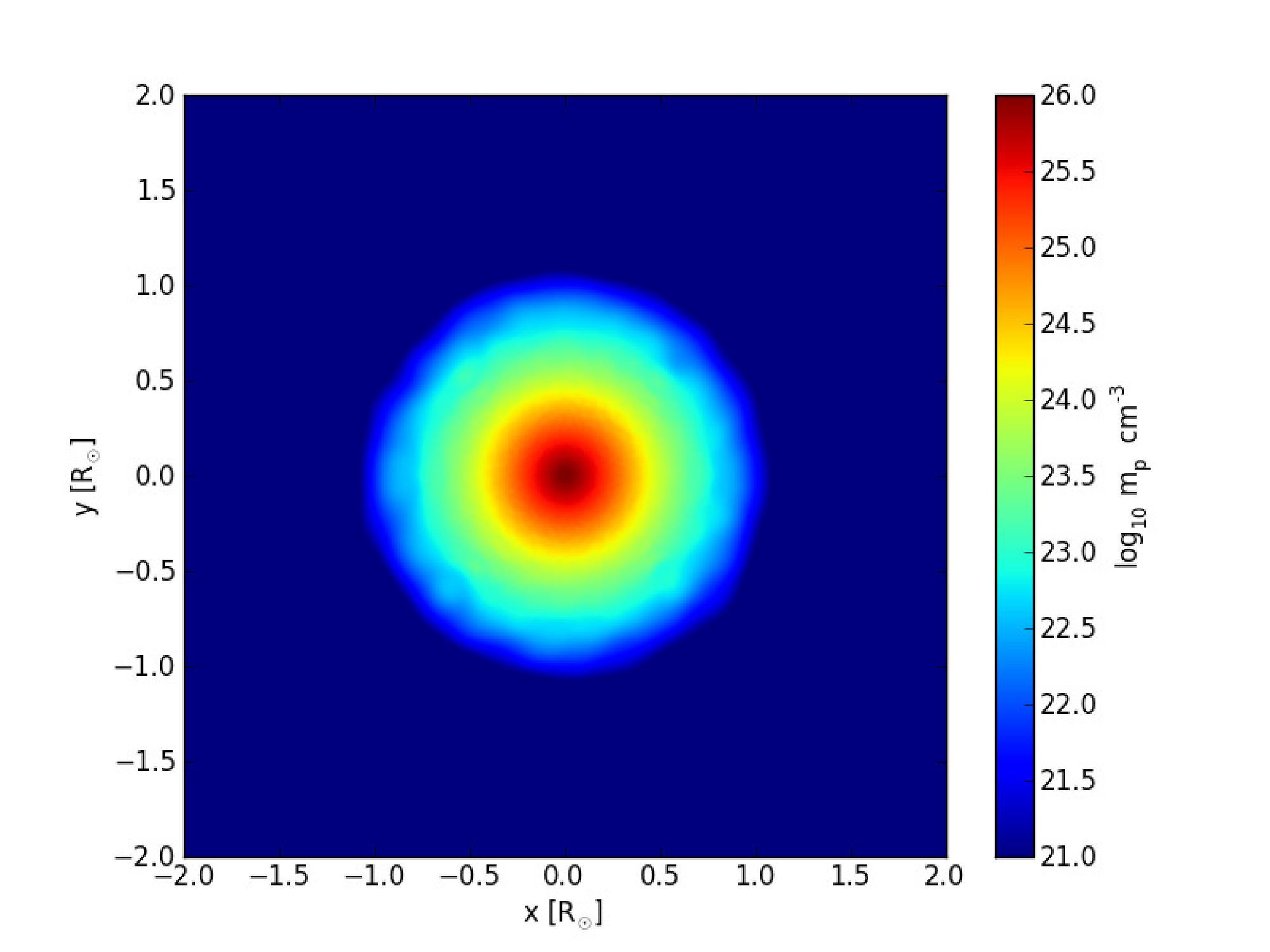, width=.49\textwidth}
 \epsfig{file=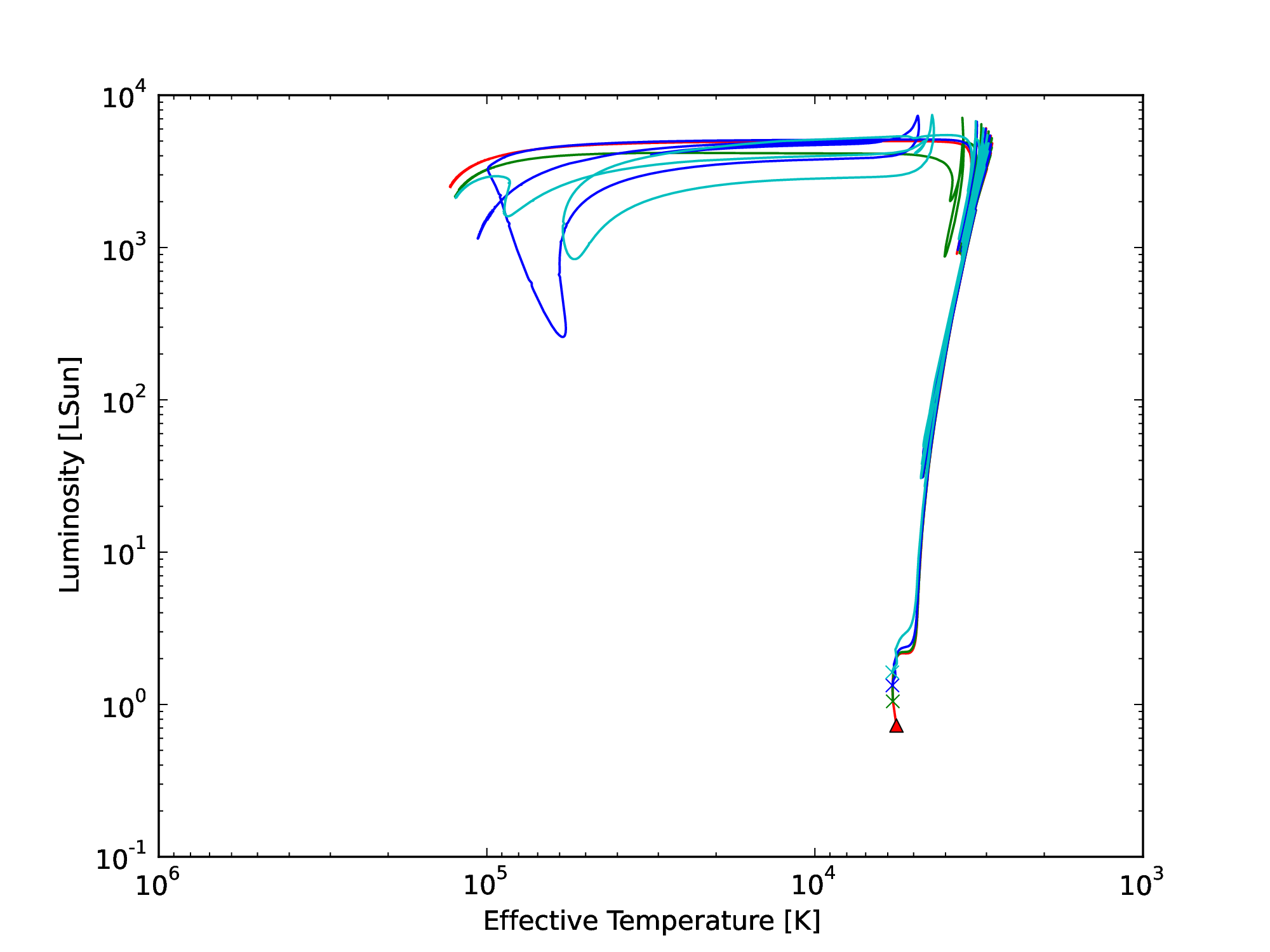, width=.49\textwidth}
 \caption{
 Conversion of a stellar evolution model to SPH and back. Left panel 
 shows a $N=10^5$ SPH realization of a $1 \Msun$ MESA stellar evolution 
 model at 5 Gyr. Right panel shows the HR diagram of the 
 evolution of a $1 \Msun$ MESA model, as well as as the evolution of a 
 MESA model derived from the SPH model, as well as two additional models 
 which where converted at 8 and 10 Gyr of age. As can be seen the 
 conversion at later evolutionary stages induces bigger deviations from 
 normal evolution, probably because of the increased density contrasts 
 (as the core becomes denser) at later times means that the particle 
 distributions induce more mixing.
 }
 \label{fig:mesasph}
\end{figure*}
  
\section{Compound solvers\label{sec:couple}}

\begin{figure}
 \centering
 \epsfig{file=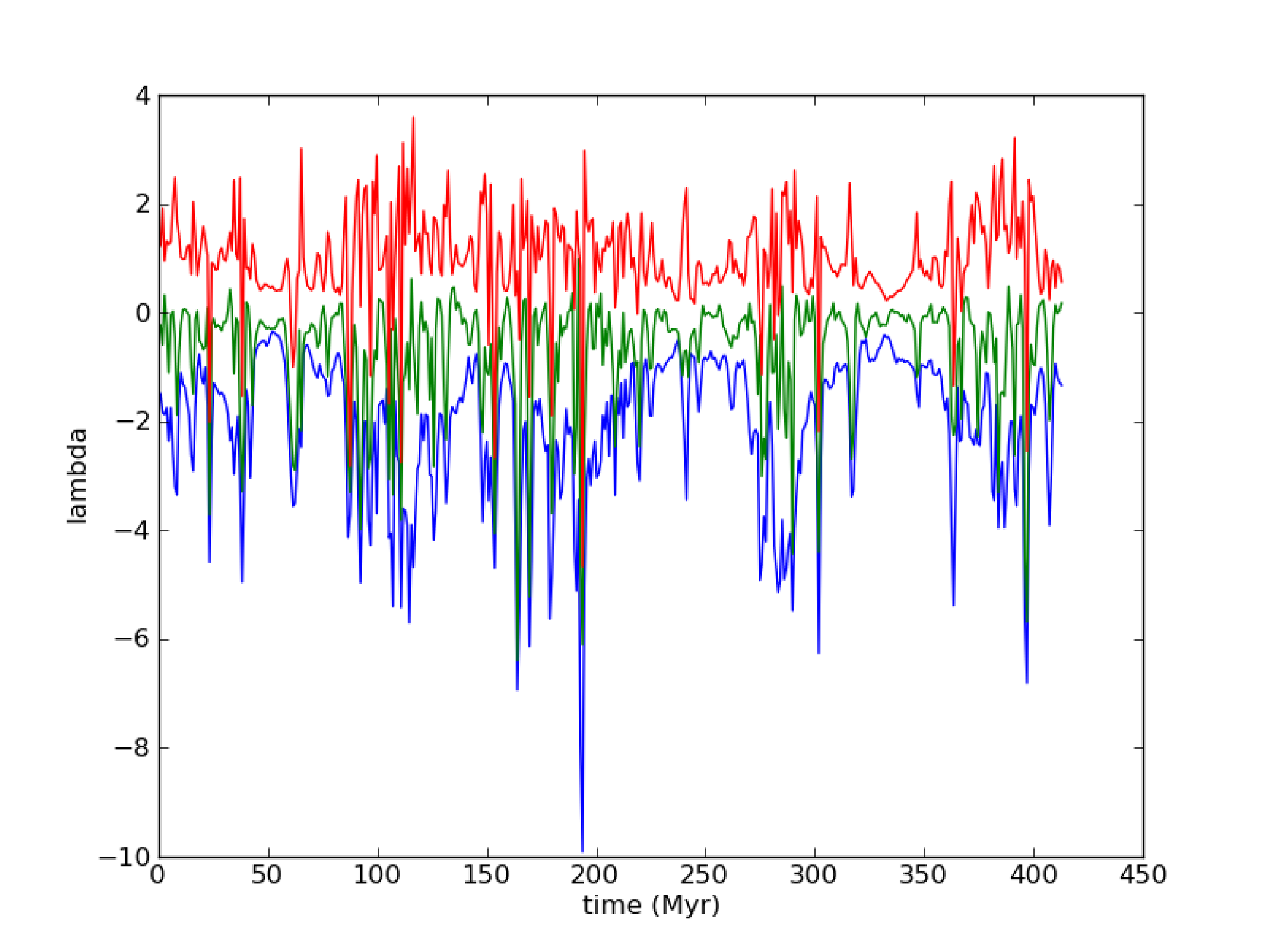, width=.39\textwidth}
 \epsfig{file=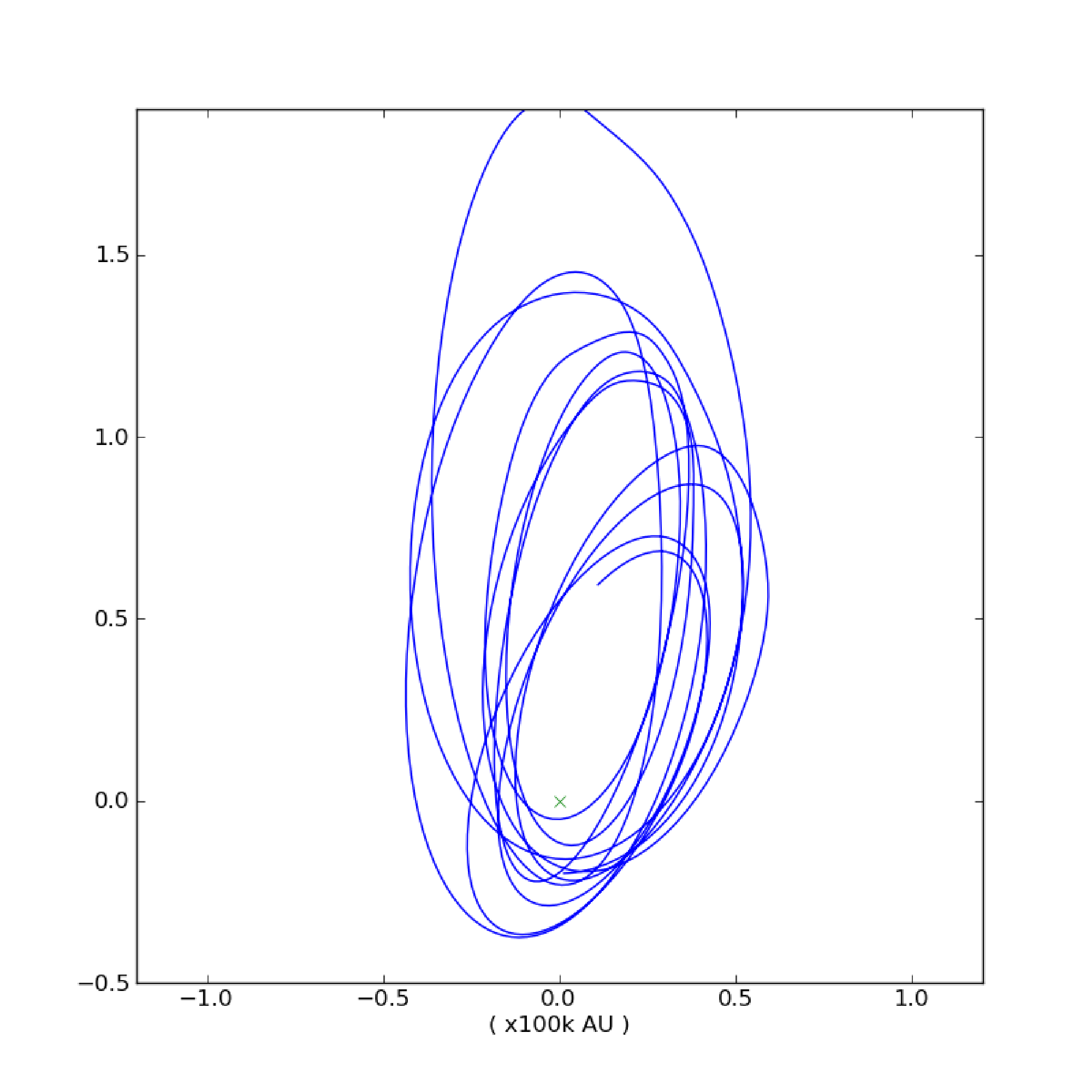, width=.39\textwidth}
 \caption{
 Integration of an Oort cloud comet in a time-dependent galaxy 
 potential. Upper panel shows the 3 eigenvalues of the tidal tensor 
 extracted from a milky way galaxy simulation, the lower panel shows an 
 integration of an Oort cloud comet subject to the corresponding tidal 
 tensor, integrated using the Bridge integrator.
 }
 \label{fig:comet}
\end{figure}

In addition to provide a unified interface to various types of codes, 
AMUSE has the objective of facilitating multiphysics simulations. For 
example, one would want to combine a gravitational dynamics code with a 
collision solver to construct a compound solver for gravitational 
dynamics that allows integration past a collision. Community codes 
can be combined within AMUSE such that the compound solver has a wider 
applicability than each by itself. The setup of AMUSE allows for this in 
a transparent manner, such that a combined solver has a similar interface 
as a plain code. The types of coupling AMUSE can be applied to fall 
roughly in the following categories~\citep{PortegiesZwart2013b}:

\begin{itemize}
\item[1] input/ouput coupling: the loosest type of coupling occurs when the 
result of one code generates the initial conditions for another code.
For example: a 3 dimensional SPH star model can be generated from a 
stellar evolution code, or vice versa (figure~\ref{fig:mesasph}),

\item[2] one way coupling: one system interacts with another (sub)system , but 
this (sub)system does not couple back (or only very weakly) to the former. 
 An example may be the stellar evolution of stars in a cluster, 
where the mass loss of the stars is important for the dynamics, but the 
dynamics of the cluster does not affect the stellar evolution (see 
section \ref{sec:secluster}),

\item[3] hierarchical coupling: one or more systems are embedded in a parent 
system, which affects the evolution, but the subsystems do not affect 
the parent or each other. An example is the evolution of cometary orbits 
of the Oort cloud in a realistic galactic potential (figure~\ref{fig:comet}),

\item[4] serial coupling: a system evolves through clearly separate regimes, 
such that different codes have to be applied in an interleaved way. An 
example may be a dense cluster where stellar collisions occur. In such a 
system, pure gravitational dynamics may be applied until the moment a 
collision between two stars is detected, at which point such a collision 
can be resolved using e.g. a full hydrodynamic code. After the hydrodynamic 
calculation the collision product is reinserted in the stellar dynamics 
code (see section~\ref{sec:evocoll}),

\item[5] interaction coupling: this type of coupling occurs when there is 
neither a clear seperation in time nor spatially. An example may be the 
coupling between the ISM and stellar dynamics, where the gas dynamics 
must account for the combined potential of the gas and stars (see 
section~\ref{sec:bridge}), 
  
\item[6] intrinsic coupling: this may occur where the physics of the problem 
necessitates a solver that encompasses both types of physics. An example 
may be magnetohydrodynamics, where the gas dynamics and the electrodynamics
of the problem are so tightly coupled it makes little sense to seperate 
the two. In such a case an integrated solver can be included in AMUSE.
\end{itemize}

With the exception of the last type of coupling, these couplings can be 
implemented in AMUSE using single component solvers. To increase the 
flexibility in combining different solvers, the interface definitions 
contain functions to transfer physical quantities. For example, the 
gravity interface definition includes functions to sample the gravity 
force at any location and the hydrodynamic interface can return the 
hydrodynamic state vector at any point. In addition, the AMUSE interface 
contains a framework to detect when a simulation code evolves outside 
its domain of validity using \emph{stopping conditions}. At least as 
important as being able to evolve the system in time, is the ability to 
actually detect when the calculation goes outside the regime where the 
solver is reliable. For gravity modules this may be for example a 
detection of close passages or collisions. For a hydrodynamics with self 
gravity this may be an instability criterion on density and internal 
energy \citep {Whitworth1998}. 

In the remainder of this section we elaborate on a number of 
examples to illustrate the coupling strategies that can be employed 
within AMUSE. We note that, while AMUSE facilitates these kinds of 
couplings, it remains the responsibility of the user to check the validity 
of the chosen coupling strategy in the regime of interest.

\subsection{Cluster evolution with full stellar evolution\label{sec:secluster}}

Stellar evolution affects the global evolution of a cluster of stars by 
dictating the mass loss of the stars. In the simplest approximation the 
stellar ejecta are assumed to escape instantaneously from the cluster. 
The only two codes needed to calculate this are then: a stellar 
evolution code to calculate the time dependent stellar mass loss, and a 
gravitational dynamics code to calculate the dynamics of the stars. The 
integration proceeds as follows: the gravitational dynamics code is 
advanced for a time $\Delta t$, at the end of this time the stellar 
evolution code is synchronized to the current simulation time and the 
stellar masses in the gravitational dynamics code are updated with the 
masses in the stellar evolution code. This approximation is good as long 
as the $\Delta t$ is smaller than the timescale on which the stars 
evolve (such that the $\Delta m$'s are small), and the mass loss from 
the cluster occurs fast enough such that at any time the gas content of 
the cluster is negligible. 

\begin{figure}
 \centering
 \epsfig{file=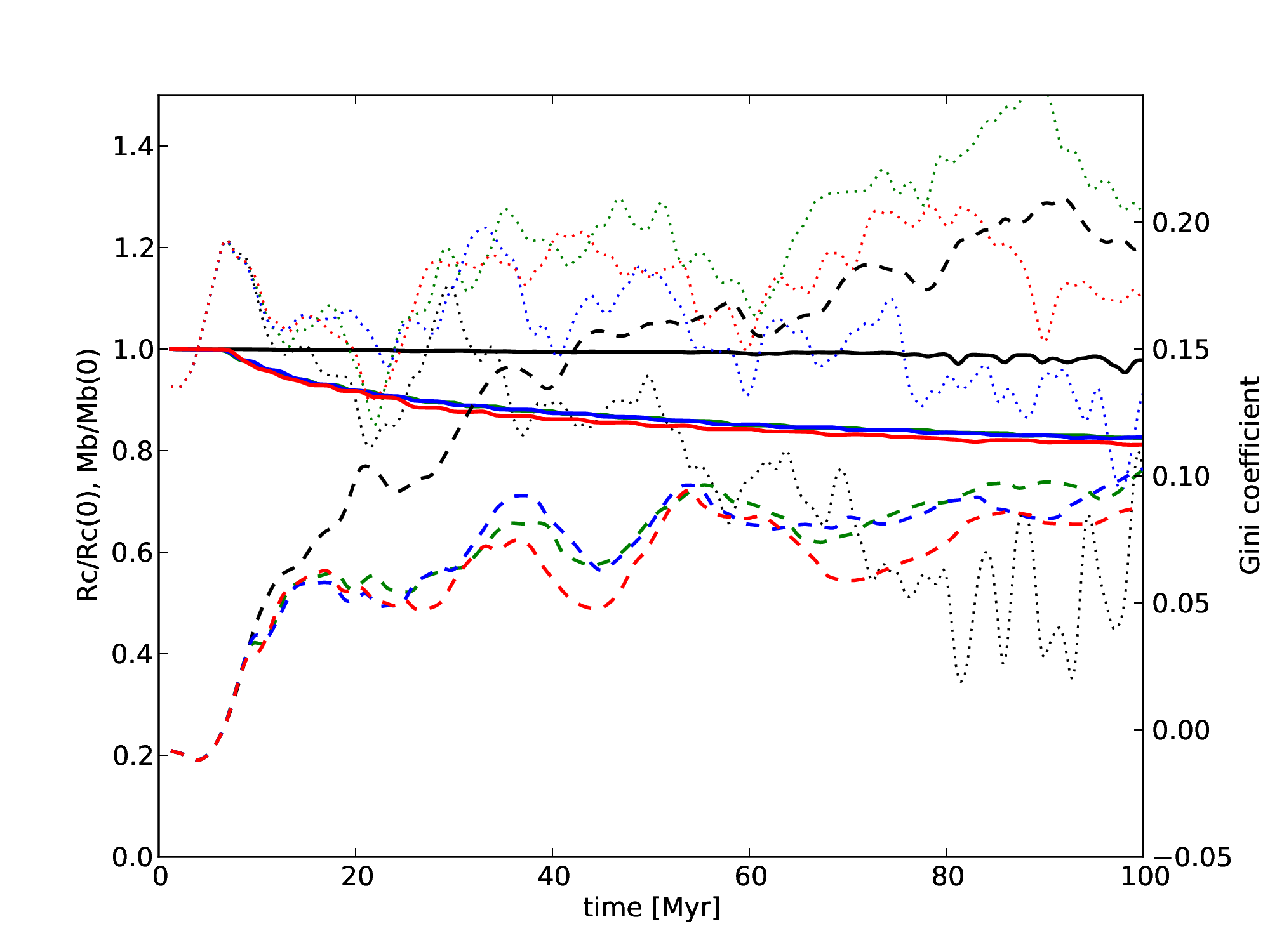, width=.4\textwidth}

 \epsfig{file=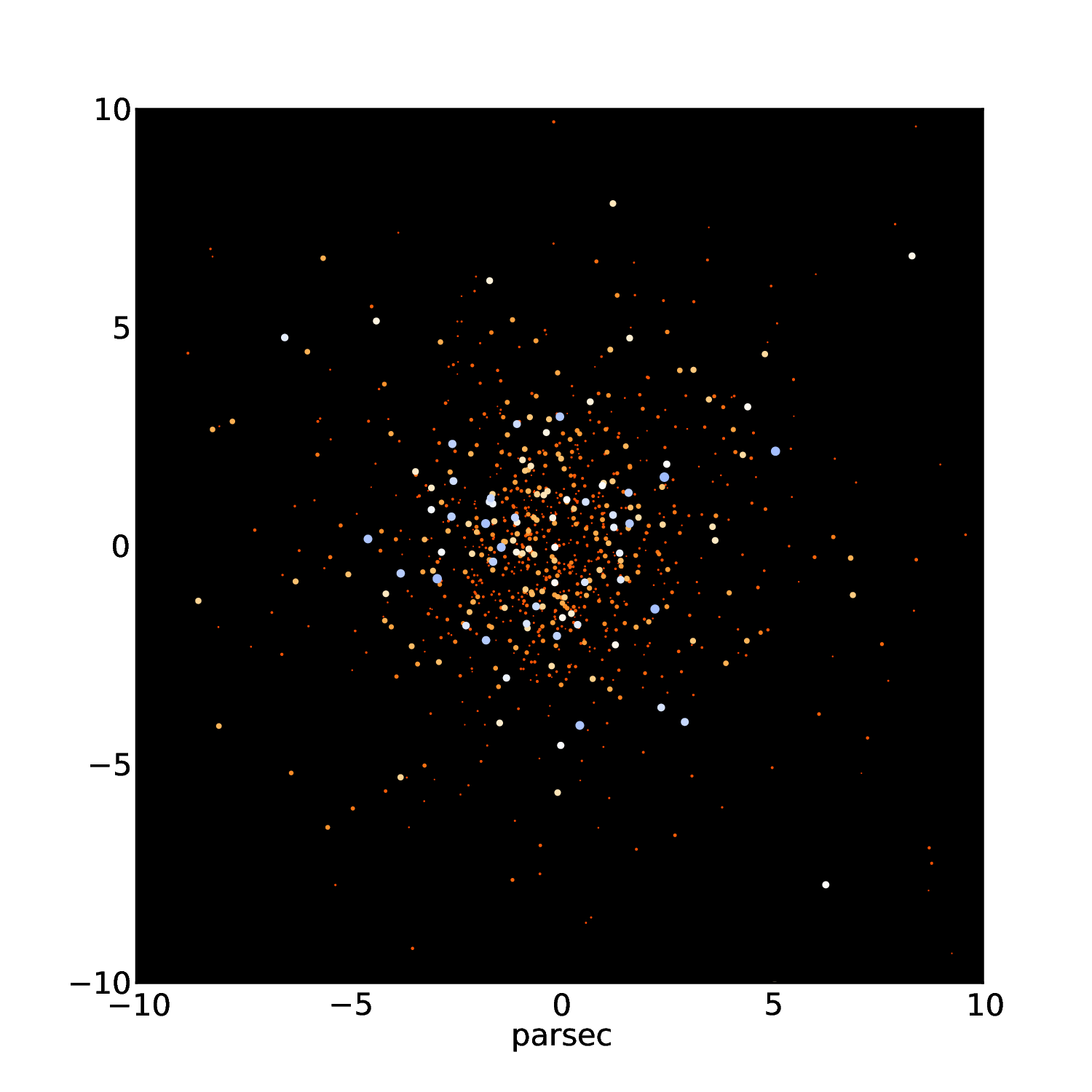, width=.4\textwidth}

 \epsfig{file=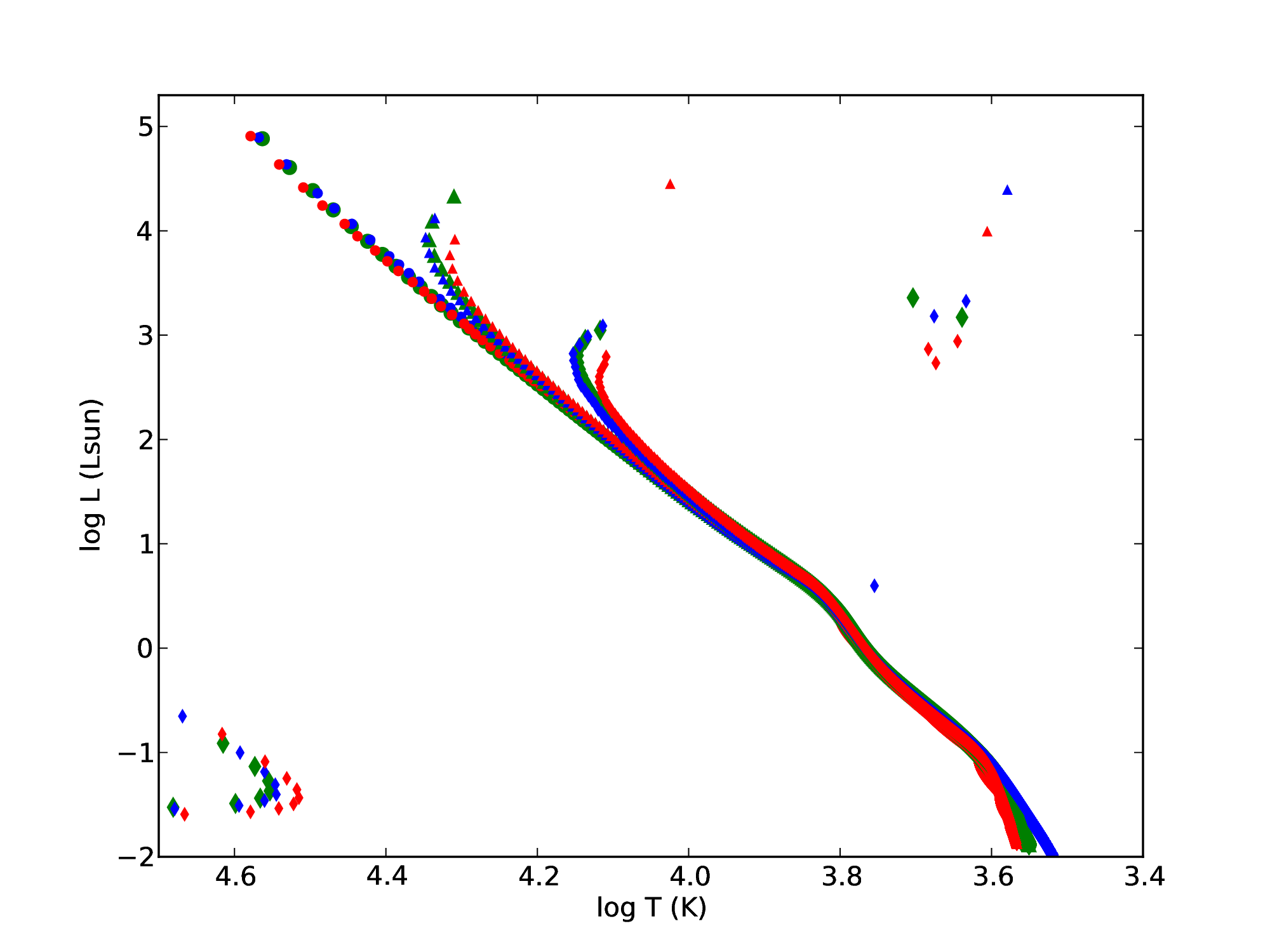, width=.4\textwidth}
 \caption{
 Cluster evolution with live stellar evolution. The upper panel shows 
 the evolution of the core radius (normalized on the initial core radius, 
 dotted lines), bound mass fraction (as fraction of the initial mass, solid lines) 
 and the evolution of the mass segregation as quantified by the \cite
 {Converse2008} ``Gini'' coefficient (dashed lines) as a function of time for 
 a $N=1000$ star cluster with a salpeter IMF in the case without 
 stellar evolution and in case the stars evolve, calculated with 
 different stellar evolution codes. The black curves are the results for 
 a run without stellar evolution, green are the results for SSE, blue 
 EVtwin and red the MESA stellar evolution code. The middle panel shows 
 the initial distribution of stars (colored according to their 
 temperature and luminosity). The bottom panel shows the 
 Herzsprung-Russell diagram for the stellar evolution as calculated 
 for the different codes (colored as top panel) for the initial 
 population (circles), after 20 Myr (triangles) and after 100 Myr 
 (diamonds).
 }
 \label{fig:secluster}
\end{figure}

We plot an example of this process \citep[which is similar to figure 6 in][]
{PortegiesZwart2009} in figure~\ref{fig:secluster}. For this example 
we evolve a plummer sphere of $N=1000$ stars with a (homogeneously 
sampled) Salpeter IMF with lower mass bound of $M=0.3 \Msun$ and upper 
mass bound of $M=25 \Msun$. The initial core radius is $R_c=2$ parsec 
and the stellar masses are assigned randomly. Runs using different 
stellar evolution codes, using SSE, EVtwin or MESA, are presented. The 
first of these is based on look-up tables and interpolation formulae, 
while the other two are Henyey-type stellar structure solvers. For 
comparison a run without any stellar evolution is also done. The full 
stellar evolution codes were run with the SSE fallback option (see 
section~\ref{sec:fallback}) with a lower limit on the timestep of the 
stellar evolution code of 1 yr as switching condition. In practice the 
switch happend around the second asymptotic giant branch stage. As can 
be seen in Figure~\ref {fig:secluster} the stellar evolution impacts the 
evolution of the cluster through the mass loss, affecting the evolution 
of the core radius and the mass segregation. Note that because the 
coupling is one way, there is no real added value in evaluating the 
stellar evolution together with the stellar dynamics, other than 
demonstrating the capabilities of AMUSE. This changes when stars are 
allowed to merge.

\subsection{Evolution with collisions\label{sec:evocoll}}

\begin{figure}
 \centering
 \epsfig{file=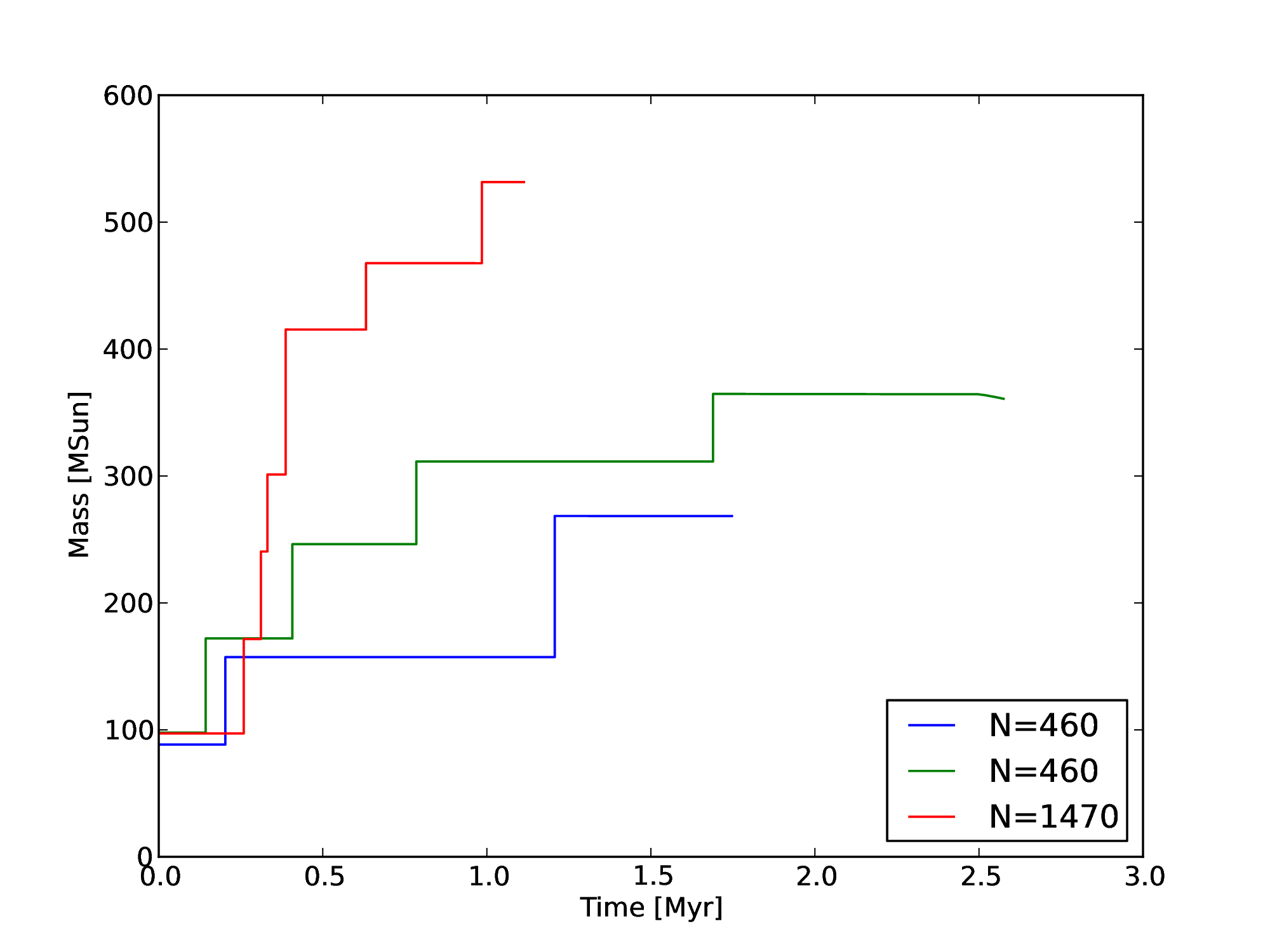, width=.49\textwidth}
 \caption{
 Evolution with stellar collisions. Shown is the mass evolution of the 
 most massive merger product for a cluster with top heavy (flat in 
 logarithm) IMF and zero metallicity, and an initial King density 
 profile with $W_0=6$ and a virial radius of 0.1 parsec for three runs 
 with the indicated number of stars (see text for details). 
 }
 \label{fig:evocoll}
\end{figure}

\begin{figure*}
 \centering
 \epsfig{file=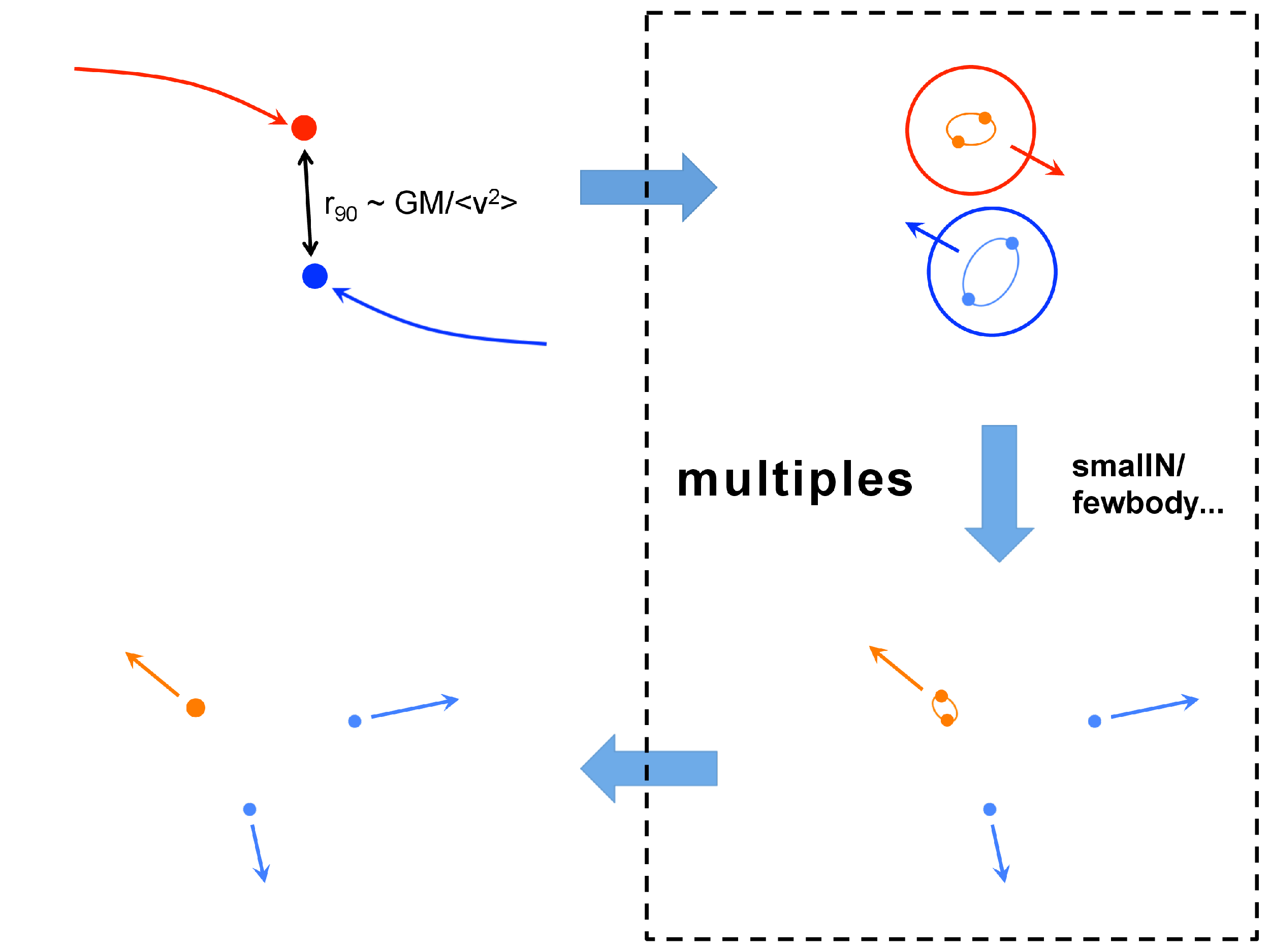,width=.7\textwidth}
 \caption{
 Resolving an interaction with the AMUSE multiples module. This diagram 
 shows the resolution of close interactions using the Multiples module. 
 Upper left: if two (possibly compound) objects pass close to each other 
 (as defined by distance $r_{90}<\frac{G M}{\langle v^2 \rangle}$) control is 
 passed to the Multiples module (right), which transforms to center of 
 mass coordinates and retrieves the internal structure of the particles. 
 It calls a sub code to resolve the collision. Once this is done, the 
 interaction products (which may not correspond to the input systems) 
 are transformed back and returned to the main code (lower left).
 }
 \label{fig:multiples}
\end{figure*}

Stellar dynamics codes often assume that collisions between stars are 
rare enough to be unimportant for the evolution of the system. For dense 
stellar systems this assumption is not correct \citep 
{PortegiesZwart1999} and a number of stellar dynamics codes allow for 
stellar collisions, e.g. Starlab \citep{PortegiesZwart1999} and Nbody 6 
\citep{Nitadori2012}. The AMUSE approach to collision handling is to 
separate this into two distinct functional tasks: 1) the detection of 
collisions and 2) the calculation of collision outcomes. Task 1) is most 
naturally handled in a stellar dynamics code and represents an example 
of the above mentioned stopping conditions: the interface to dynamics 
code defines a collision radius $R_{\rm coll}$ - if the code detects a 
passage of two particles such that they pass closer than $R_{\rm coll}$, 
the code flags a `collision' stopping condition (such a collision event 
may be a  true collision or a close passage between two bodies in the 
system) and returns control to the AMUSE framework. The actual 
calculation of the outcome of a collision is a completely different 
problem and the appropriate method to resolve this is strongly problem 
dependent. For a lot of applications a simple `sticky particle' 
approximation (merging two particles while conserving momentum) may be 
sufficient, whereas in other cases (e.g. runaway collisions in dense clusters) 
a more precise characterisation of the internal structure of the 
collision product is necessary and in this case the collision has to be 
calculated using more sophisticated codes, e.g. entropy mixing (using 
MMAS) or a self-consistent SPH stellar collision simulation.

Figure~\ref{fig:evocoll} shows an example of the stellar evolution in a 
star cluster with stellar collisions. For this calculation a model for a 
dense Pop III star cluster was evolved. Because of the high stellar 
density and the adapted top-heavy IMF a run-away merger process occurs. 
The codes that where used in this calculation were: MESA for stellar 
evolution with zero metallicity stellar models, ph4 for the 
gravitational dynamics and MMAMS was used to resolve the structure of 
merger products. Note that in this case the fallback option for stellar 
evolution is not available - the models were actually stopped as soon as 
a resulting stellar model did not converge.

\subsection{Multiples\label{sec:multiples}}

The pattern of collision detection and resolution can also be applied 
to close passages and low multiples interactions in purely gravitational 
dynamics in the collisional regime. In this case the collision does not 
involve physical collisions between stars, but close passages, and the 
systems involved may be multiple stellar systems. The advantage of 
resolving these seperately lies in the short timescales of the 
interactions - handling these by a seperate code can be more efficient, 
or these interactions may require a different integrator. 

Such interactions are handled the same way as the physical 
collisions above: a parent code detects a close passage (using an 
interaction radius instead of the physical radius), flags a stopping 
condition and returns control to the AMUSE framework. Within AMUSE the 
\texttt{multiples} module implements this type of solver (see fig.~\ref 
{fig:multiples}). The AMUSE framework identifies the particles involved 
and handles them over to the resolving subcode, which may be for example 
a Kepler solver, smallN (a code which uses algorithmic regularization) 
or Mikkola (in case relativistic corrections are important). This code 
runs until the low N interaction is finished and returns the 
`collision' products, which may be single stars, stable binaries or 
hierachical multiples on outgoing trajectories. An additional step may 
be required to scale back the position to the original interaction 
radius, since the interactions are assumed to occur instantaneously. 

\subsection{Bridge integrator\label{sec:bridge}}

\begin{figure*}
 \centering
 \epsfig{file=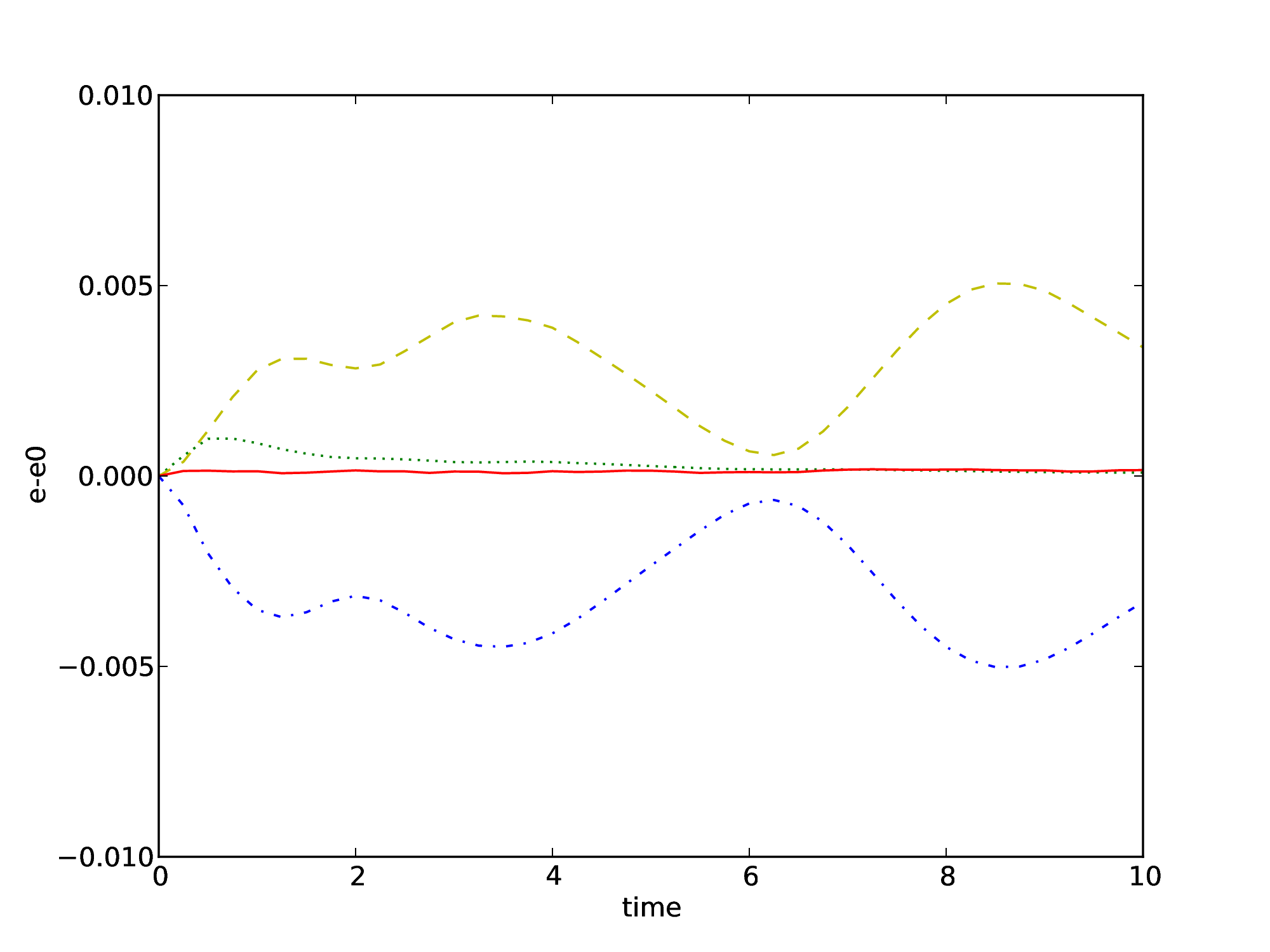, width=.49\textwidth}
 \epsfig{file=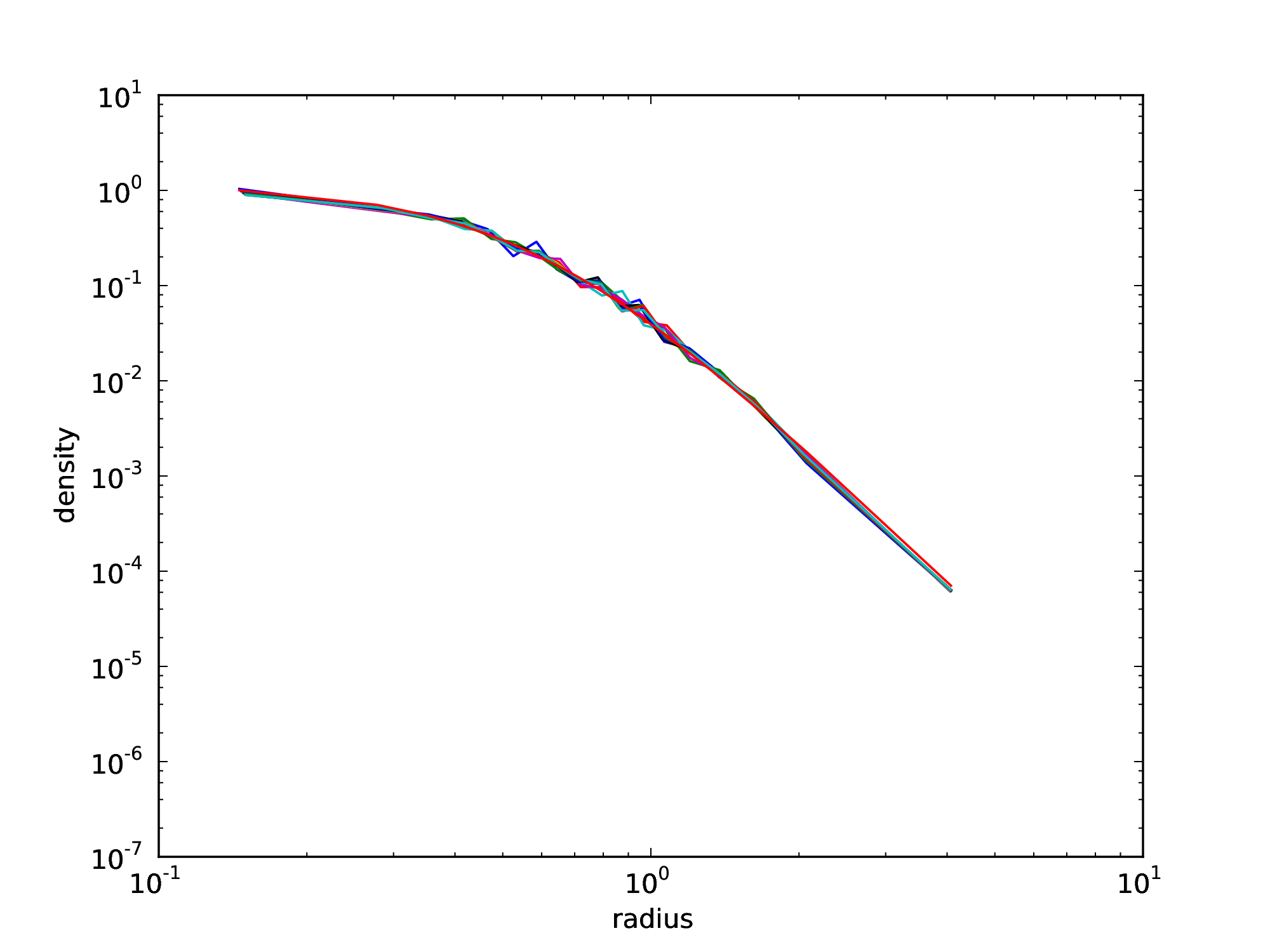, width=.49\textwidth}
 \epsfig{file=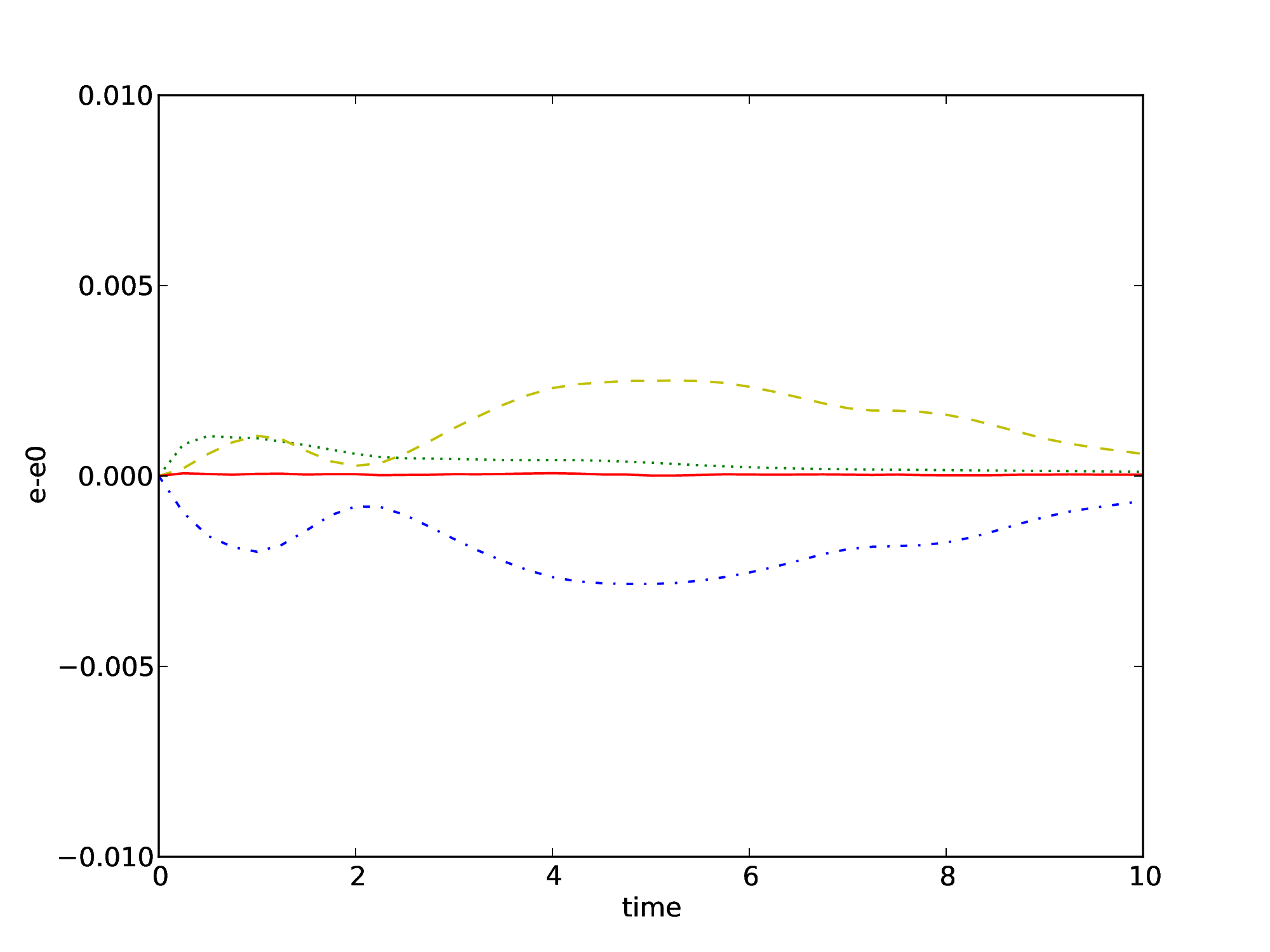, width=.49\textwidth}
 \epsfig{file=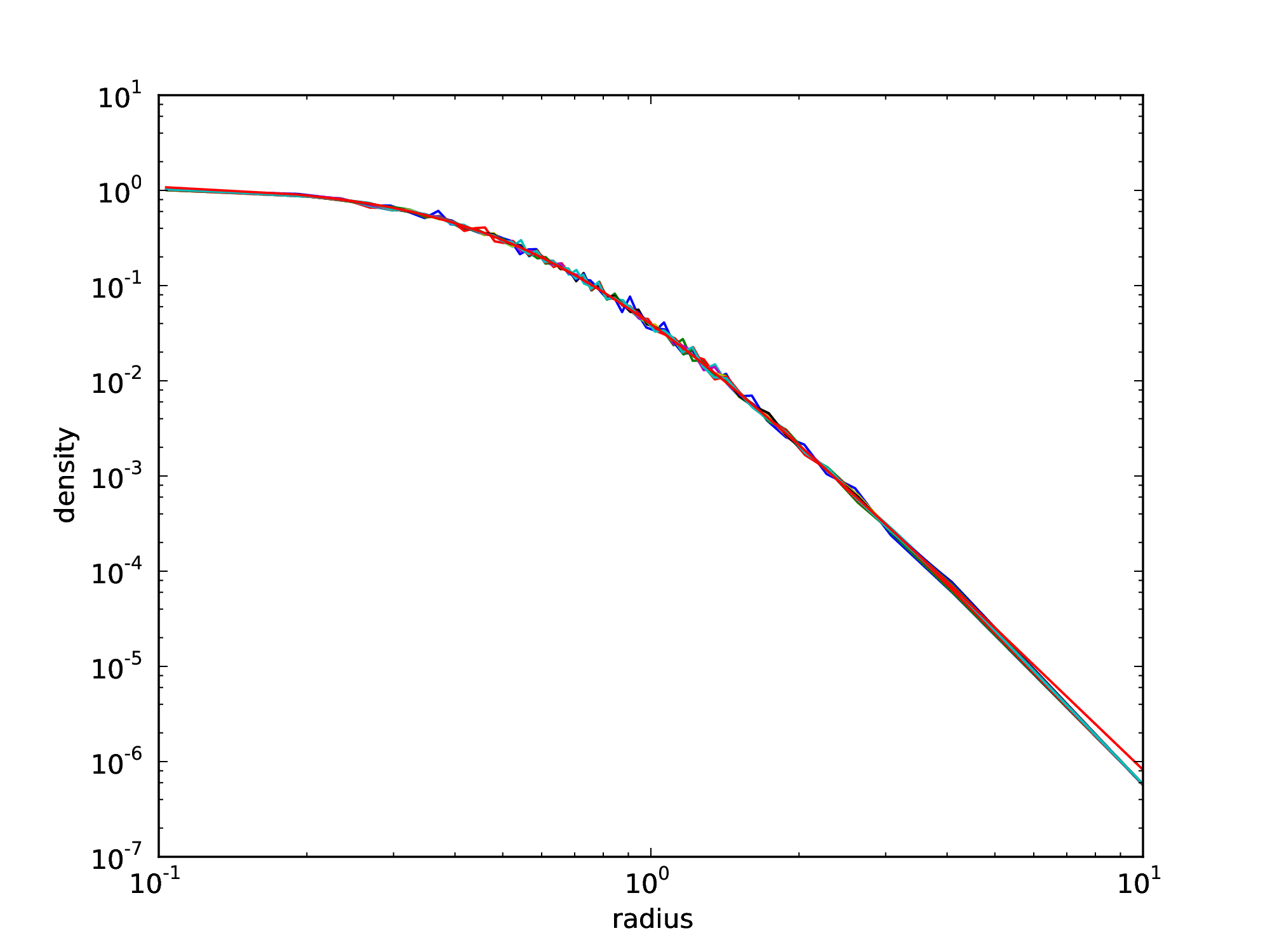, width=.49\textwidth}
 \caption{ 
 Stable gas plummer sphere test for Bridge. The test consists of 
 evolving a gas Plummer sphere for 10 Nbody times where the hydro forces 
 and self gravity are split over two instances of Fi using Bridge. Left 
 panels show the deviation of the potential (yellow dashed), kinetic 
 (green dotted), thermal (blue dash-dotted) and total energy (red drawn) 
 as a function of time (initial energy values are -0.5, 0, 0.25 and 
 -0.25 respectively). Right panels show the gas density profiles at 10 
 equally spaced times. Upper panel show results for $N=10^4$ particles, 
 while lower panels are for $N=10^5$.
 }
 \label{fig:plummer}
\end{figure*}

The Bridge integrator \citep{Fujii2007} provides a symplectic mapping 
for the gravitational evolution in cases were the dynamics of a system 
can be split in different regimes. 
The compound nature of the Bridge integrator 
makes it an ideal target for inclusion within AMUSE and we have 
implemented a generalized Bridge-type gravitational integrator in AMUSE 
\citep[see also][]{PortegiesZwart2013b}. 

The Bridge formulation can be derived from an Hamiltonian splitting 
argument, as used in planetary dynamics to derive symplectic integrators
\citep{Fujii2007, WisdomHolman1991, Duncan1998}. Consider the Hamiltonian  
\begin{eqnarray}
\label{eq:H}
H & = & \sum_{i \in A \cup B} \frac{p_i^2}{2 m_i} + 
        \sum_{i \ne j \in A \cup B} \frac{G m_i m_j}{\| {r_i-r_j} \| }     
\end{eqnarray}
of a system of particles consisting of subsystems A and B. 
This can be divided into three parts,
\begin{eqnarray}
\label{eq:splitH}
H & = & \sum_{i \in A} \frac{p_i^2}{2 m_i} + 
        \sum_{i \ne j \in A} \frac{G m_i m_j}{\| {r_i-r_j} \|} + \nonumber \\
  & &   \sum_{i \in B} \frac{p_i^2}{2 m_i} +
        \sum_{i \ne j \in B} \frac{G m_i m_j}{\|{r_i-r_j} \|} + \\
  &   & \sum_{i \in A, j \in B} \frac{G m_i m_j}{\| {r_i-r_j} \| } \nonumber \\ 
  & = & H_A + H_B + H^{int}_{A,B} \nonumber   
\end{eqnarray}
with $H_A$ and $H_B$ the Hamiltonians of subsystems A and B
respectively and the cross terms are collected in $H^{int}$. The formal 
time evolution operator of the system can then be written as:
\begin{eqnarray}
e^{\tau H} & = &
e^{\tau/2 H^{int}} e^{\tau \left(H_A+H_B\right)} e^{\tau/2 H^{int}} \nonumber \\
& = & e^{\tau/2 H^{int}} e^{\tau H_A} e^{\tau H_B} e^{\tau/2 H^{int}}.
\end{eqnarray}
Here the $e^{\tau/2 H^{int}}$ operator consists of pure momentum kicks 
since $H^{int}$ depends only on the positions. The secular evolution 
part $e^{ \tau (H_A+H_B)}$ splits since $H_A$ and $H_B$ are independent. 
The evolution of the total system over a timestep $\tau$ can thus be 
calculated by mutually kicking the component systems A and B. This 
involves calculating the forces exerted by system A on B and vice versa 
and advancing the momenta for a time $\tau/2$. Next the two systems are 
evolved in isolation (each with a suitable method) for a time $\tau$, after 
which the timestep is finished by another mutual kick. 

Within AMUSE, the gravitational dynamics interfaces (table~\ref{tab:grav}) 
provide for an \texttt
{evolve\_model} method that can be regarded as an implementation of the 
secular time evolution operators $e^{\tau H}$. The momentum kicks, on the other 
hand, are reasonably fast to implement within the framework in Python 
once the forces are known, and for this the gravitational dynamics 
interface provides a utility function \texttt {get\_gravity\_at\_point} 
to query the gravitational forces on arbitrary positions.

Note that the above formulation provides fully symplectic evolution. 
However in practice the implementation may not be symplectic, because 
the user is free to choose - and will do so often considering 
performance - non-symplectic codes for the component integrators 
and/or approximate methods to calculate the acceleration of the kick 
operators. Note also that the procedure is not restricted to two systems 
- the formulation extends to multiple systems by either splitting the 
Hamiltonian into more parts or applying the above split recursively, and
the implementation in AMUSE allows for an arbitrary number of systems. 

\subsubsection{Gravity-hydrodynamics coupling with Bridge\label{sec:btest}}

\begin{figure}
 \centering
 \epsfig{file=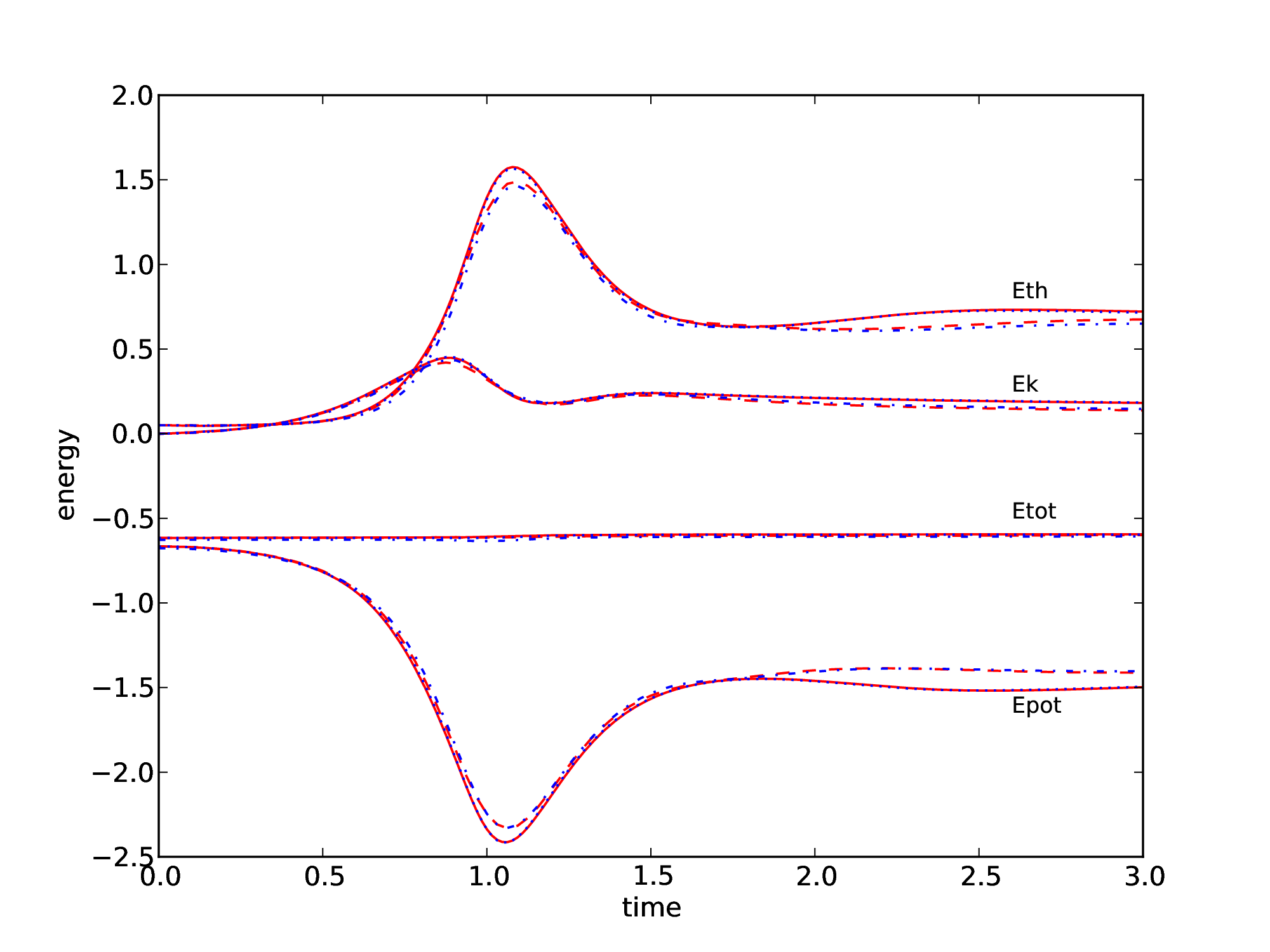, width=.49\textwidth}
 \caption{
 Energy plot of the Evrard collapse test. Plotted are the kinetic, 
 potential, thermal and total energy as a function of time for runs 
 with Gadget (red dashed: $N=10^4$, red drawn: $N=10^5$), and  
 for Gadget hydro bridged with Octgrav gravity (blue dash-dotted: $N=10^4$, Blue 
 dotted: $N=10^5$). 
 }
 \label{fig:evrard}
\end{figure}

\begin{figure}
 \centering
 \epsfig{file=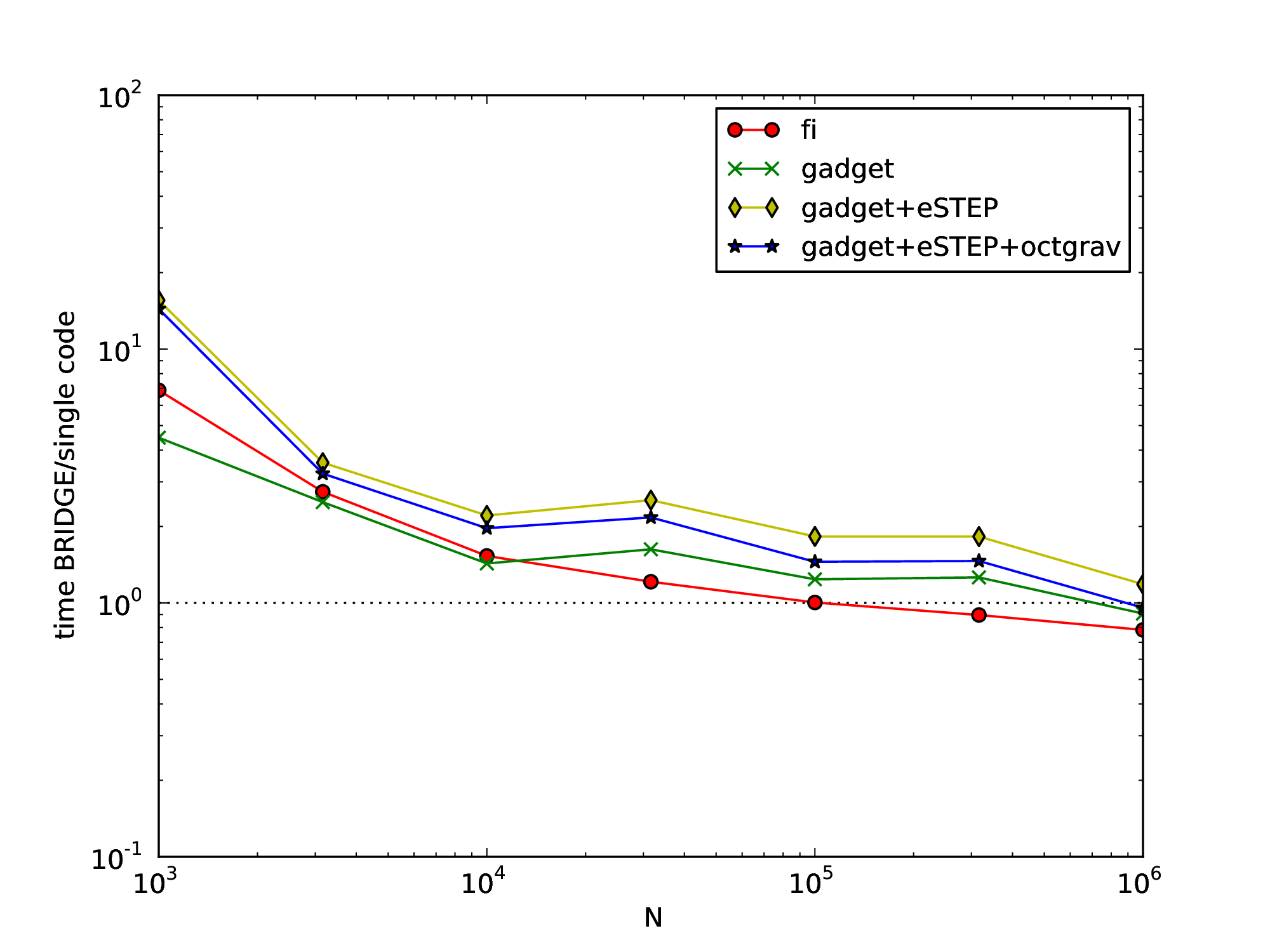, width=.49\textwidth}
 \caption{
 Performance of AMUSE split solvers vs monolithic solvers. Plotted is 
 the ratio of wallclock times using the AMUSE split solver and using the 
 corresponding TreeSPH code (see text for details) as a function of the 
 number $N$ of gas particles.
 }
 \label{fig:splitperf}
\end{figure}

As an illustration of the Bridge integrator and of the tests necessary to
verify a coupling strategy within AMUSE, we will present  
tests of the coupling between gravitational and hydrodynamic SPH 
systems. The dynamical equations for SPH evolution can also be derived 
from a Hamiltonian formalism and thus the Bridge formalism directly 
carries over to a split between purely gravitational and SPH particles~ 
\citep{Saitoh2010}. We test the Bridge integrator for an equilibrium 
configuration using a plummer sphere composed of gas, and for the 
classic Evrard SPH test. In these cases there is no clear spatial or 
scale separation, and thus this represents a challenging test of the 
efficiency of the AMUSE framework. A Bridge type integrator for SPH can 
be implemented in a computationally efficient way within a monolithic 
code if the time step requirement for the hydrodynamics is  more 
stringent than for the gravitational dynamics \citep[][]{Saitoh2010}. 
Below we also test the performance of an implementation within the AMUSE
framework.

In the test presented in figure~\ref{fig:plummer} we evolve a gas 
plummer sphere for 10 Nbody times using Bridge with two different 
instances of Fi: one for the hydrodynamics and one for the self-gravity. 
The gas initially is at rest and thermally supported and should remain 
stable. This configuration tests the ability of Bridge to maintain such 
a stable configuration in spite of the alternation of evolution 
operators. As can be seen the Bridge integrator conserves energy and 
maintains the equilibrium gas distribution satisfactorily to within 
$\sim 0.1\%$. A slight relaxation of the initial conditions is visible.

The second Bridge test consists of a dynamic test where we conduct the 
classic \citep{Evrard1988} test of a collapsing gas sphere. We compare 
the results of the test using a conventional TreeSPH code (Gadget) with 
a Bridge solver where different codes are used to calculate the 
hydrodynamics and the self-gravity (in this case the self-gravity of the 
SPH code is turned off). As can be seen in figure~\ref{fig:evrard} 
essentially the same results are obtained using the Bridge solver and 
the monolithic solver.

We also examine the performance of the hydrodynamics-gravity coupling 
using the Bridge integrator. \cite {Saitoh2010} showed that for an 
implementation of Bridge within a single code, performance gains were 
possible due to the fact that the gravitational and hydrodynamic force 
evaluations were better matched with their respective timestep criteria. 
Although performance is not the prime motive for AMUSE, this kind of 
coupling represents a challenge due to the strong coupling between the 
different sub systems, so it is instructive to see how the performance 
of AMUSE compares with using a monolithic code.

The simulations performed start out with a plummer sphere of mixed gas 
and star particles, with $1000$ equal mass star particles and a varying 
number of gas particles. The run time is compared with a monolithic code 
(either Fi or Gadget) performing the same calculation. Because these 
codes are targeted to collisionless dynamics, the stellar gravitational 
interactions were smoothed using a smoothing length $\epsilon=0.01$ (in 
Nbody units).

\begin{figure*}
 \centering
 \epsfig{file=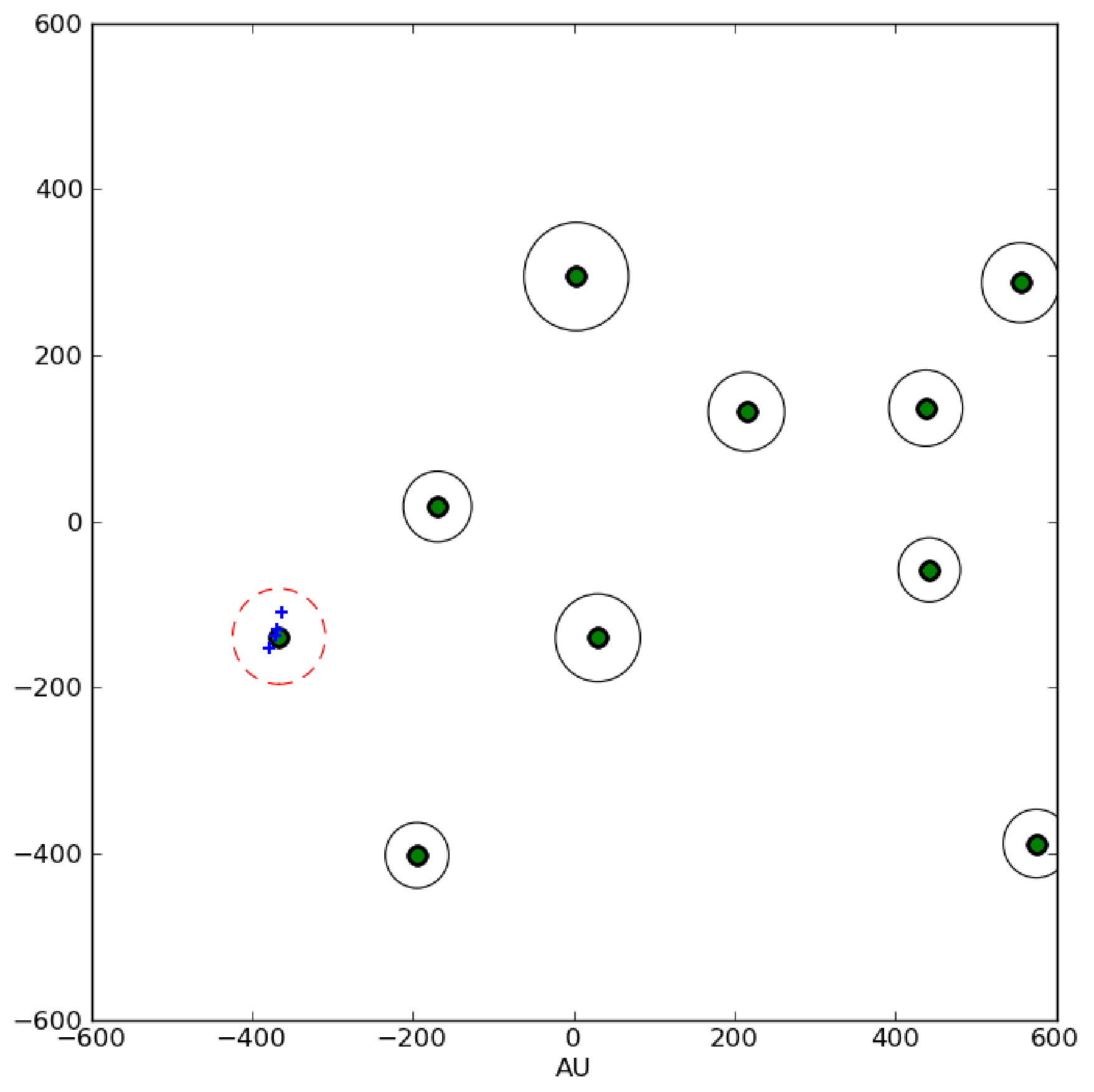, width=.39\textwidth}
 \epsfig{file=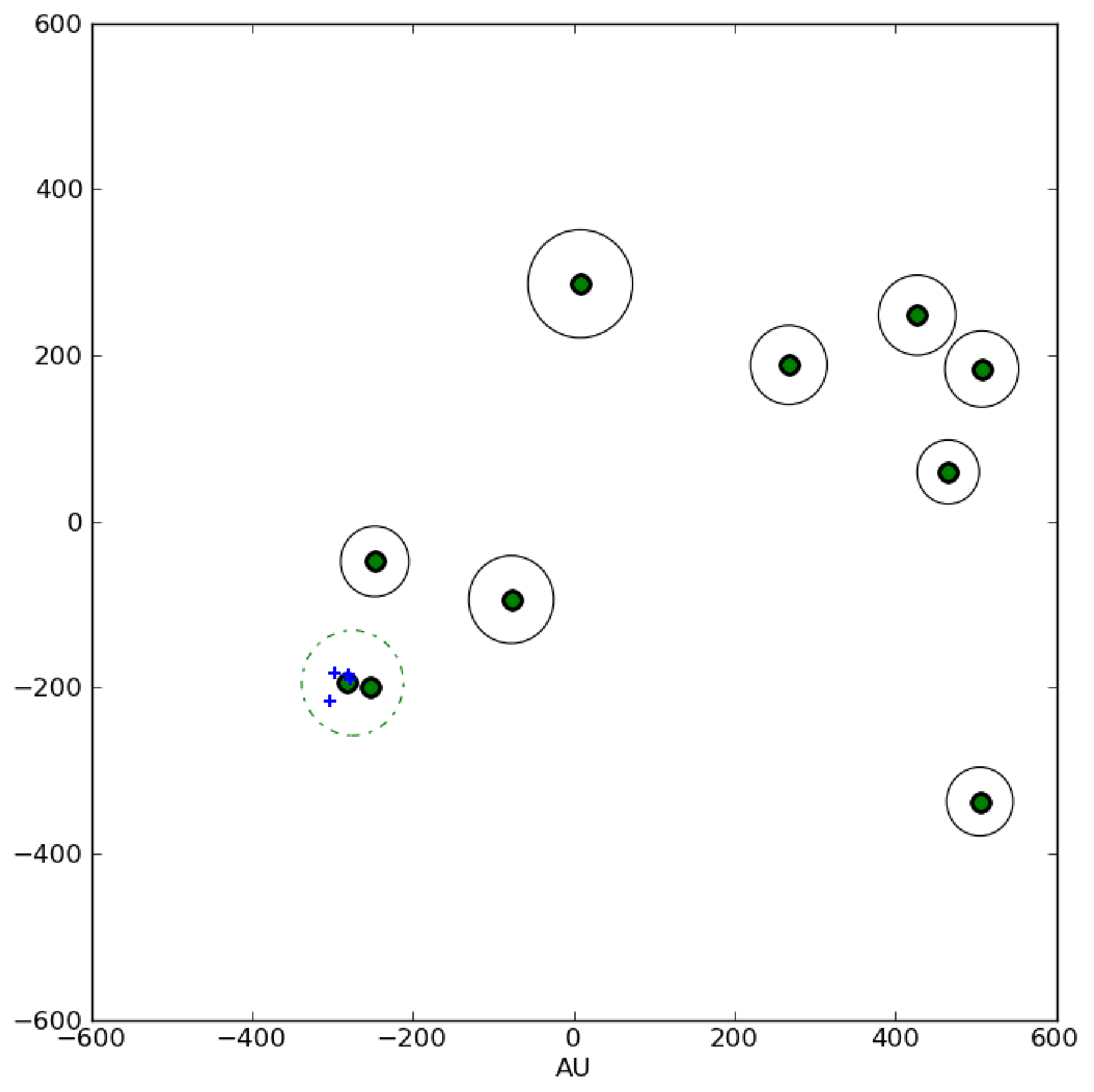, width=.39\textwidth}
 \epsfig{file=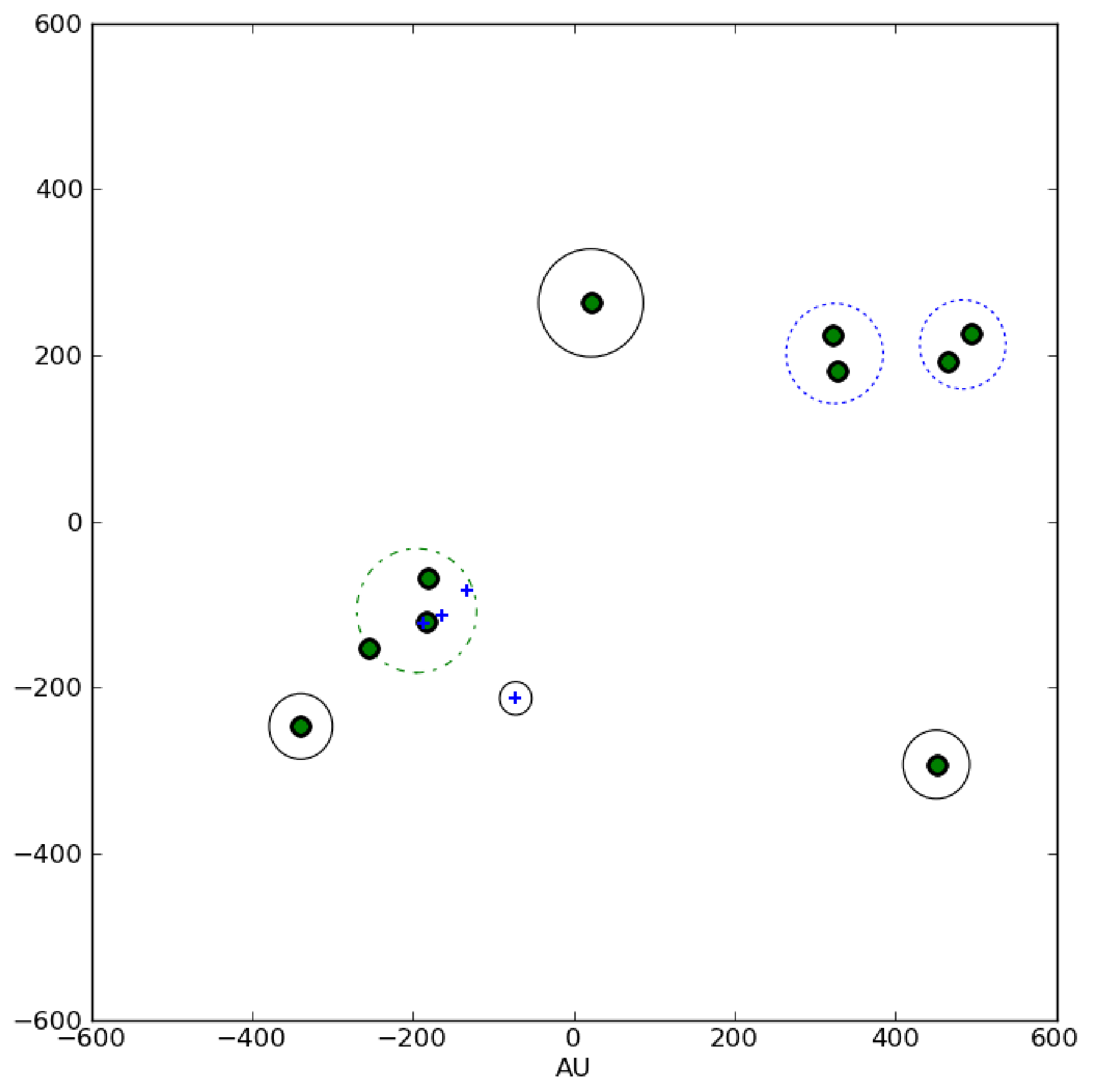, width=.39\textwidth}
 \epsfig{file=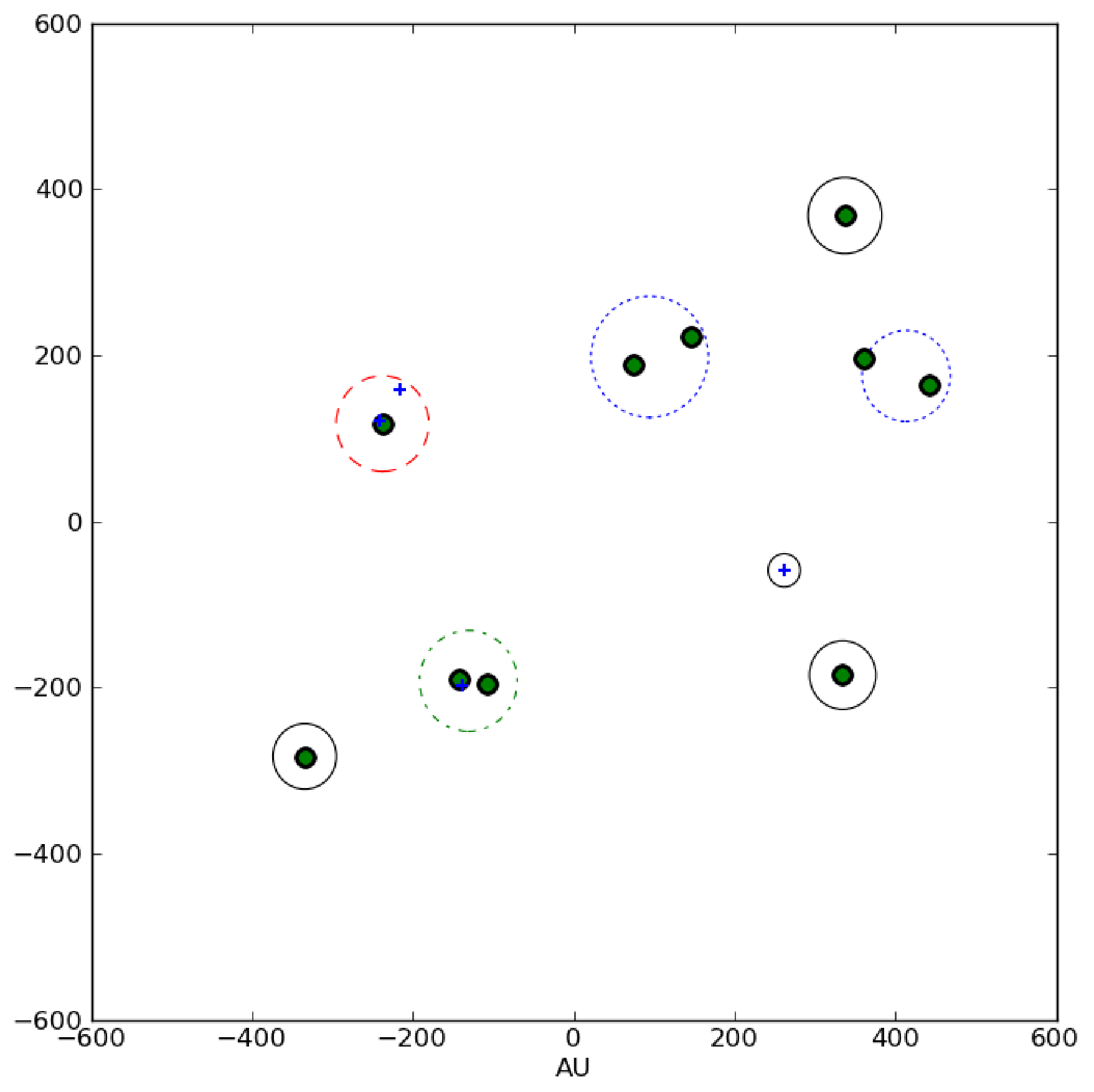, width=.39\textwidth}
 \caption{
 Solver for planetary interactions. Plotted are different frames of 
 a test simulation where circles are drawn around each center of mass node 
 in the parent integrator (Hermite0), where the linestyle indicates the 
 type of subnode: black solid indicates  a single particle in the parent 
 code, blue dotted a two-body Kepler solver, green dash-dotted a general 
 N-body code (Huayno) and red dashed a symplectic planetary 
 integrator (Mercury). The green dots are stellar systems, the plusses are 
 planets.
 }
 \label{fig:hymn}
\end{figure*}
 
Figure~\ref{fig:splitperf} shows the timing results for 4 different 
simulation sets, which are labelled with the code used: Fi on a 4 
core workstation, Gadget using 4 MPI processes on one workstation, 
Gadget split over 2 workstations using eSTeP and Gadget split over 2 
workstations using eSTeP while using the GPU code Octgrav on a different 
machine to calculate gravity. As can be seen in the figure the overhead 
of using AMUSE is large for small N. This is to be expected: each call 
in the framework is transported using either MPI or eSTeP, which is 
considerably more costly than an in-code function call. However, 
increasing the number of particles the relative overhead of the AMUSE 
framework goes down and for $N=10^6$ the AMUSE split solvers are 
competetive with the monolithic code. The reason for this is that the 
communication scales at most linearly with $N$, while the scaling of the 
computation goes as $N\ {\rm log}\ N$. Noting the two runs with eSTeP, 
we see that this imposes an extra overhead. However if running on 
multiple machines entails a more efficient use of resources (here 
exemplified by using a GPU enabled treecode to calculate the gravity) 
spreading out the computation on machines with different architectures 
can also increase the efficiency of the computation (although for this 
particular problem not enough to offset the communication cost).

\subsection{Strong and weak coupling of planetary systems in clusters\label{sec:planetcluster}}

The Bridge integrator is suitable for problems where one or a small 
number of composite particles is embedded in a larger system. A limitation of 
the normal bridge is the fact that the interactions between a composite particle and
the rest of the system are calculated using 2nd order leap-frog splitting.
When most or all `particles' in a larger system are compound, say a 
cluster of stars where the stars are multiple systems or have planets 
orbiting them, the Bridge integrator is equivalent to a 2nd order 
leap-frog implemented in Python. Furthermore, in this case it is possible that compound 
systems will encounter each other and interact, so provisions for this 
have also to be made.

For such a case one can formulate a variant of bridge where all 
particles, including the compound particles, are evolved in a parent 
code (which can be e.g. a high order Hermite code). The compound 
particles have a representation in the parent code in the form of a 
center of mass particle with an associated interaction radius. This 
center of mass particle is evolved in the parent code, while the 
compound subsystem is evolved by a different code, with perturbations 
due to the rest of the system. The perturbations consist of 
periodic velocity kicks. Subsystems are allowed to merge if they pass 
close to each other (using the collision stopping condition of the 
parent code) and can be split if seperate associations within the 
subsystem are identified. Such a method is implemented in the Nemesis 
integrator (Geller et al., in prep.). Figure~\ref {fig:hymn} shows an example, where we evolve a 
random configuration of stars (the initial condition is chosen planar 
for clarity, so it does not represent any realistic object), with 
initially one compound system (the sun with the outer gas giants). As 
the stars evolve,  the integrator detects close passages between 
systems, identifies subsystems and decides dynamically the most suitable 
integrator for such a system, choosing a Kepler solver if the subsystem 
consists of two bodies, the solar system integrator Mercury when the 
heaviest particle is more than a factor 100 heavier than the second most 
massive particle or the Huayno integrator otherwise. 

Although the Nemesis integrator and the Multiples module (section~\ref 
{sec:multiples}) have similar goals, they differ in the handling of 
compound systems. Interactions in the Nemesis code proceed on the same 
timeline as the parent system (it is not assumed to be resolved 
instantaneously) and feel (albeit in a perturbation sense) the large 
scale density distribution, while for Multiples the interactions are 
resolved in isolation and instantaneously. This does result in more 
stringent timestepping constraints for the Nemesis integrator, though.

\subsection{Gravitational dynamics and hydrodynamics with stellar feedback\label{sec:gravgassse}}

\begin{figure}
\centering
\epsfig{file=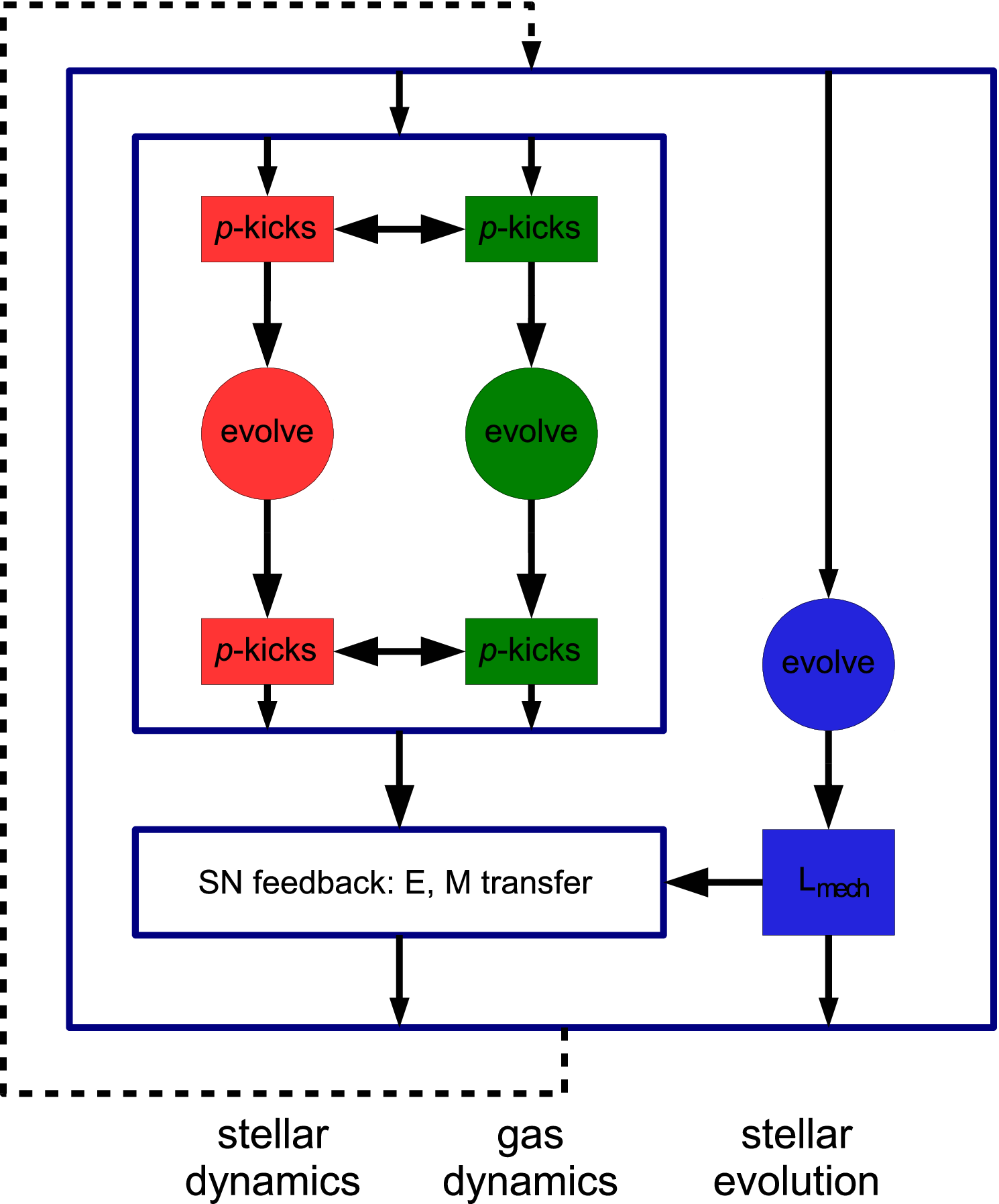, width=.49\textwidth}
\caption{AMUSE gravitational/hydrodynamic/stellar evolution integrator. 
This diagram shows the calling sequence of the different AMUSE elements in
the combined gravitational/hydro/stellar solver during a time step of the 
combined solver. Circles indicate calls to the (optimized) component solvers, 
while rectangles indicate parts of the solver implemented in Python 
within AMUSE~\citep[from][]{Pelupessy2012}. }
\label{fig_int}
\end{figure}

The cluster simulations with stellar evolution of section~\ref 
{sec:secluster} assumed the clusters consisted of stars only. This is 
not a good approximation for the earliest phase after the formation of a 
cluster, when the cluster is still embedded in the gas cloud from which 
it formed. In this case the interplay of stellar evolution, 
hydrodynamics and stellar dynamics becomes more complicated because 
young massive stars inject energy into the surrounding gas medium by 
stellar winds and supernovae. This has a drastic effect on the gas 
dynamics, which in turn affects the stellar dynamics.

The coupling between hydrodynamics and gravity is a special case of 
Bridge (see section~\ref{sec:bridge}). The feedback from stellar winds 
and supernovae can be implemented as energy source terms for the gas 
dynamics, where the mechanical luminosity $L_{\rm mech}$ is determined 
by the stellar evolution and parametrized mass loss rates and terminal 
wind velocities \citep[e.g.][]{Leitherer1992, Prinja1990}. The calling 
sequence of a combined gravitational/hydrodynamic/stellar evolution 
integrator is shown in figure~\ref{fig_int}. This integrator was used to 
study the early survival of embedded clusters (see section~\ref
{sec:embed})

\subsection{Radiative hydrodynamics\label{sec:radhydro}}

Another example of a close coupling is the interaction of radiation and 
hydrodynamics. In the general case, this is an example of a type 
of problem where a dedicated coupled solver may be necessary, but in 
many astrophysical applications an operator-splitting approximation is  
reasonably efficient. The operator splitting for system governed by 
radiative hydrodynamic equations can be effected in a 
`leap-frog' fashion: to advance a given system from time $t$ to 
$t+\Delta t$, a hydrodynamic solver is first advanced for a half step $\Delta t/2$ 
and then communicates its state variables at $t+\Delta t/2$ to the 
radiative transfer code. The radiative transfer code proceeds evolving
the radiation field, internal energy, ionization, etc, taking a 
full integration step. The internal energy from the radiative transfer 
code (at $t+\Delta t$) is then used to update internal energy of the 
hydrodynamics code, which then advances to $t+\Delta t$, completing the 
timestep. This approximation holds as long as the effects from the finite
speed of light can be ignored, and as long as $\Delta t$ is small enough 
that the reaction of the gas flow to the radiation field can be followed.
 We illustrate this simple scheme with an integrator based on 
an analytic cooling prescription (sec.~\ref{sec:therm}) and an 
integrator using a full radiative transfer code (sec.~\ref{sec:iliev2}).

\subsubsection{Thermally unstable ISM\label{sec:therm}}

\begin{figure*}
 \centering
 \epsfig{file=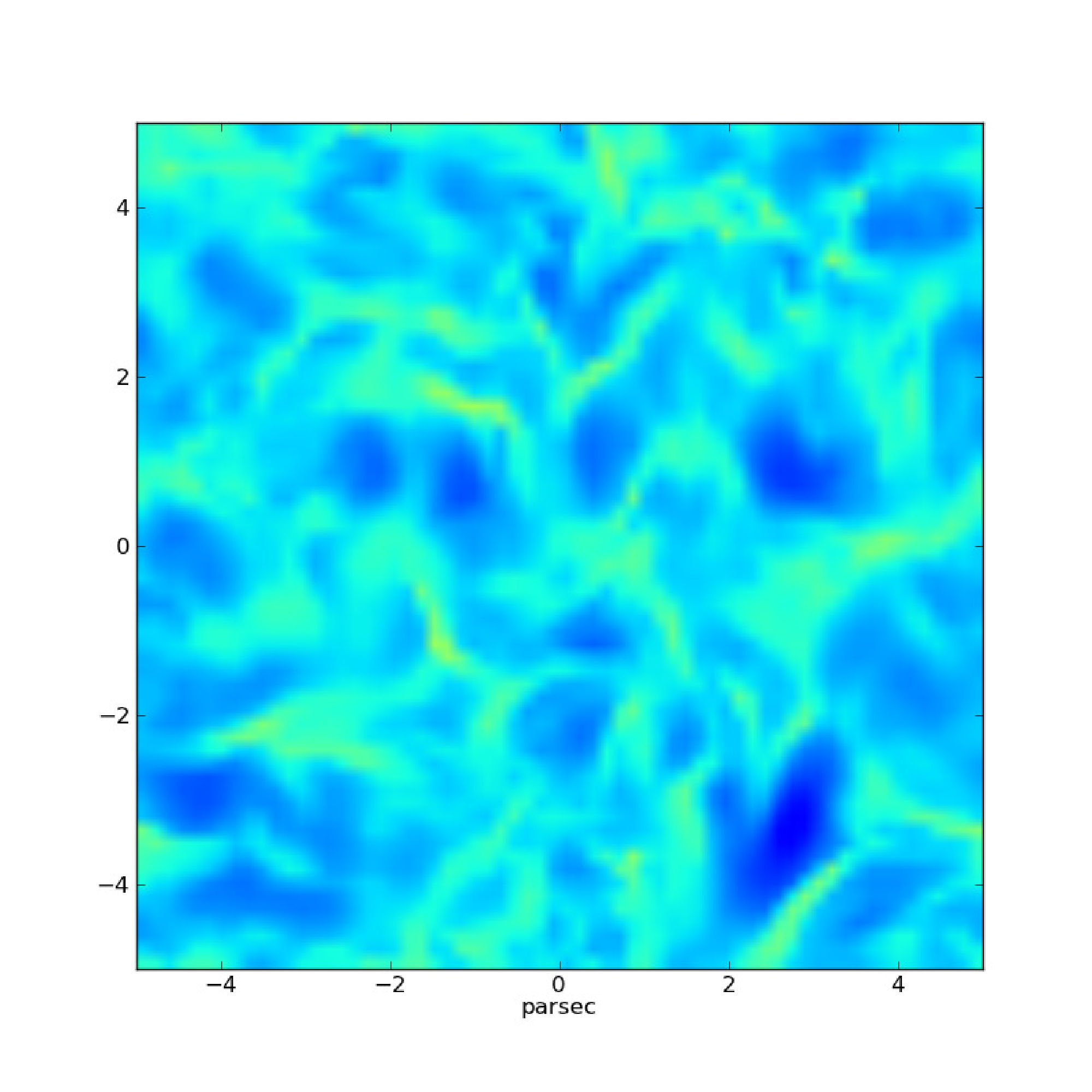, width=.49\textwidth}
 \epsfig{file=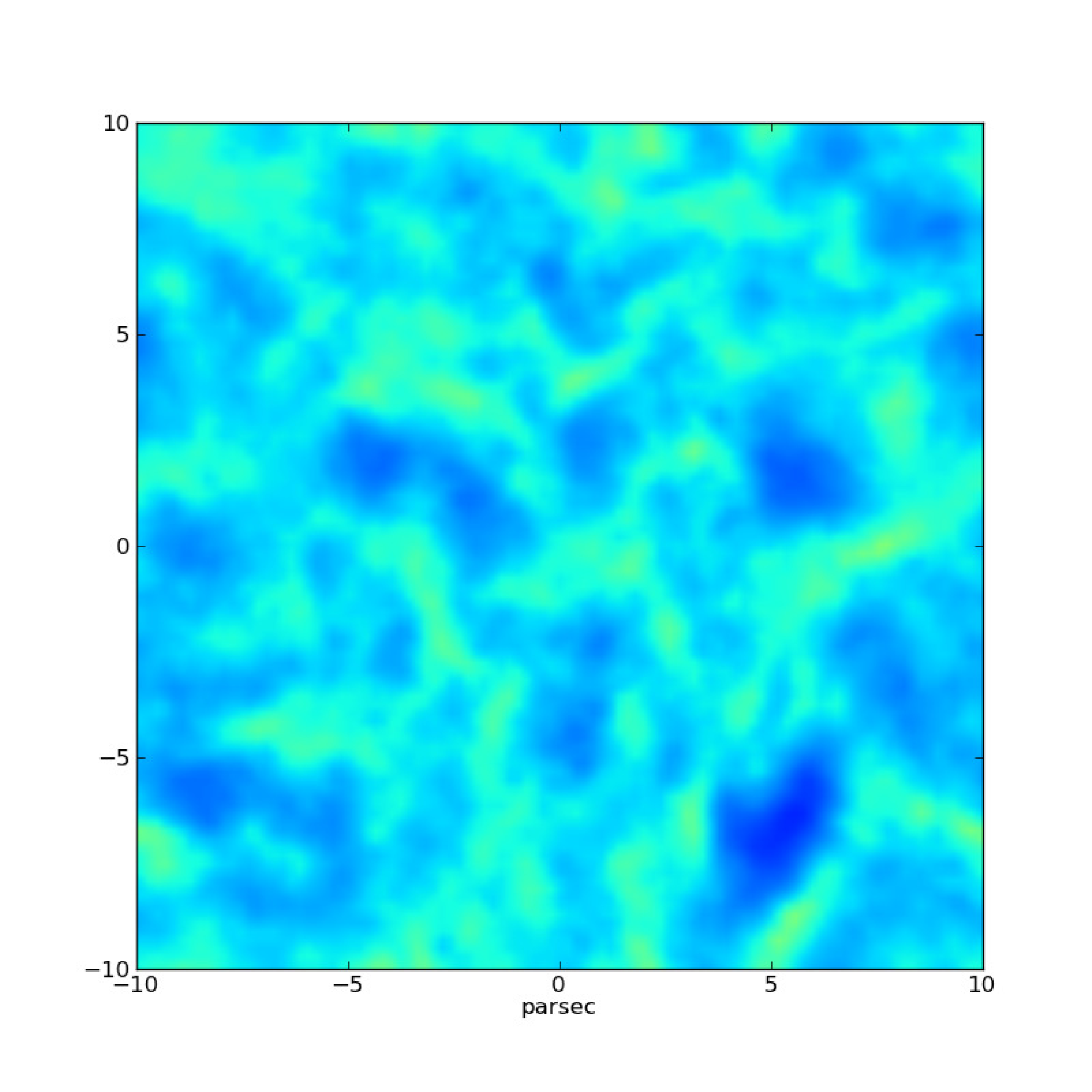, width=.49\textwidth}
 \epsfig{file=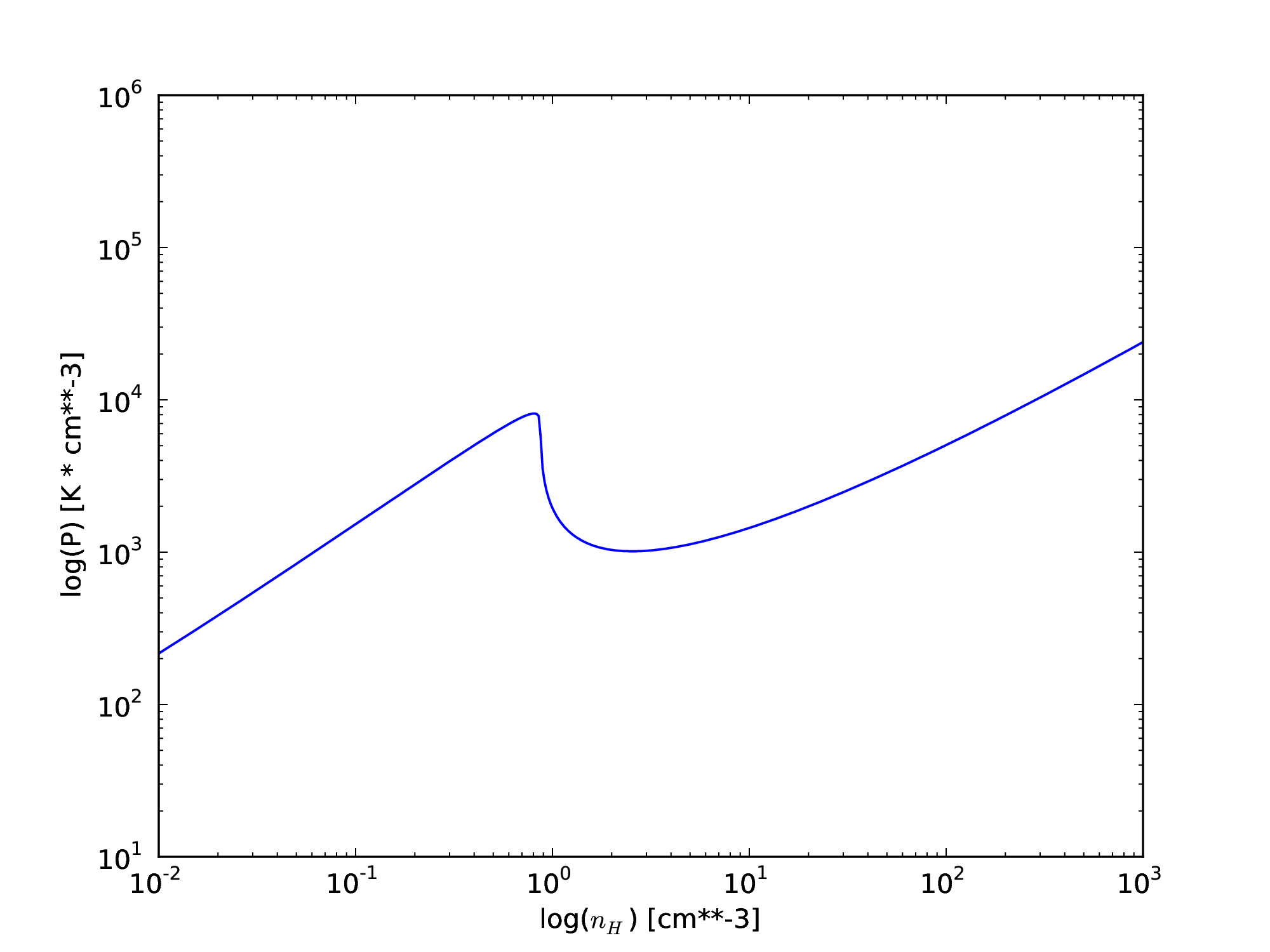, width=.49\textwidth}
 \epsfig{file=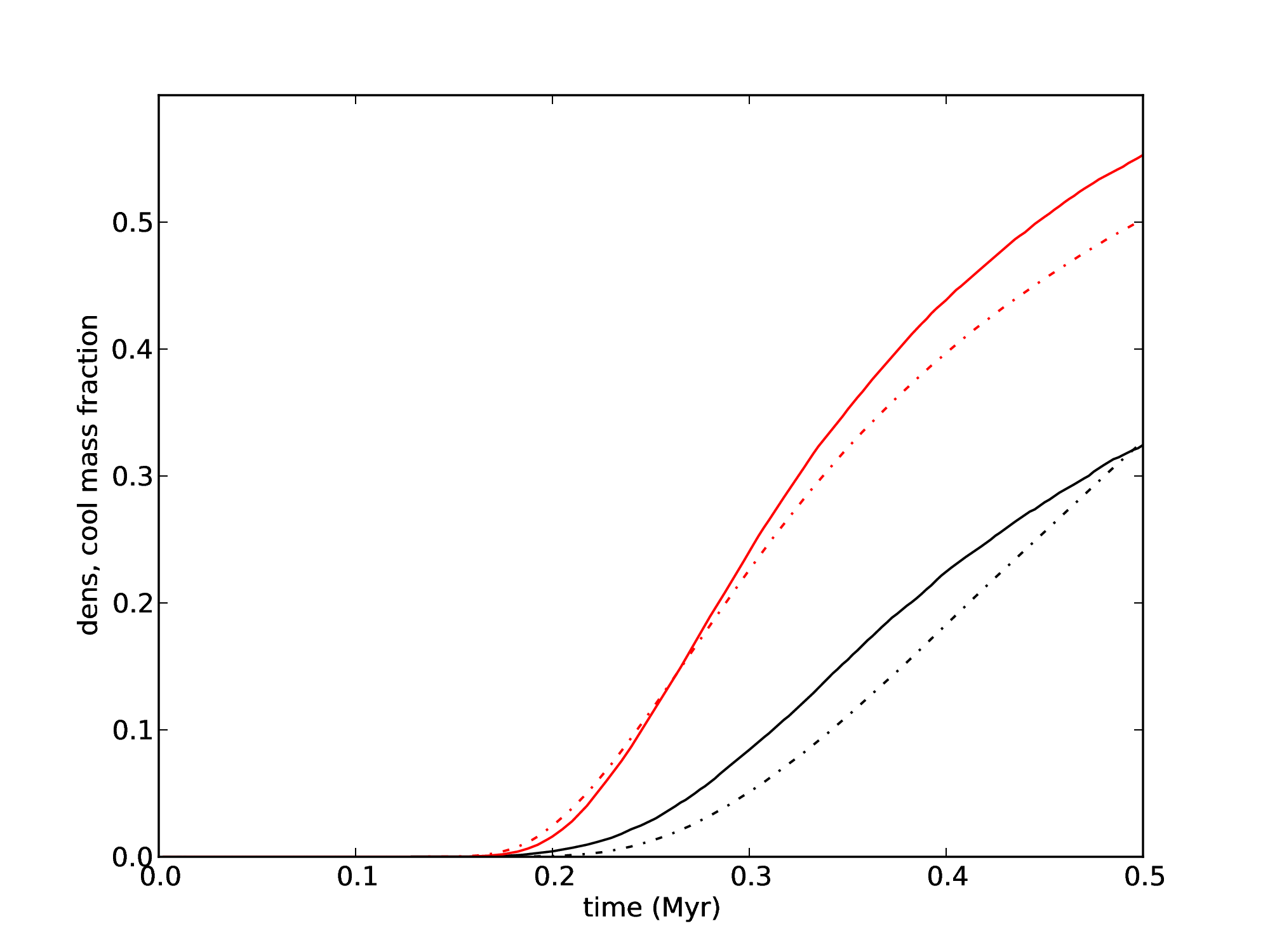, width=.49\textwidth}
 \caption{
Simulations of a thermally unstable ISM model. Upper panels show slices 
of the density field for the Athena grid code (left) and the SPH code Fi 
(right) after 0.5 Myr of evolution. In both cases the thermal evolution 
is calculated using constant heating and the simple analytic \cite
{Gerritsen1997} cooling curve, resulting in a thermally unstable model. 
This can be seen in the (equilibrium) density-pressure relation (lower 
left panel). In the lower right panel we show the evolution of the cool (
$T<100$ K, blue lines) and dense ($\rho>10$ cm$^{-3}$, red lines) gas 
fractions, where the solid lines show the results for Athena, 
dash-dotted those of Fi. Both models show broadly similar behaviour, although 
clearly differences can be seen, which must be traced to the use of different 
numerical hydrodynamic solvers.
}
 \label{fig:ism}
\end{figure*}

Probably the simplest case of this coupling concerns the cooling of the 
ISM in optically thin atomic cooling lines in the low density limit. In 
this case the radiative coupling can be described with a cooling 
function $\Lambda$, which is a function of the temperature $T$ for which 
various tabulations exist. Here we will use a simple analytic approximation 
\citep{Gerritsen1997},
\begin{eqnarray}
\Lambda(T) & = & 10^{-21} \times \left( 10^{(-0.1-1.88*(5.23-log(T))**4)} + \right. \nonumber \\ 
 & & \ \ \ \left. 10^{(-3.24-0.17*|(4-log(T)|)**3)}\right) {\rm \ erg\ cm}^3/{\rm s} \nonumber
\end{eqnarray}
with a constant heating $\Gamma=0.05 \times 10^{-23} n\ {\rm ergs}/{\rm 
cm}^3/{\rm s}$. While highly idealized, this results in a classic \cite 
{Field1965} two phase instability, as can be seen in the equilibrium 
density pressure relation in figure~\ref{fig:ism}. The dynamical effect 
of this instability in a turbulent ISM can be explored with the 
operator splitting radiative hydro scheme. An example of 
this is shown in figure~\ref {fig:ism}, where we have set up a periodic 
box 10 parsec accross with initially a uniform density $n=1.14\ {\rm 
cm}^{-3}$, temperature $T=8000\ K$ and a divergence free turbulent 
velocity spectrum with velocity dispersion $\sigma=8\ {\rm km}/{\rm s}$. 
We compare two completely different hydrodynamic methods, namely the SPH 
code Fi and the finite volume grid solver Athena. Because the initial 
conditions of the former consist of particles, while the latter is a 
grid code, some care must be taken in generating equivalent initial 
conditions. Here the initial hydrodynamic state grid for Athena was 
generated by sampling the hydrodynamic state from the SPH code. The 
thermal evolution was solved in Python by a simple ODE solver for the 
thermal balance. As can be seen in the snapshots shown in figure~ \ref 
{fig:ism} the two methods give qualitatively similar results, but 
clearly the chosen numerical method affects the details.

\subsubsection{Dynamic Iliev tests\label{sec:iliev2}}

\begin{figure*}
 \centering
 \epsfig{file=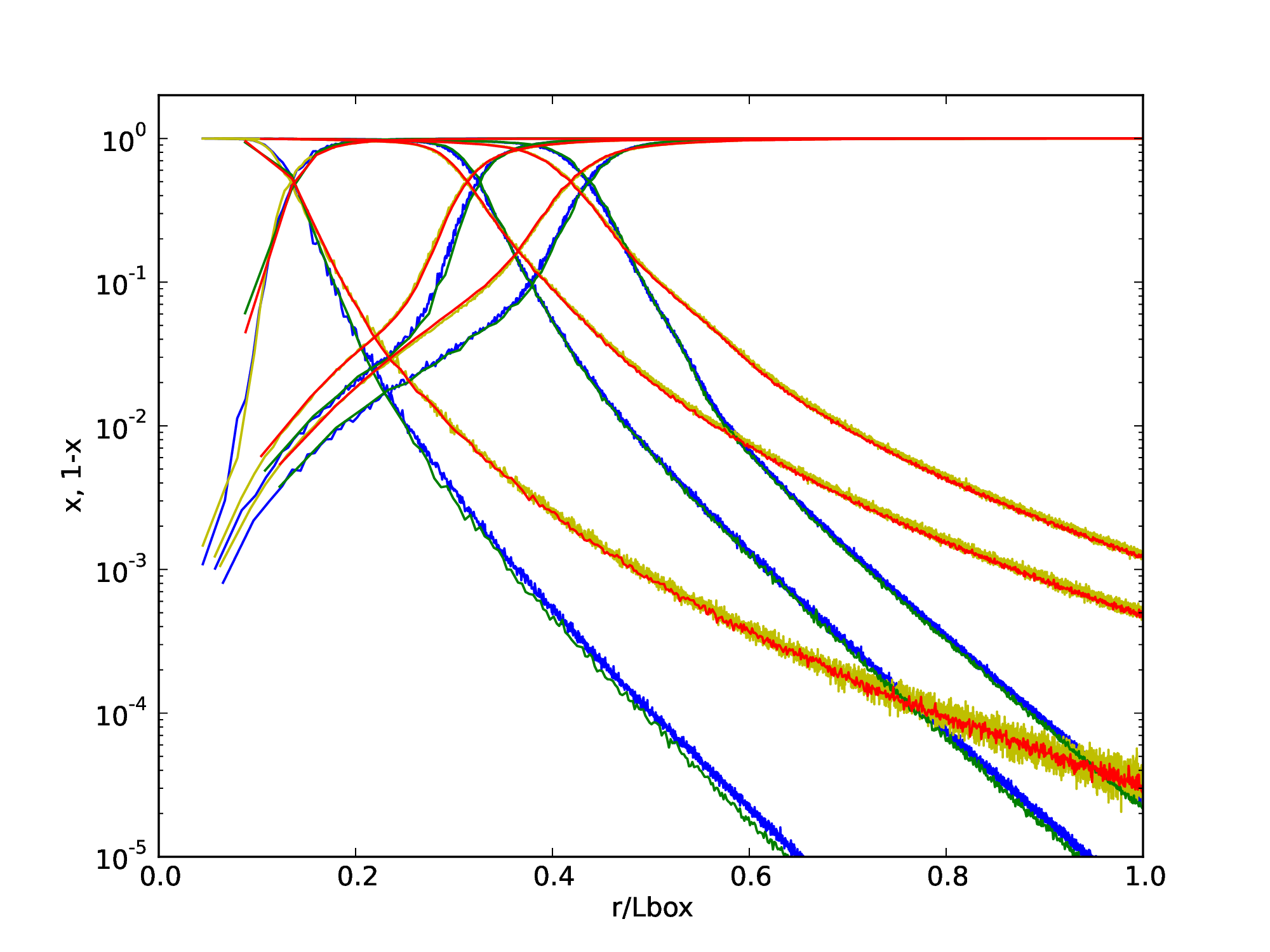, width=.45\textwidth}
 \epsfig{file=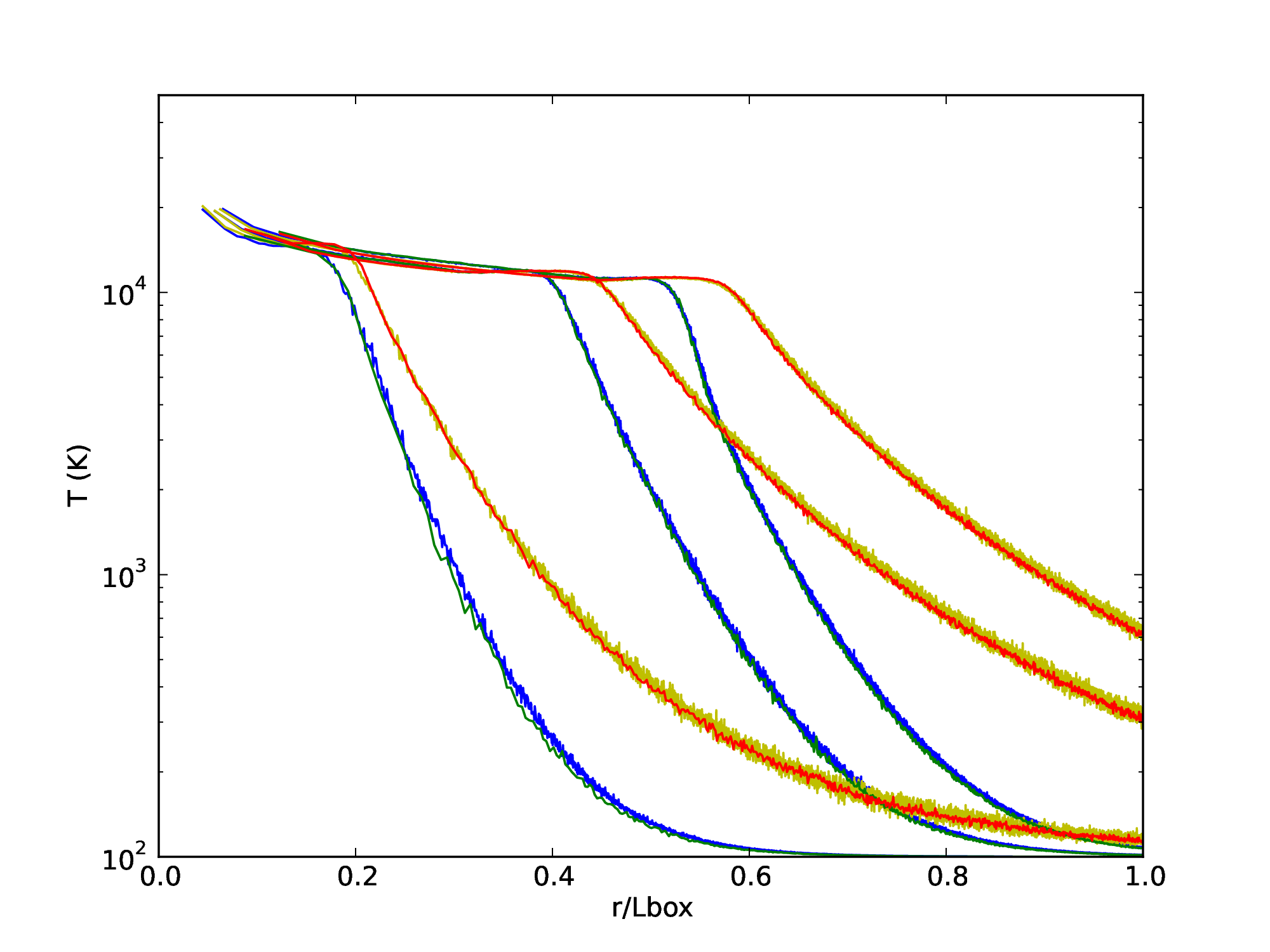, width=.45\textwidth}
 \epsfig{file=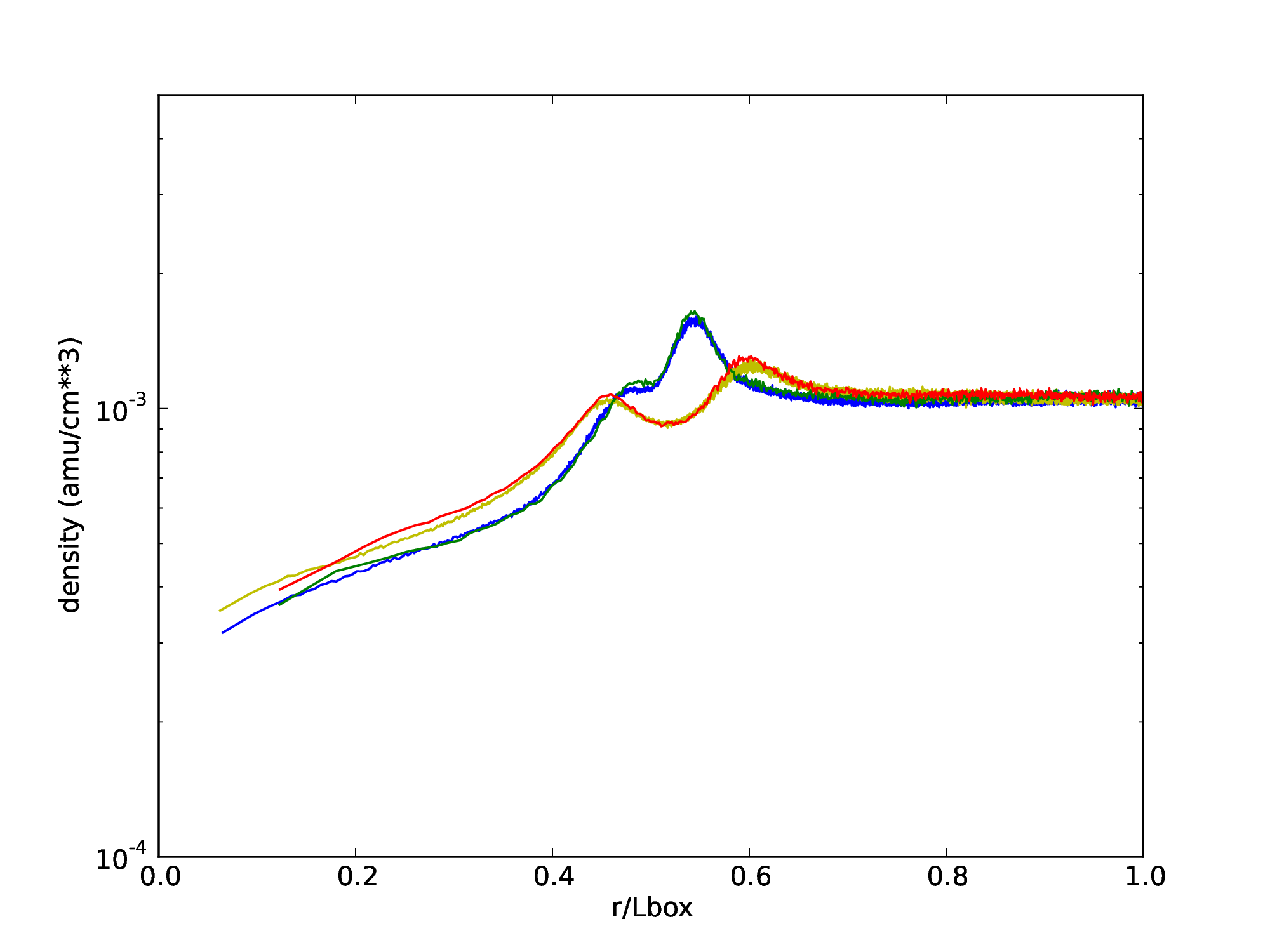, width=.45\textwidth}
 \caption{
 Test 5 of \cite{Iliev2009}: ionization front expansion in an initially 
 homogeneous medium. Shown are plots of the ionization fraction $x$ (top left panel), 
 temperature T (top right) and the density profile (bottom) as a function of the distance
 (normalized on the box size $L_{\rm box}$) to a source with a luminosity $L=5 
 \times 10^{48} {\rm s}^{-1}$ after 10, 30 and 200 Myr of evolution (for 
 density only the 200 Myr case is shown). Blue line: Fi with SimpleX, green: 
 Gadget2 with SimpleX, yellow: Fi with SPHRay, red: Gadget2 with SPHRay.
 }
 \label{fig:iliev2}
\end{figure*}

The leap-frog coupling scheme described above for optically thin cooling 
and heating can be applied also for the more computationally expensive 
case of the transport of ionizing radiation in an optically thick 
medium. In this case, instead of a simple ODE solver using only local 
information, the full radiation transport must be solved for the 
internal energy integration. Thus a radiative hydrodynamics solver can 
be constructed within AMUSE by coupling seperate hydrodynamic and 
radiative transfer solvers.

In figure~\ref{fig:iliev2} we show the result of an application of this 
coupling. Here we perform a test described in a comparison paper of 
different radiative-hydrodynamics codes~\citep  {Iliev2009}. This test 
\citep[test 5 of][]{Iliev2009} involves the calculation of the HII front 
expansion in an initially homogeneous medium. We evaluated the four 
different possible combinations of the two SPH codes (Fi and Gadget2) 
and the two radiative transfer codes (SPHRay and SimpleX) currently 
interfaced in AMUSE. Figure~\ref {fig:iliev2} shows the resulting 
evolution of the ionization structure, temperature and density for the 4 
different solvers. The resulting HII front expansion falls within the 
range of solutions encountered in \cite{Iliev2009}. Both radiative codes 
show different behaviour in the fall-off of the ionization and 
temperature. This is due to differences in the treatment of the high 
energy photons in SPHRay and SimpleX. A similar scatter in the profiles 
ahead of the ionization front is encountered in \cite {Iliev2009}. Note that
the hydrodynamic code chosen has relatively little impact on the resulting
profiles, since both codes are SPH codes. A more quantitive examination 
of the radiative hydrodynamics coupling within AMUSE is in preparation
(Clementel et al.).

\section{Applications\label{sec:res}}

A number of studies have been conducted using AMUSE, amongst which an 
examination of the origin of the millisecond pulsar J1903$+$0327, a 
pulsar with a companion in an unusually wide and eccentric orbit \citep 
{PortegiesZwart2011}. Here, AMUSE was used to resolve triple 
interactions. \cite{Whitehead2011} examined the influence of the choice 
of stellar evolution model on cluster evolution using AMUSE. The Huayno 
integrator~\citep{Pelupessy2012c} was developed within AMUSE. \cite 
{Kazandjian2012} studied the role of mechanical heating in PDR regions. 
Here AMUSE was used as a driver to conduct parameter studies. In \cite 
{PortegiesZwart2013} AMUSE was used in order to resolve stellar 
evolution and planetary dynamics after the common envelope phase in 
binary systems with planets. Whitehead et al (2013, in prep.) used AMUSE 
to examine stellar mass loss and its effect on cluster lifetimes. We will 
examine three other applications in some detail below, as they provide
instructive case studies. 

\subsection{Globular clusters in realistic galaxy potentials}

\begin{figure}
 \centering
 \epsfig{file=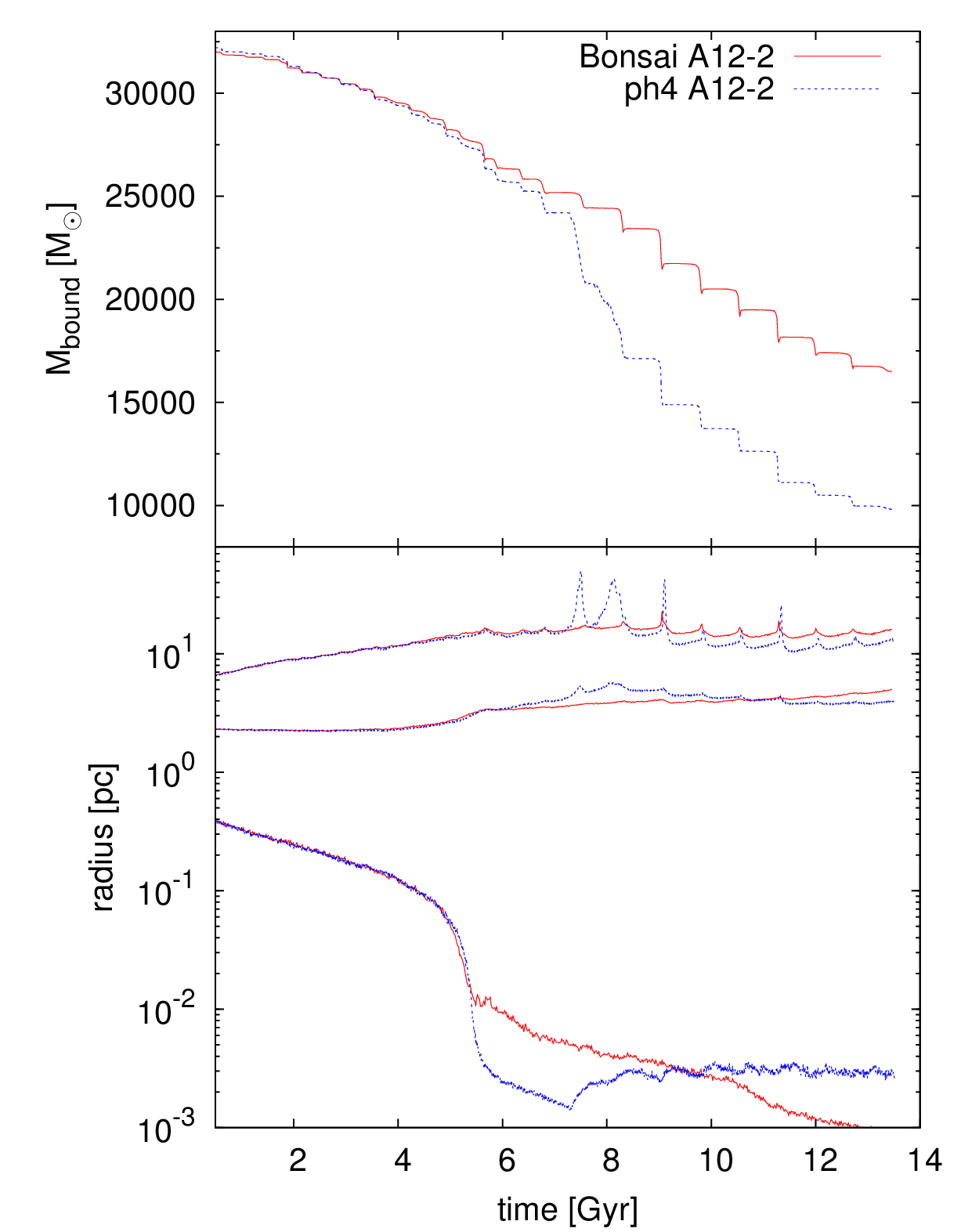, width=.49\textwidth}
 \caption[]{  
  Simulations of a globular cluster: comparison of Tree and direct Nbody 
  codes. Evolution of the bound mass (top panel) and the 90\%, 50\% and 
  1\% Lagrangian radii (top to bottom, bottom panel) of a simulated star 
  cluster subject to a galactic tidal field extracted from a 
  cosmological simulation (model A12-2 from Rieder et al. (2012, 
  submitted)). The cluster was simulated with the GPU treecode Bonsai 
  (red, solid curves) and Hermite direct N-body code ph4 (blue, dashed 
  curve). From Rieder et al. (2012).
  } \label{fig:TreeCodevsDirect}
\end{figure}

Rieder et al. (2012, submitted) studied the evolution of globular 
clusters in cosmological density fields. For this study realistic tidal 
fields extracted from the CosmoGrid simulation \citep{Ishiyama2013} were 
used as the time varying background potential against which globular 
cluster models were evolved using the Bridge integrator. Technically this 
means that a minimal gravity interface - only consisting of
\texttt {get\_gravity\_at\_point} and \texttt{get\_potential\_at\_point} and 
an \texttt{evolve\_model} method which interpolates against precomputed 
tidal field tensors - is constructed for use in the Bridge integrator.

Fig.\,\ref{fig:TreeCodevsDirect} presents the mass and the Lagrangian 
radii of a simulated cluster as a function of time, for the two codes 
Bonsai and ph4.  For most of the simulations in the paper Bonsai, a 
treecode running on the GPU, was used, because the main interest of the 
authors was in the survivability of the cluster, and not in the details 
of the internal structure of the clusters. However, in order to validate 
the use of the tree code, the simulations were tested against runs with 
the direct Hermite N-body code ph4, only changing the core integrator. 
As can be seen in Figure~\ref{fig:TreeCodevsDirect} the difference in 
mass evolution between ph4 and Bonsai remains quite small until about 
5Gyr. After this moment, both clusters go into core collapse. Until 
about 8.5Gyr, the ph4 cluster displays much higher mass loss than the 
Bonsai cluster as it expands following core collapse. After 8.5Gyr, both 
codes again show similar behaviour. The Lagrangian radii of the clusters 
are nearly equal until core collapse occurs at about 5Gyr. After this, 
the core collapse is initially deeper in ph4, while after 8.5Gyr Bonsai 
reaches the same depth. From these results, the authors inferred that 
the Bonsai simulations were not as well suited for determining the 
internal structure evolution of the star clusters as ph4 would be. 
However, the effect of the external cosmological tidal field remains 
largely unchanged between ph4 and Bonsai. Since the main interest in the 
study was the mass evolution of the clusters due to the tidal field, 
Bonsai was considered adequate to give an indication of the effect of 
tidal fields on the cluster mass loss.

\subsection{Embedded cluster evolution\label{sec:embed}}

\begin{figure*}
 \centering
 \epsfig{file=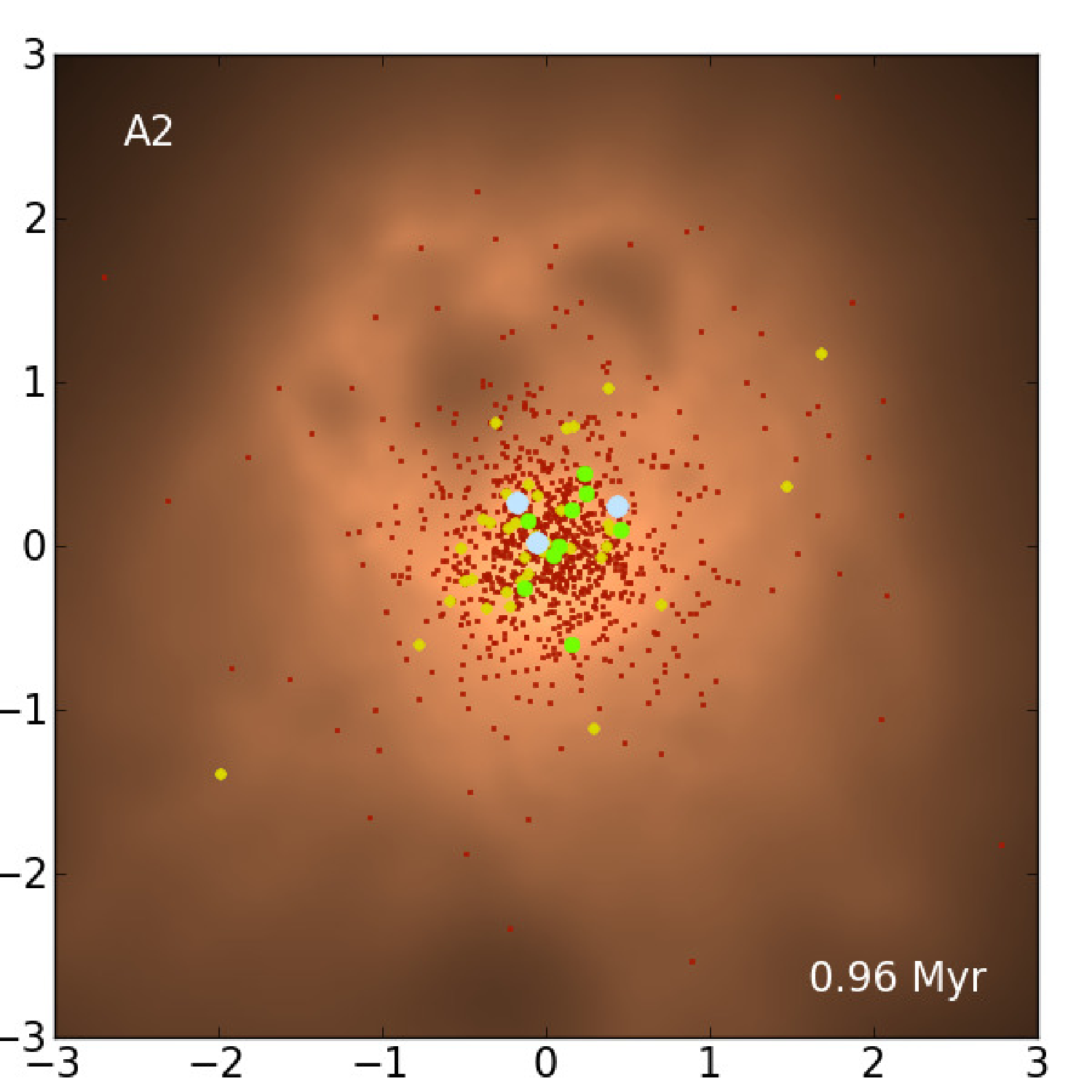, height=.39\textwidth}
 \epsfig{file=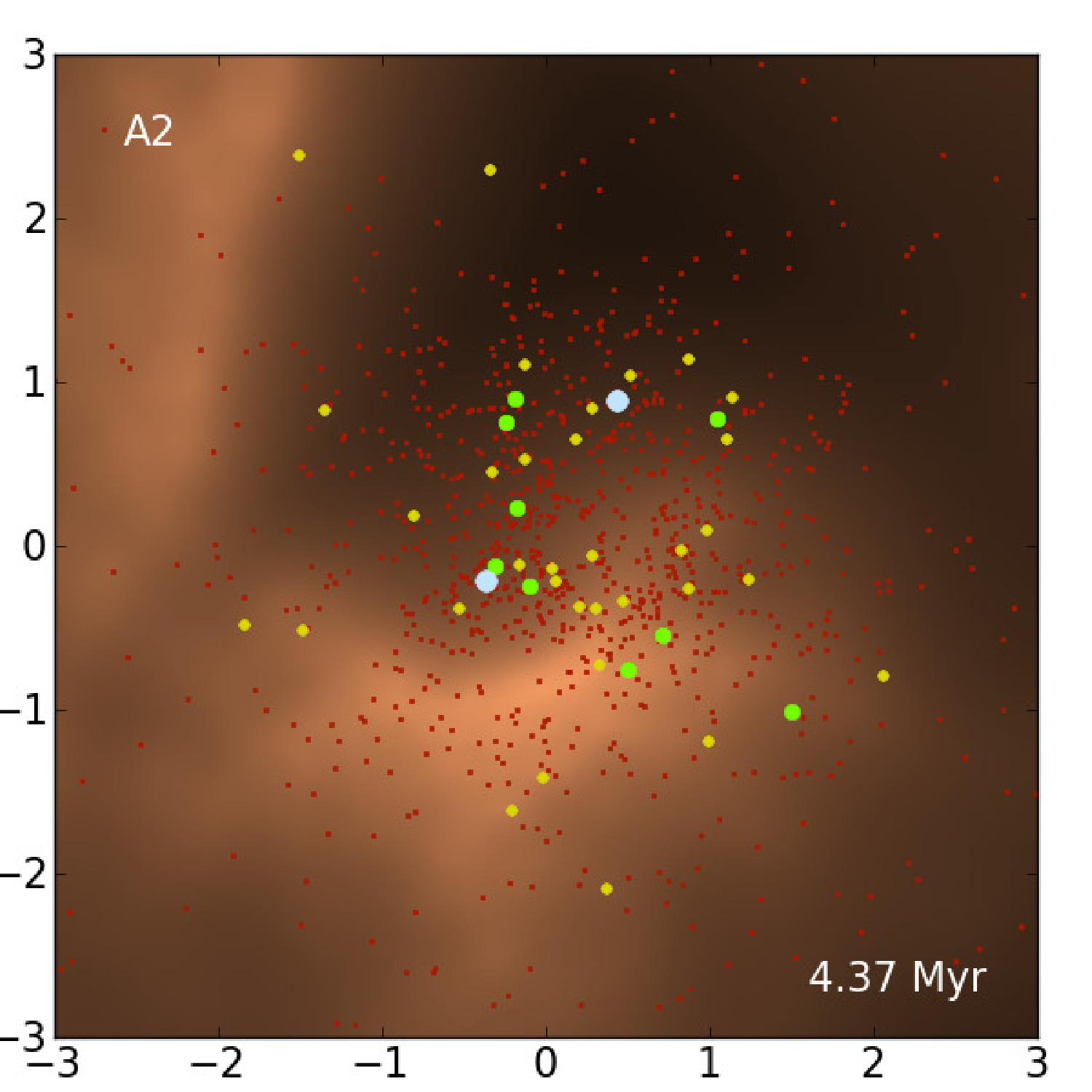, height=.39\textwidth}
 \epsfig{file=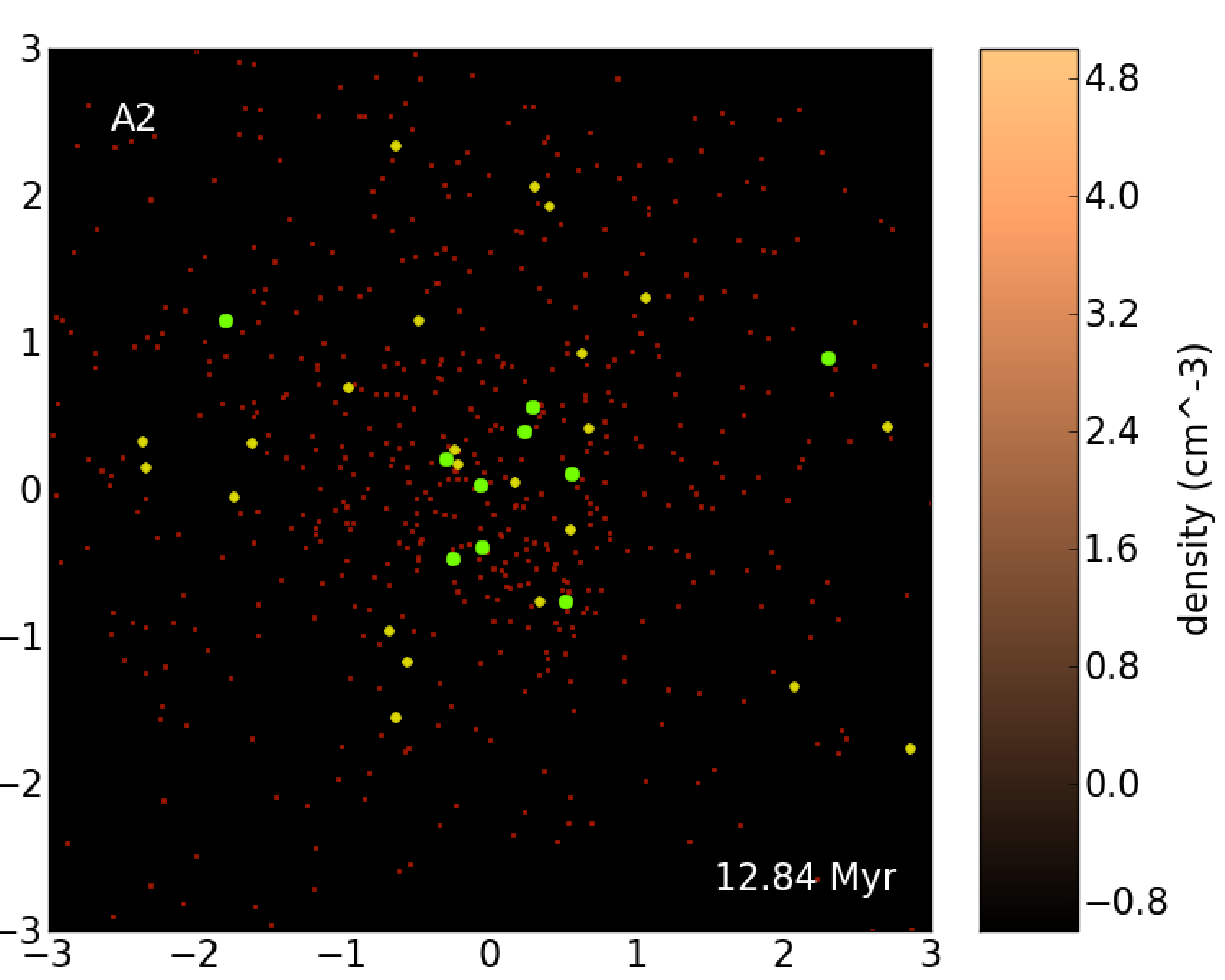, height=.39\textwidth}
 \caption{
Evolution of the stellar and gas distribution of an embedded cluster, 
including stellar evolution and feedback from stellar winds and 
supernovae. Snapshots are labelled with their time 
in the lower right corner. Shown as a density plot is a slice through 
the midplane of the gas density. Stars are divided in 4 mass bins: 
$m_\star \le 0.9 \Msun$  (smallest red dots), $0.9 \Msun \le m_\star \le 
2.5 \Msun $  (intermediate yellow dots), $2.5 \Msun \le m_\star \le 10 
\Msun$ (intermediate green dots) and $m_\star \ge 10 \Msun$ (large light 
blue dots). From \cite{Pelupessy2012}.
}
 \label{fig:clustergas}
\end{figure*}

Stellar clusters form embedded in the natal cloud of gas with star 
formation efficiencies (the fraction of mass that end up in stars) of 
about 5\%-30\% \citep{Williams1997}. The energy output from all the 
massive stars typically exceeds the total binding energy of the embedded 
star cluster. The loss of gas from the embedded proto-cluster may 
cause the young cluster to dissolve~\citep{Hills1980, Lada1984}. The 
survivability of the cluster depends on the efficiency with which the 
radiative, thermal and mechanical energy of the stellar outflows couple 
to the inter-cluster gas, and these are determined by stellar evolution 
processes. Therefore in order to fully represent the evolution from its 
embedded to a gas free and more evolved state one needs to take into 
account various physical ingredients. This problem was examined by \cite
{Pelupessy2012}. AMUSE was used because of the need of combining gas 
dynamics with high precision N-body dynamics and stellar evolution. The 
combined solver described in section~\ref{sec:gravgassse} was developed 
for this purpose. The resulting script consists of a number of 
components that can be reused in other settings (see fig.~\ref
{fig_int}), and are also flexible in the choice of core integrators. In 
this case the hydrodynamics of the gas was calculated using an SPH code 
(Gadget), gravitational dynamics was calculated using a 4th order 
Hermite N-body solver (PhiGRAPE) while the cross coupling was done using 
a tree- gravity solver (Octgrav). The mechanical luminosity of the stars 
was calculated using the results of the stellar evolution code SSE, 
using the stellar radii and temperatures and empyrical relations for the 
terminal wind velocities \citep {Leitherer1992, Prinja1990}, this - 
together with the supernova energy and mass loss determined the energy 
and mass injection into the cluster medium. Figure~\ref{fig:clustergas} 
shows snapshots of an example simulation (model A2 of \cite 
{Pelupessy2012}). 

\subsection{Circumbinary disks}

\begin{figure}
 \centering
 \epsfig{file=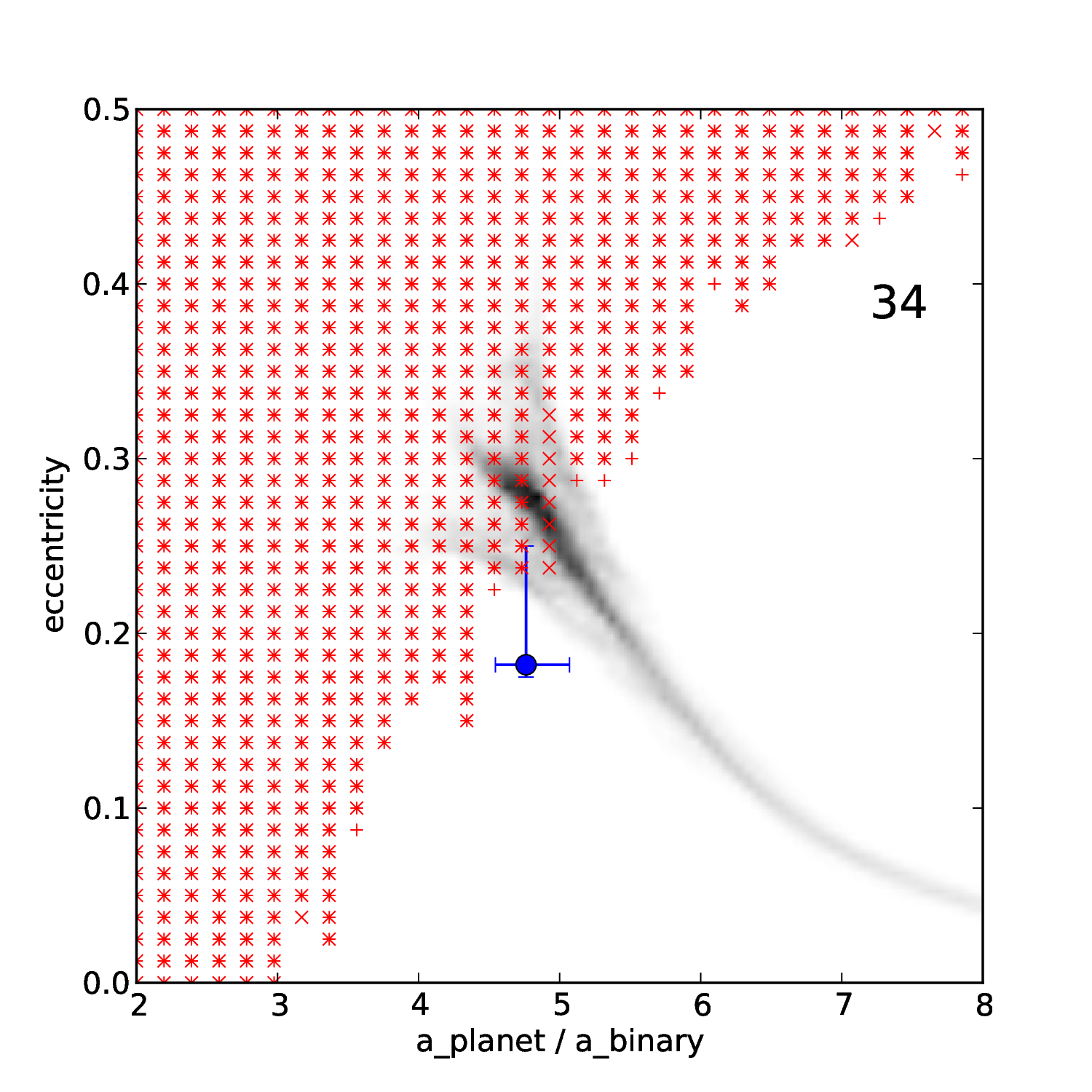, width=.49\textwidth}
 \caption{
 Circumbinary disk simulation for a model matching the parameters of 
 Kepler 34, from \cite{Pelupessy2013}. Shown is the distribution of gas 
 in the semi-major axis - eccentricity plane with the orbital parameters 
 of the observed planet around Kepler 34 as the blue dot (with the range 
 of semi-major axis and eccentricity encountered during long term 
 integrations given by the error bars). The stability of planetary 
 orbits is indicated by the red crosses and plusses, where these 
 indicate the unstable models from a grid of models calculated with AMUSE.
}
 \label{fig:circumbin}
\end{figure}

\cite{Pelupessy2013} examined the formation of planets around binary 
stars. This study used AMUSE to answer two different questions, namely: 
what is the hydrodynamic reaction of a disk centered on the components 
of a binary and secondly what is the (approximate) region of stability 
(in semi-major axis - eccentricity space) for planets around 
circumbinary systems. For the hydrodynamic simulations the binary 
parameters were chosen to match a set of recently detected binary 
systems with planetary systems orbiting both components~\citep{Doyle2011, 
Welsh2012} and for a set chosen from a survey of eclipsing binaries 
\citep{Devor2008}. The gas turned out to settle in a disk with increasing 
eccentricity towards the center of the disk (fig.~\ref{fig:circumbin}). 
To explore the question whether planets formed from the gas disk would be 
in stable orbits, a grid of three-body models was run for each system, 
where for varying planetary semi-major axis, eccentricity and 
pericentric angle a three-body initial condition was integrated for 1 
Myr using a high precision N-body code (Huayno). These runs are also 
summarized for Kepler 34 in figure~\ref{fig:circumbin}. Each of these 
models take a non-trivial amount of time (in the order of 30 min). In 
order to speed up the calculation these jobs where farmed out over 
different work-stations using the AMUSE remote running functionality 
(see section \ref{sec:jobserver}).

\section{Discussion\label{sec:disc}}

The AMUSE framework provides an environment for conducting multi-physics 
computational experiments in Astrophysics. We have presented the design, 
the core modules and an overview of the different coupling 
strategies that can be employed within AMUSE. We have presented 
applications of the AMUSE framework that show different aspects of the 
usage of the framework. AMUSE is an easy platform for computational experiments,
where the barrier of using different codes is lowered by standardizing IO 
and the calling sequence. Coupling different codes and physical processes or 
regimes is made transparent by automatic unit conversion and restricting 
the interface communication to physically relevant quantities. Finally,
the remote callability of the interfaces allows for scalable computing.

An additional important aspect of AMUSE is that it encourages 
reproducibility in computational experiments: the source of the 
framework and the community codes are distributed in an open source 
package. The scripts describing the computational experiments can be 
easily communicated and by allowing to easily change between different 
solvers and numerical methods it enables routine cross verification of 
calculations. 

The current design of AMUSE has a number of limitations which must be 
kept in mind when using the framework and designing numerical 
experiments. Currently AMUSE uses a centralized message system, where 
all the communication passes through and is initiated by the master user 
script. This means that ultimately the scripts will be limited by the 
communication bandwidth of the machine where the user script is started. 
Even if the communication only concerns, for example, a gravity and a 
hydro-code started on different machines exchanging information, this 
communication has to pass through the master script's machine, and will 
block the script from performing other taks. This can be mitigated in 
the current design by encapsulating such tasks and sending them off as 
one unit and by using threading in the master script. A different 
solution we will explore in the future is to implement communication 
channels using tangential communication paths. The fact that the 
communication is always initiated by the master script imposes a 
limitation on the algorithms implemented in AMUSE. It is for example 
difficult to implement a coupling using adaptive individual 
time-stepping. In the future, a call-back mechanism, whereby a community 
code can put in requests for communication, could solve this.

In spite of these limitations there is a large class of problems to 
which AMUSE can be applied productively. It is instructive to compare 
the AMUSE approach to coupling codes with two conventional ways in which 
codes are coupled: namely using simple command line scripting, using 
e.g. UNIX pipes, and native implementations where codes are coupled on 
the source code level. Command line scripting is used often in 
astrophysical simulations, and mostly for input/output coupling. Looking 
back at the other couplings in section~\ref {sec:couple}, it is clear 
that these are progressively more cumbersome to implement using pure 
command line scripting, needing more and more ad-hoc constructs. On the 
other hand, writing a native solver is the preferred method for tightly 
coupled physics. However as one progresses towards more loosely coupled 
domains this becomes more cumbersome from the viewpoint of modularity 
and flexibility.

\subsection{Testing and verification\label{sec:test}}

The source code includes more than 2065 automatic tests. These unit 
tests test aspects of the core framework as well as the community 
interfaces. Coverage of the tests (the fraction of the code that is 
executed during the tests) is 80\%. The source code is maintained 
in an SVN\footnote{Apache Subversion, \texttt{subversion.apache.org}} 
repository and the test suite is executed for every change in the 
source code base, and daily on a selection of (virtual) test machines. The 
tests cover the base framework code, support libraries and community code 
interfaces, where for the latter the base functionality of the interfaces is 
checked.

In addition, to assess the utility of AMUSE for realistic scientific 
problems, the code is also validated against published results. Most of 
the framework code in AMUSE can be tested with the unit tests, but these 
only test if a small part of the code works as planned. The unit test do 
not tell the user anything about the accuracy or the validity of an 
AMUSE solver. Validation against existing problems provides more 
insight. For these validation tests we select test problems which only 
use physics implemented in AMUSE, provide a challenging test of one or 
more modules, and are well defined and possible to implement using 
easily generated data and initial conditions, but not too expensive 
computationally. Examples of such test problems are: 
\cite{Bonnell2003}: cluster formation, \cite{Fujii2007}: 
galaxy/ cluster interaction, \cite{Glebbeek2009}: evolution of 
runaway stellar collision products, \cite{Iliev2009}: radiative
comparison tests, \cite{Mellema2002}: cloud$/$shock interactions.

Finally, we want to caution that the above tests do not mean that 
the framework will give correct and meaningful results under all 
circumstances. Developing a new application for AMUSE entails careful 
testing, especially when new physical regimes are probed.

\subsection{Performance\label{sec:perf}}

A general statement about the performance of AMUSE is difficult to make, 
since such a wide variety of types of numerical experiments can be 
conducted. The design of AMUSE does take into account efficiency, using 
array operations as much as possible for communication, basing storage 
in Python on Numpy arrays and minimizing the use of relatively expensive 
data structures (such as Python dictionaries). The community solvers on 
the other hand are usually well optimised, and a large class of problems 
spend only a tiny fraction of their time in the framework. In these 
cases the performance of the framework is certainly not a concern. 
Experiments in section~\ref{sec:bridge} \cite[and also 
in][]{PortegiesZwart2013b} show that even in relatively tightly coupled 
problems, where intensive communication is needed between the component 
solvers, good performance (compared to monolithic solvers) is possible. 
As we mentioned above however, the current design does impose 
limitations for very large problems.

\subsection{Extending AMUSE\label{sec:extend}}

The main framework and community modules are production ready, however 
AMUSE is foreseen to grow over time with new codes and capabilities. 
AMUSE is freely downloadable and can easily be adapted for private use 
or extensions may be submitted to the repository, a practice we want to 
encourage by providing a mechanism whereby at the end of an AMUSE script 
the framework provides the references for the community codes used.

In our experience writing an interface to a new code, which also 
involves writing tests and testing and debugging the interface, 
represents a modest amount work. While every code is different and has 
its own peculiarities, it is typically something that can be finished 
during a short working visit or small workshop. Defining an interface 
for a new domain can take longer, as these need refinement over time.

\section{Conclusions\label{sec:concl}}

We present the Astrophysical Multipurpose Software Environment (AMUSE). 
AMUSE as an environment presents a wide variety of astrophysical codes 
using homogeneous interfaces, simplifying their use. It allows the 
design of computational experiments with multiple domains of physics 
spanning a wide range of physical scales using heterogeneous computing 
resources. It allows for rapid development, encouraging the creation of 
simple scripts, but its core design is suitable for deployment in high 
performance computing environments. It enables cross verification of 
simulations by allowing experiments to be repeated trivially easy with 
different codes of similar design, or using wholly different numerical 
methods. It fosters reproducibility in numerical experiments by reducing 
complicated coding to self-contained and easily portable scripts. 

\ \\ \\
{\bf acknowledgements}
This work was supported by the Netherlands Research Council NWO 
(grants \#643.200.503, \#639.073.803 and \#614.061.608) and by 
the Netherlands Research School for Astronomy (NOVA). We would like 
to thank the following people for contributing figures:
J\"urgen J\"anes (figure~\ref{fig:se_fallback}), Steven Rieder 
(figure~\ref{fig:TreeCodevsDirect}) and Nicola Clementel 
(figure~\ref{fig:iliev2}).

\bibliographystyle{aa}
\bibliography{ms}

\end{document}